\documentclass[11pt, a4paper, nofootinbib,aps,showpacs]{revtex4-1}

 \usepackage[bottom]{footmisc}

\usepackage[T1]{fontenc}
\usepackage[latin9]{inputenc}
\usepackage{mathrsfs}
\usepackage{bm}
\usepackage{amsmath}
\usepackage{amsthm}
\usepackage{amssymb}
\usepackage{stmaryrd}
\usepackage{mathtools}

\edef\ordinarycolon{\mathchar\the\mathcode`: }
\edef\ordinaryequals{\mathchar\the\mathcode`= }

\usepackage{breqn}
\catcode`^=7

\AtBeginDocument{%
  \catcode`^=12
  \def\coloneqq{%
    \mathrel{{\mathop\ordinarycolon}\mkern-1.2mu{\ordinaryequals}}%
  }%
}

\makeatletter

\allowdisplaybreaks

\let\cat@comma@active\@empty

\usepackage[capitalize]{cleveref}

\usepackage{color}
\RequirePackage[dvipsnames,usenames]{xcolor}

\DeclareMathOperator*{\argmin}{arg\,min}

\DeclareMathOperator*{\supp}{supp}

\newif\ifnotes
\notestrue

\ifnotes
 \newcommand{\dhwc}[1]{{\color{Red}{\bf{DHW COMMENT: #1}}}}   

 \newcommand{\ak}[1]{{\color{Orange}{{AK: #1}}}}
\else
 \newcommand{\dhwc}[1]{ }
 \newcommand{\dhws}[1]{ }
 \newcommand{\ak}[1]{ }
\fi

\newcommand{\todoPostArxiv}[1]{}

\renewcommand{\mid}{\vert}

\makeatother

\newcommand{\ba}{\begin{eqnarray}}
\newcommand{\ea}{\end{eqnarray}}
\newcommand{\eqn}[1]{\begin{align*}#1\end{align*}}
\newcommand{\eq}[1]{\begin{align}#1\end{align}}
\newcommand{\erf}[1]{Eq.~\eqref{#1}}

\newcommand{\p}{p}
\renewcommand{\r}{r}
\newcommand{\rr}{r}
\newcommand{\map}{\pi}
\newcommand{\Map}{\pi}

\newcommand{\Cin}{{\mathrm{IN}}}
\newcommand{\Cout}{{\mathrm{OUT}}}

\newcommand{\W}{{\mathcal{Q}}}

\newcommand{\EF}{{\mathcal{E}}}
\newcommand{\R}{{\mathbb{R}}}
\newcommand{\B}{{\mathbb{B}}}

\newcommand{\Z}{{\mathbb{Z}}}

\newcommand{\PP}{{\proc}}
\newcommand{\CC}{{C}}
\newcommand{\KK}{{K}}
\newcommand{\WWW}{{\mathcal{Q}}}
\newcommand{\WW}{{\mathcal{L}}}
\newcommand{\cs}[1]{\mathsf{#1}}   

\newcommand{\vl}{{\vec{\lambda}}}

\newcommand{\hn}{{{m}}}

\newcommand{\subWW}{\WW}
\newcommand{\EPbase}{\sigma}
\newcommand{\EP}[1]{\EPbase_{#1}}
\newcommand{\EPmin}[1]{{\EPbase^{min}_{#1}}}
\newcommand{\subEP}[1]{{\EPbase}_{#1}}

\newcommand{\subEPmin}[1]{{\EPbase^{min}_{#1}}}

\newcommand{\subEF}[1]{\EF_{#1}}

\newcommand{\qcPP}{q^{c}}

\newcommand{\DDbase}{D}
\newcommand{\KKbase}{S}
\newcommand{\IIbase}{\mathcal{I}}

\newcommand{\SSS}{S}
\newcommand{\DD}{D}

\newcommand{\KKf}[2]{\KKbase(#1\Vert #2)}
\newcommand{\DDf}[2]{\DDbase(#1\Vert #2)}
\newcommand{\DDb}[2]{\DDbase\big(#1\Vert #2\big)}
\newcommand{\II}{\IIbase}

\newcommand{\IIKKf}[2]{\II(#1\Vert #2)}
\newcommand{\IIDDf}[2]{\mathcal{D}(#1\Vert #2)}
\newcommand{\ktlntwo}{k_B T \ln 2}

\newcommand{\pa}{\mathrm{pa}}
\newcommand{\pag}{{\pa(g)}}
\newcommand{\ppag}{p_\pag}
\newcommand{\qpag}{q_\pag}

\newcommand{\proc}{}

\newtheorem{theorem}{Theorem}

\newtheorem{corollary}[theorem]{Corollary}

\newtheorem{definition}{Definition}
\newtheorem{example}{Example}

\newtheorem{proposition}[theorem]{Proposition}

\begin{document}

\title{The Stochastic Thermodynamics of Computation}

\author{David H. Wolpert}

\affiliation{Santa Fe Institute, Santa Fe, New Mexico \\
Complexity Science Hub, Vienna\\
Arizona State University, Tempe, Arizona,\\
\tt{http://davidwolpert.weebly.com}}

%

\begin{abstract}
One of the central concerns of computer science is how the resources needed to perform a given computation depend on that computation. Moreover, one of the major resource requirements of
computers --- ranging
from biological cells to human brains to high-performance (engineered) computers ---
is the energy used to run them, i.e., the thermodynamic costs of running them. 
%
Those thermodynamic costs of performing a computation have been a long-standing focus of research in physics, going back (at least) to the early work of Landauer, 
in which he argued that the thermodynamic 
cost of erasing a bit in any physical system is at least $kT \ln[2]$.

One of the most prominent aspects of computers is that they are inherently \textit{non}-equilibrium
systems. However, the research by Landauer and co-workers 
was done when non-equilibrium statistical physics 
was still in its infancy, requiring them to rely on equilibrium statistical physics.
This limited the breadth of issues this early research could address, leading them to
focus on the number of times a bit is erased during a computation --- an 
issue having little bearing on the central concerns of computer science theorists.
 
Since then there have been major breakthroughs in nonequilibrium statistical physics,
leading in particular to the new subfield of ``stochastic thermodynamics''.
These breakthroughs have allowed 
us to put the thermodynamic analysis of bit erasure on a fully formal (nonequilibrium)
 footing. They are also 
allowing us to investigate the myriad aspects of the
relationship between statistical physics and
computation, extending well beyond the issue of how much work is required to erase a bit.

In this paper I review some of this recent work on the ``stochastic thermodynamics
of computation''.
After reviewing the salient parts of information theory, 
computer science theory, and stochastic thermodynamics, 
I summarize what has been learned about the entropic costs of performing a broad
range of computations, extending from bit erasure to loop-free circuits to logically reversible circuits
to information ratchets to Turing machines.
These results reveal new, challenging engineering problems for how to design
computers to have minimal thermodynamic costs. They also
allow us to start to combine computer science theory and stochastic thermodynamics
at a foundational level, thereby expanding both.

\end{abstract}

\maketitle

%
%
%

\nocite{demaine2016energy,bennett1989time}

\section{Introduction}

\subsection{Computers ranging from cells to chips to human brains}

Paraphrasing Landauer, real world computation involves thermodynamic costs~\cite{landauer1991information} --- and
those costs can be massive. Estimates are that computing accounts
for $\sim 4\%$ of current total US energy usage in 2018 --- and all of that energy
ultimately goes into waste heat.

Such major thermodynamic costs arise in naturally occurring, biological computers
as well as the artificial computers we have constructed. Indeed,
the comparison of thermodynamic costs in artificial and biological computers
can be fascinating. For example, a large fraction of the energy budget of a cell
goes to translating RNAs into sequences of amino acids (i.e., proteins), in the
cell's ribosome. The thermodynamic efficiency of this computation --- the amount
of heat produced by a ribosome per elementary operation  --- is many orders of magnitude
superior to the thermodynamic efficiency of my current artificial
computers~\cite{kempes2017thermodynamic}. Are
there ``tricks'' that cells use to reduce their costs of computation
that we could exploit in our artificial computers?

More speculatively, the human brain consumes $\sim 10-20\%$  of all the calories
a human being requires to live~\cite{West2017}. 
The need to continually gather all those extra calories is a massive evolutionary
fitness cost that our ancestors faced. 
Does this fitness cost explain why human levels of intelligence
(i.e., of neurobiological computation) are so rare in the history of life on earth?

These examples involving both artificial and biological computers
hint at the deep connections between computation and statistical physics. Indeed,
the relationship between computation and statistical physics has arguably been
a major focus of physics research at least since Maxwell's demon was introduced. 
A particularly important advance was made in the middle of the last century
when Landauer (and then Bennett) famously argued 
that erasing a bit in a computation results in a entropic cost of at least $kT \ln[2]$, 
where $k_B$ is Boltzmann's constant and $T$ is the temperature of a heat bath connected
to the system.

This early work was grounded in the tools of equilibrium
statistical physics. However, computers are highly \textit{non}equilbrium
systems. As a result, this early work was necessarily semiformal,
and there were many questions it could not address.

On the other hand, in the last few decades there have been major breakthroughs 
in nonequilibrium statistical physics.  Some of the most important
of these breakthroughs now allow us to analyze the thermodynamic behavior of any system 
that can be modeled with a time-inhomogeneous continuous-time
Markov chain (CTMC), even if it is open, arbitrarily
far from equilibrium, and undergoing arbitrary external driving. 
In particular, we can now decompose the time-derivative of the (Shannon) entropy of such a system into 
an ``entropy production rate'', quantifying the rate of change of the total entropy
of the system and its environment, minus a ``entropy flow rate'', quantifying the 
rate of entropy exiting the system into its environment. Crucially, the entropy production
rate is non-negative, regardless of the CTMC. So if it ever adds a nonzero amount to system
entropy, its subsequent evolution cannot undo that increase in entropy. (For this reason 
it is sometimes referred to as \textit{irreversible} entropy production.) This
is the modern understanding of the second law of thermodynamics, for systems undergoing
Markovian dynamics. In contrast to entropy production,
entropy flow can be negative or positive. So even if entropy flow increases system entropy
during one time interval (i.e., entropy flows into the system), 
often its subsequent evolution can undo that increase.

This decomposition of the rate of change of entropy is the starting point
for the field of ``stochastic thermodynamics''. However, the decomposition
actually holds even in scenarios that have nothing to do with
thermodynamics, where there are no heat baths, work reservoirs, or the like coupled to the system.
Accordingly, I will refer to the dynamics of the entropy production rate,
entropy flow rate, etc. as the ``entropy dynamics'' of the physical system. In addition,
integrating this decomposition over a non-zero time interval, we see that the
total change in entropy of the system equals the total entropy flow plus the total
entropy production. Since entropy production is non-negative, this means that the minimal
value of total entropy flow is the total change in entropy, which is sometimes
called the ``Landauer cost''.\footnote{While I will focus on CTMCs in this review,
the reader should be aware that nonequilibrium statistical physics extends beyond
Markovian systems. See for example~\cite{parrondo2015thermodynamics} and references
therein, in which there is no restriction to Markovian systems.}

\subsection{The effect of constraints on how we are allowed to perform a computation}

These recently derived equations for the entropy dynamics of physical systems allow us to revisit the entire topic that was historically referred to as
``the thermodynamics of computation'', but now in a richer, and fully formal manner.
For example, as discussed in \cref{sec:bit_erasure} in the case of bit erasure,
we now understand that any function $ f : X \subseteq \Z \rightarrow X$
can be implemented in a thermodynamically
reversible manner, with no (irreversible) entropy production. Importantly, this thermodynamic
reversibility holds even if $f$ is logically irreversible (as bit erasure is).
This result holds for functions ranging from bit erasure, where $X = \{0, 1\}$) 
(see \cref{sec:bit_erasure}), to
simple algebraic functions like $y = x \mod 3$ (where $X = \Z$)
to the input-output function of finite deterministic automaton (where $X$
is the joint state of all allowed input strings to the automaton and
the internal state of the automaton). It even holds
for partial functions representing the input-output
behavior of universal Turing machines, in which $X$ encodes
the set of all finite input bit strings.\footnote{Recall that a map from $X \rightarrow Y$
is a \textbf{total} function if it is defined for all of its inputs, and otherwise it is
a \textbf{partial} function.}  
Importantly, this thermodynamic reversibility holds even if f
is logically irreversible (as bit erasure is).

These results are now well understood. However, they all assume that
we can use an unconstrained physical system to implement $f$, with no
restrictions on how the Hamiltonian couples the components of the full system-state, $x$.
The richness of the current research on the thermodynamics of computation 
arises from the fact that if we wish to implement the function $f$ in a 
real-world physical system, then \textit{for
practical reasons}, we rarely have the freedom to exploit an arbitrary Hamiltonian,
that couples the components of $x$ in an arbitrary manner.
Instead, in practice $f$ must be physically implemented by iterating one or 
more interconnected functions, $\{g_i\}$, each of which is constrained to lie in some highly restricted set of functions.

One example of this is where $f$ is the function implemented by 
a particular Boolean circuit and the $g_i$ are Boolean gates in the circuit. Another
example is where $f$ is the function implemented by 
a particular deterministic finite automaton. In this example there is a single
$g_i$, specifying how the internal state of the automaton is updated based
on its current state and the current input character, and that function is iterated
using the successive characters of the input string until the end of the string is reached.
The crucial point is that that $g_i$ can only couple a very limited number of the components
of the full input string $x$ at a given time.

Most of conventional computational complexity theory is,
in essence, the investigation of how the constraints on the functions $\{g_i\}$ affect
the capabilities and resource demands of the function $f$ constructed
from those functions~\cite{moore2011nature,arora2009computational,savage1998models,hopcroft2000jd,sipser2006introduction}. 
In particular, in general for any given function $f$ we want to implement, and any fixed
set of functions $\{g_i\}$ we are allowed to couple together to implement $f$, 
there are many (typically infinitely
many) different ways of using those $\{g_i\}$ to implement $f$. For example,
in general, for any fixed function $f$ from one finite space to
another, there are an infinite number of circuits, all using
the same set of gates $\{g_i\}$, that all implement $f$. However, 
there are hard constraints on the relationship among 
the various resources  (e.g., the number of gates,
the maximal depth through the circuit from the inputs to the outputs, etc.)
that each of those circuits requires. A central concern
of computational complexity theory is understanding those constraints
on the resource requirements, and how they changes as we vary $f$.

Similarly, many of the interesting issues in the thermodynamics
of computation concern how the constraints on the functions $\{g_i\}$ affect the \textit{thermodynamic} resource demands of any physical system  constructed
from those functions (in other words, their entropic costs). In the words of~\cite{Boyd2018thesis},
``different implementations of the same computation (input to output) can have radically different thermodynamic consequences''. 
We can illustrate such consequences by again considering the fact that in general
there are an infinite number of circuits, all using
the same set of gates $\{g_i\}$, that implement the same function $f$. In general, for any fixed
distribution over the inputs to the function $f$, those circuits will all differ
in both their Landauer costs and their entropy production. 
This results in a highly nontrivial optimization problem, of finding
the circuit with least entropic costs to perform a given function $f$.

This shared concern with constraints in both computational complexity theory and the thermodynamics of 
computation suggests that we can partially synthesize the two fields. In particular,
it may be possible to relate the resource demands they separately investigate to each other,
both in general, and in the context of particular computational machines
more specifically.

An example of what such a synthesis might look like, described below in 
\cref{sec:TM_me_old}, involves the 
entropy dynamics of Turing machines. Loosely speaking,
the Kolmogorov complexity of a given bit string $\sigma$, with respect to a given Turing
machine $M$, is defined as the shortest length of any bit string $x$ that we can input 
to $M$ that results in $M$ computing the output $\sigma$ and then halting. 
Similarly, we can define the ``Kolmogorov
work'' of $\sigma$ as the smallest amount of \textit{thermodynamic work}, $W$, such that 
there is some input string $x$ 
to $M$ that results in $M$ computing $\sigma$ while requiring work $W$. In general,
both the Kolmogorov complexity and
the Kolmogorov work of $\sigma$ will vary depending on the Turing machine one
uses to compute $\sigma$. However, it turns out that for any Turing machine $M$, 
the Kolmogorov complexity for $M$
of computing $\sigma$ and the Kolmogorov work for $M$ of computing $\sigma$ have a simple
(though uncomputable!) relationship with one another.

%
%

\subsection{Roadmap}

I begin with background sections, first presenting notational
conventions, then reviewing the relevant
parts of information theory, and then reviewing relevant computer science theory.
These are followed by a section that summarizes the modern understanding
of the dynamics of entropy in arbitrary dynamical systems. I then briefly
discuss the modern understanding of the relationship between logical
and thermodynamic (ir)reversibility. (In short, the laws of physics 
no more forbid the erasure of 
a bit in a thermodynamically reversible manner than they
forbid the compression of a piston of gas in a thermodynamically reversible manner.)

Computers, by definition, are open systems. Accordingly,
``where one draws the dotted line'', saying which precise
physical variables are the ``computer'', and which are not, can affect the entropic costs
one ascribes to the computer. In particular, in order to be able to use
a computer multiple times, and have both the dynamics of the logical variables and
the  thermodynamic variables behave the same way each time one runs the computer,
we need to reinitialize variables internal to the computer when it finishes its computation.
Accordingly, we need to carefully specify how the entropic costs associated
with the reinitialization of variables are allocated among the computer and
the offboard systems interacting with the computer.
(Some of the primary sources of confusion in the literature arise with papers that
are not fully explicit about how they do this.) 
The convention I adopt in this paper for how to do this is
presented in \cref{sec:accounting_section}.
%
%
%
%
%
%

The following sections consider the entropic costs of various
information processing systems. These sections are ordered from the simplest forms of information
processing to progressively more complicated forms. Those who are interested
in the costs of full computational machines, in the computer science sense
of the term, can skip \cref{sec:overwrite_inputs},
which consider simpler kinds of information processing systems, jumping 
from \cref{sec:accounting_section} directly to \cref{sec:subsystem}.


In \cref{sec:overwrite_inputs}, I review the entropic costs 
of a physical process that implements a single, arbitrary conditional distribution 
over its state space, without any constraints on its rate matrices,
Hamiltonians, etc. (Bit erasure is a special example of such a process.)

The analysis to this point in the paper concerns information processing systems
that have no \textit{a priori} constraints on their operating behavior. However, 
subcomponents of full computers only have access to a limited
number of degrees of freedom of the overall computer. These provide major
constraints on how any full computer is allowed to transform its input into
an output. (Indeed, full computers, in the Chomsky hierarchy sense, are \textit{defined} by the
constraints in how they transform their inputs into their outputs.)
General results concerning the relationship between such constraints on
how a subcomponent of a computer can be coupled to the rest of the computer 
on the one hand, and the consequences for the entropy dynamics of the overall
computer on the other hand, are reviewed in \cref{sec:subsystem}.

These results are the foundation of a recent
analysis of the entropy dynamics of ``straight line'' circuits, that have
no loops or branching structure. I review this analysis in \cref{sec:circuit_entropy_dynamics}.

There have been confusing claims in the literature concerning the
possible thermodynamic benefits of implementing a computation using a circuit
made out of logically reversible gates, as opposed to implementing that same
computation using a circuit made out of arbitrary (in general logically irreversible) gates.
In \cref{sec:fredkin}, I apply the convention for
how to allocate entropic costs that was introduced in \cref{sec:accounting_section},
to clarify some of these claims.

%
In \cref{sec:sfa_costs} I review some
work on the Landauer costs of finite automata that only run
once. 


Then in \cref{sec:thermo_mealy} I review some recent
results on the thermodynamic costs of information ratchets, which
can be viewed an extension of what are called ``transducers'' in computer science. 
These results range from analyzing detailed physical models of systems that act as 
information ratchets~\cite{mandal2012work,mandal2013maxwell} 
to a relatively large body of literature on infinite-time limiting behavior
of the entropy dynamics of information ratchets~\cite{boyd2016identifying,boyd2017correlation,boyd2017transient,boyd2017leveraging}.

After this, in \cref{sec:TMs_us} 
I review recent research on the entropy dynamics of Turing Machines (TMs)
that uses modern nonequilibrium statistical physics. I also review some earlier work that
(because it was done before modern nonequilibrium statistical physics was developed)
had to use \textit{equilibrium} statistical physics to analyze TMs.

Either implicitly or explicitly, most of the work in the literature
on the entropy dynamics of computers, stretching back to the early work of Landauer but also including 
modern work, has assumed that a computational machine is implemented 
as the discrete-time dynamics induced by an underlying CTMC, e.g., a master equation. (In particular,
this is the case for all systems analyzed in stochastic thermodynamics). In
\cref{sec:hidden} I discuss
a recently discovered surprising limitation of this assumption. Specifically, any non-trivial 
computation over a set of ``visible'' physical states cannot be implemented
as the discrete-time dynamics induced of \textit{any} underlying CTMC. For such 
``visible'' discrete-time computation to arise, the underlying CTMC must couple the visible states
to a set of ``hidden'' states. In general, the number of hidden
states needed will depend on the computation being implemented.
Moreover, the continuous-time dynamics of the joint visible-hidden
system has to proceed through a nonzero, countable number of successive time-intervals
which are demarcated by the raising or lowering of infinite energy barriers. 
The number of such intervals will depend on both the visible computation being
implemented and the number of hidden states.

In essence, any non-trivial visible computation is embedded in a larger,
hidden computation. 
To give a simple example, any CTMC that implements a simple bit
flip has to have at least one hidden state in addition to the two visible
states of the bit, and requires at least three hidden time-intervals to complete.

In the last section, I present some possible future directions
for research. In particular, I describe a set of ``computer-science-style'' 
open questions that concern the thermodynamic properties of computational machines
rather than the other kinds of resources (amount of memory, number of
timesteps, etc.) traditionally considered in computer science.

%

\subsection{Topics not covered in this paper}

In this paper I will review the literature on the dynamics of entropy in physical
systems that implement the computational machines considered in conventional computer science
theory~\cite{hopcroft2000jd}. 
To clarify the precise topics considered in this paper, it is worth listing some of the
related topics that I do \textit{not} consider. 

I do not consider all computational machines, but only a few of the most common
ones. So for example, I do not consider the thermodynamics of running push-down automata. More broadly,
I do not consider the thermodynamics of running
analog computers (see~\cite{diamantini2016landauer} for work on this topic).

I also do not review any work that 
analyzes abstract dynamical systems (e.g., cellular automata) using mathematical techniques that just
happen to have been developed in the statistical physics literature.
The focus of this paper is not the \textit{mathematical techniques} of statistical
physics, but rather the \textit{actual} statistical physics of physical systems. (See~\cite{baez2012algorithmic}
for some interesting recent work on applying statistical physics
techniques to analyze abstract dynamical systems.) 

In this paper I will not consider the thermodynamics
of quantum mechanical information processing, or indeed any aspect of quantum
computation. See~\cite{nielsen2010quantum} for a good overview
of quantum computation in general,~\cite{strasberg2017quantum,goold2016role}
specifically for overviews of the thermodynamics of quantum information 
processing,~\cite{deffner2017quantum} for a review of quantum speed 
limits, and~\cite{gour2015resource,sparaciari2017resource,krummintroduction} for
work on the related topic of quantum mechanical ``resource theory''.

However, it should be noted that even if
quantum computers become a reality, they may have limited
impact on the thermodynamic costs of real-world computers. One reason
for this is that quantum computers as currently understood could only
reduce the costs (and increase the speed) 
of an extremely limited set of possible computations, like factoring products
of prime numbers. More importantly, many of
the major challenges facing modern computing, both in terms of speed and
thermodynamic costs, do not arise in the operation of the CPU (which is where
quantum computation might be helpful), but rather in the i/o. 
In applications ranging from widely popularized applications of deep learning to search engines
to simple back-office database management, modern computation 
exploits \textit{massive}, heterogeneous data sets. Perhaps the primary
difficulty faced by computers running these applications is in accessing
that data and performing low-level (e.g., cache-based) processing of that data,
not in the elaborate transformation of that data within a CPU. Yet as currently
envisioned, quantum computers would not help reduce the thermodynamic costs
of all this i/o.

Similarly, I do not consider the fundamental computational 
limitations of physical systems in our universe. Some
good discussions of this topic can be found in \cite{barrow2011godel,pour1982noncomputability,moore1990unpredictability,lloyd2000ultimate,Lloyd2017}.

I will also not consider the (copious) literature on the thermodynamics of biochemical processes in cells
that can be viewed as noisy types of information processing.
(See~\cite{ouldridge2017fundamental,ouldridge2017importance,ouldridge2017thermodynamics,sartori2014thermodynamic,sartori_thermo_error_correction2015,mehta2012energetic,mehta2015landauer} for some work on
this topic.) The key criterion for including some literature in the current paper is whether it concerns
thermodynamic properties of one of the types of computational machine considered in
the computer science literature. 
(See~\cite{prohaska2010innovation,bryant2012chromatin,benenson2012biomolecular} for 
some work on the preliminary issue of how closely the standard computational
machines considered in computer science can be modeled with biochemical systems
in the first place, and
see~\cite{angluin2006stably,chen2014deterministic,dong2012bisimulation,soloveichik2008computation,thachuk2012space} 
for work concerned with modeling such
machines with arbitrary chemical reaction networks, irrespective of whether they
arise in naturally occurring biological systems.)

In addition, there have been recent papers, building on~\cite{goldt_seifert_learning2017},
that have analyzed the thermodynamics of systems that
gather information about their noisy environment and build models from
that data, mostly using neural networks. These too are out of scope, since
their focus is not on computation per se.
%
%

Furthermore, I do not consider the joint thermodynamics of control systems
interacting with their environment, since that would require analyzing
the environment as well as the controller, along with their coupling. 
Some papers that analyze the
thermodynamics of control systems interacting with their environment
are~\cite{touchette2004information,touchette_information-theoretic_2000,barato2017thermodynamic,sagawa2008second,sagawa_nonequilibrium_2012,wilming2016second,large2018stochastic,barato2017thermodynamic,gingrich2016near,horowitz2017information}. In addition, of course, 
there is a huge literature on the special type of control system
designed to extract work from an equilibrium heat bath, i.e., the thermodynamics of Maxwell's
demon. Some representative papers on this topic, showing how to use modern
nonequilibrium statistical physics to resolve most (arguably all)
of the mysteries concerning Maxwell's demon, are~\cite{parrondo2015thermodynamics,barato2013autonomous,horowitz_multipartite_2015,mandal2012work}.

In addition, in all the analysis below I identify the \textit{microstates} 
of a physical system with
the logical variables of the associated computation (i.e., with the so-called ``information-bearing
degrees of freedom''~\cite{benn73,benn82}). In real-world computers of course,
it is rarely the case that microstates directly code for logical
variables. Instead the logical variables are typically identified with 
macrostates, usually defined by coarse-graining the space of microstates.
(Some authors prefer to characterize the logical variables as
an intermediate level of coarse-grained ``meso-states''.)
The analysis below can be extended to apply to such scenarios (see~\cite{wolpert_arxiv_beyond_bit_erasure_2015,riechers_thermo_comp_book_2018}
for examples). However, this requires 
extra notational complexity, and in many cases also requires
extra assumptions (concerning the microstate dynamics among the macrostates),
and so is not pursued here.
%

I  focus here on the integration of topics in \textit{theoretical} computer science and 
the theory of nonequilibrium statistical physics. I will not consider the thermodynamic
properties of real-world artificial computers in any detail.
However, it is important to note that several major engineering communities
are working to reduce the entropic cost of such computers. 
At present, these efforts are based on a phenomenological characterization of the thermodynamics
of computation. These have led to techniques like approximate computing (in which
noise is added to the computation to reduce the heat it 
produces~\cite{khadra2017introduction,kugler2015good,palem2014inexactness})
adaptive slowing (in which subcomponents of the computer are slowed down to reduce the heat
they produce) and automatic code reconstruction to minimize energy 
consumption~\cite{lacomis_forrest_my_book_2018} (using techniques similar to those in optimizing compilers). 
It is intriguing to note that at a high level, these macroscopic
phenomena of real-world computers parallel analogous phenomena that are central
to the theoretical analyses reviewed below, even though those analyses concern
behavior on the far smaller energy scale of $k_B T$.

Even within these strong restrictions on the subject matter of this paper,
there are directly relevant results in the literature that are not discussed. 
In particular,~\cite{grochow_wolpert_sigact2018} presents preliminary theorems concerning the extension
of computational complexity theory~\cite{moore2011nature} 
to include thermodynamic costs, in addition to the more conventional costs considered in computer
science. \cite{grochow_wolpert_sigact2018} goes on to introduce a number of nontrivial open questions. That paper
is aimed at the computer science theory community however, and without
a lot more background than what is presented in \cref{sec:comp_machines}
below, it would be opaque to most physicists. This is why that paper is not discussed here. 


\section{Terminology and general notation}
\label{sec:notation}

In this section I introduce some of the terminology and notation I will use. I also
define ``islands'', which will be needed in much of the analysis concerning
entropy production.

\subsection{Notation}

I take the binary values to be $\B \equiv \{0, 1\}$. As usual, for any set $A$, 
$A^{*}$ is the set of all finite strings of elements from $A$.
Given a directed acyclic graph (DAG) $\Gamma$, 
I indicate its root nodes as $R(\Gamma)$. I write the Kronecker delta function as
\eq{
\delta(a, b) =
\begin{cases}
      & \text{1 if }a = b  \\
      & \text{0 otherwise}
\end{cases}
}
So in particular, the vector indexed by values $b$ that equals $1 / 0$ depending on 
whether $b$ equals some particular scalar $a$ is written as $\delta(a, .)$.

I identify $1$ as the Boolean TRUE, and $0$ as FALSE.
In agreement with this, I write the indicator function for any Boolean function $f(z)$ as
\eq{
{\bf{I}}(f) =
\begin{cases}
      & \text{1 if }f(z) = 1   \\
      & \text{0 otherwise}
\end{cases}
}

I write $\ell(\vec{\lambda})$ (or sometimes $|\vec{\lambda} |$) to 
mean the length of any finite
string $\vec{\lambda}$. So given any integer $m$, $P(\ell(\vec{\Lambda}) = m)$ means
$P(\vec{\lambda} : \ell(\vec{\lambda}) = m)$. As usual, the concatenation
of a string $\vec{\lambda}'$ after a string $\vec{\lambda}$ is written
as $\vec{\lambda} \vec{\lambda}'$. In addition, I use the common notation
that $\lceil \cdot \rceil$ and
$\lfloor \cdot \rfloor$ are the ceiling and floor operators, respectively.

In general, random variables are written with upper
case letters, and instances of those random variables are written with the corresponding lower case letters. When the context makes the meaning clear, I will
often also use the upper case letter indicating a random variable, e.g., $X$, 
to indicate the set of possible outcomes of that random variable.
For any distribution $\p(x)$ defined over the set $X$,
and any $X'\subseteq X$, I write $\p(X') = \sum_{x\in X'} \p(x)$.
Given some set $X$, I write the set of all distributions over the elements of $X$ whose support is restricted to some subset $X' \subseteq X$ as $\Delta_{X'}$.
Finally, given any conditional distribution $\map(y \mid x)$ and a distribution $\p$ over
$X$, I write $\map \p$ for the distribution over $Y$
induced by $\map$ and $\p$,
\eq{
(\map \p)(y) := \sum_{x\in X} \map(y\vert x) \p(x)
\label{eq:matrixnotation}
}

\subsection{Islands}
\label{sec:islands}

Computational machines are most often defined in terms of single-valued, logically-irreversible state-update
functions. Often that update function $f$ operating over the entire domain $X$ can 
be expressed as a set of
distinct functions $\{f_i\}$, ``operating in parallel'', each with its own distinct domain and 
distinct image. Intuitively, one might expect that one should 
analyze the entropic costs of such an update function $f$ by appropriately averaging
the entropic costs of each of the distinct parallel functions $\{f_i\}$.
That is indeed the case, as elaborated below. In this subsection I introduce some of the associated terminology.


Suppose that a given physical process implements a single-valued
function $f$ over a space $X$. The \textbf{islands}
of $f$ are defined as the pre-images of $f$, i.e., the sets $\{f^{-1}(x) : x \in X\}$~\cite{wolpert2018exact}.
I will write the set of islands of a function $f$ as $L(f)$.
%
%
%
As an example, the logical AND operation,
\[
\map(c \vert a,b) = \delta(c, a \wedge b)
\]
has two islands, corresponding to $(a,b) \in \{\{0,0\},\allowbreak \{0,1\},\allowbreak \{1,0\}\}$ and $(a,b) \in \{\{1,1\} \}$, respectively.
I write the set of all distributions over the elements of
an island $c \in L(f)$ as $\Delta_c$. I make the obvious definitions that
for any distribution $p(x)$ and any $c \in L(f)$, the associated distribution
over islands is
$p(c)= \sum_{x\in c} p(x)$. As shorthand, I also write 
$p^c(x) = p(x\vert X\in c) = p(x){\bf{I}}(x \in c)/p(c)$.



Note that the islands of a dynamic process depends on how long it runs. For example,
suppose $X = \{a, b, c\}$, $\map$ is a single-valued, deterministic
function, and $\map(a) = a, \map(b) = a$, while $\map(c) = b$. Then $\map$ has two islands,
$\{a, b\}$ and $\{c\}$. However if we iterate $\map$ we have
just a single island, since all three states get mapped under $\map^2$ to the state $a$.


In the more general case where the physical process implements an arbitrary
stochastic matrix $\map$ that may or may not be single-valued, we extend the definition of islands
to be
the transitive closure of the equivalence relation,
\eq{
x \sim x' \Leftrightarrow \exists x'' : \map(x''|x) > 0, \map(x'' \mid x') > 0
}
%
(Note that $x$ and $x'$ may be in the same island even if there is no $x''$
such that both $P(x'' \mid x) > 0$ and $P(x' \mid x) > 0$, due to the transitive
closure requirement.) Equivalently, the islands of $\map$ are a partition $\{X^i\}$ of $X$
such that for all $X^i$, $x \not \in X^i$, 
\eq{
\supp_{x' \in X} \map(x' \mid x) \;\cap \bigcup_{x'' \in X^i} \supp_{x' \in X} \map(x' \mid x'') \;=\; \varnothing
}

Although the focus of this paper is computers, which are typically viewed as
implementing single-valued functions, 
most of the results below also hold for physical processes that implement
general stochastic matrices as well, if one uses this more general definition
of islands. Note that for either single-valued or non-single-valued stochastic matrices $\map$,
the islands of $\pi$ will in general be a refinement of the islands of $\pi^2$, since
$\pi$ may map two of its islands to (separate) regions that are mapped on top of one another by $\pi^2$.
This is not always the case though, e.g., it is not the case if $\map$ permutes its islands.


\section{Information theory}
\label{sec:info_notation}

In this section I review the parts of information theory that are relevant
for this paper, and specify the notation I will use. I also introduce
some new information theoretic concepts, which arise naturally in the analysis
of the entropy dynamics of circuits presented below in \cref{sec:circuit_entropy_dynamics}.

The Shannon entropy of a distribution over a set $X$, 
the Kullback-Leibler (KL) divergence between two distributions
both defined over $X$ (sometimes called the ``relative entropy'' of those two distributions), 
and the cross-entropy between two such distributions, respectively, are defined as
\eq{
\SSS(\p(X)) &= -\sum_{x \in X} \p(x) \ln \p(x) \\
\DDf{\p(X)}{\r(X)} &= \sum_{x \in X} \p(x) \ln \frac{\p(x)}{\r(x)} \\
\KKf{\p(X)}{\r(X)} &= \SSS(\p(X)) + \DDf{\p(X)}{\r(X)} \label{eq:3} \\
& = -\sum_{x \in X} \p(x) \ln {\r(x)} 
}
(I adopt the convention of using natural logarithms rather than logarithms base 2
for most of this chapter.) I sometimes refer to the second arguments of KL
divergence and of cross-entropy as a \textbf{reference distribution}. Note that the
entropy of a distribution $p$ is just the negative of the KL divergence from $p$ to the uniform
reference distribution, up to an overall (negative) constant.

The conditional entropy of a random variable $X$ conditioned on 
a variable $Y$ under joint distribution $\p$ is defined as
\eq{
     \SSS(\p(X \mid Y)) &= \sum_{y \in Y} \p(y) \SSS(\p(X \mid y))  \nonumber \\
        &= -\sum_{x \in X, y \in Y} \p(y) \p(x \mid y) \ln \p(x \mid y)
}
and similarly for conditional KL divergence and conditional cross-entropy. 
The \textit{chain rule} for entropy~\cite{cover_elements_2012} says that
\eq{
\SSS(\p(X \mid Y)) + \SSS(\p(Y)) = \SSS(\p(X, Y))
}
Similarly,
given any two  distributions $p$ and $r$, both defined over $X \times Y$,
the conditional cross entropy between them 
equals the associated conditional entropy plus the associated conditional KL divergence:
\eq{
\KKf{\p(X \mid Y)}{\r(X \mid Y)} &= \SSS(\p(X \mid Y)) + \
    \DDf{\p(X \mid Y)}{\r(X\mid Y)}  \\
& = -\sum_{x \in X, y \in Y} \p(x, y) \ln {\r(x \mid y)} 
}

The mutual information between two random variables $X$ and $Y$ jointly
distributed according to $\p$ is defined as
\eq{
 I_\p(X ; Y) &\equiv \SSS(\p(X)) + \SSS(\p(Y)) - \SSS(\p(X, Y)) \\
 	&= \SSS(\p(X)) - \SSS(\p(X \mid Y))
\label{eq:mut_alt}	
}
(I drop the subscript $p$ where the distribution is clear from context.)
The \emph{data processing inequality} for mutual information~\cite{cover_elements_2012}
states that if we have random variables $X$, $Y$, and $Z$, and 
$Z$ is a stochastic function of $Y$, then $I(X;Z)\le I(X;Y)$.

Where the random variable is clear from context, I sometimes simply write $\SSS(\p)$, $\DDf{\p}{\r}$, and $\KKf{\p}{\r}$. I also sometimes abuse notation, and (for example) if $a$ and
$b$ are specified, write $\SSS(A = a \mid B =b)$ to
mean the conditional entropy of the random variable $\delta(A, a)$ conditioned on the 
event that the random variable $B$ has the value $b$.
When considering a set of random variables, I usually index them and their outcomes with subscripts, as in $X_1, X_2, \ldots$ and $x_1, x_2, \ldots$.  I also use notation like $X_{1,2}$ to indicate the joint random variable $(X_1,X_2)$.  


One extension of mutual information to more than two random variables
is known as total correlation, or \textbf{multi-information}~\cite{watanabe1960information}:
\eq{
 \II(\p(X_{1}; X_2 ;\ldots)) 
    \equiv  \bigg[ \sum_i \SSS(\p(X_i)) \bigg]  -\SSS(\p(X_{1,2,\ldots}) ) 
 }
which when the context is clear I abbreviate as $\II(\p(X_{1,2,\ldots}))$.
I sometimes use this same notation when $X$ has just two components, in which case multi-information
is the same as mutual information. Like mutual information, 
multi-information is always non-negative~\cite{watanabe1960information}.
The multi-information of a distribution $\p$ over $X_{1,2,\ldots}$ is a measure of the amount of information I can learn from the random variables $X_1, X_2, \ldots$ considered together, that I cannot extract from them considered in isolation from one another. 
In other words, it is a measure of the strength of the statistical dependencies of the variables $X_1, X_2, \ldots$, under $\p$. 

I sometimes use superscript conditioning bars to indicate conditional 
versions of these information theoretic quantities.
In particular, given two joint distributions $\p^{a,b}$ and $\r^{a,b}$ over a product space $X^{a,b} = X^a\times X^b$,
I sometimes write the conditional KL divergence between $\p$ and $\r$ of $X^a$ given $X^{b}$ as
\eq{
\DDb{ \p^{a \mid b} }{ \r^{a \mid b}}  &= \sum_{x^a, x^{b}} \p(x^a, x^{b}) \ln \frac{\p(x^a \mid x^{b})} {\r(x^a \mid x^{b})}
}

I write $S(\map \p)$ to refer to the entropy of distributions over $Y$ induced by $\p(x)$ and the conditional distribution $\map$, as defined in \erf{eq:matrixnotation}.
I use similar shorthand for the other information-theoretic quantities,
$\DDf{\cdot}{\cdot}$, $\KKf{\cdot}{\cdot}$ and $\II(\cdot)$.
In particular, the \emph{chain rule}  for KL divergence and the \emph{data-processing
inequality} for KL divergence, respectively, are~\cite{cover_elements_2012}:
\begin{enumerate}
\item  For all distributions $\p, \r$ over the space $X^a \times X^{b}$,
\eq{
&\DD\big(p(X^a, X^b) \mid\mid \r(X^a, X^b)\big) \nonumber \\
& \qquad\qquad =
   \DD\big( p(X^b) \mid\mid \r(X^b)\big) + \DD\big(p(X^a \mid X^b) \mid\mid \r(X^a \mid X^b)\big)
}
\item For all distributions $\p, \r$ over the space $X$ and conditional distributions $\map(y \mid x)$,
\eq{
\DDf{ \p }{ \r } &\ge \DDf{\map \p }{ \map \r}
}
\end{enumerate}
(Note that by combining the chain rule for KL divergence with the chain rule for entropy,
we get a chain rule for cross entropy.)

Some of the results reviewed below are formulated in terms of
the \textbf{multi-divergence} between two probability distributions over the same multi-dimensional space. This is
an information-theoretic measure recently introduced in~\cite{wolpert2018exact}, 
which can be viewed as an extension of multi-information to include a reference distribution. 
It is defined as follows:
\begin{align}
& \IIDDf{\p(X_1; X_2,\ldots)}{\r(X_{1}; X_2;\ldots)} \nonumber \\
& \equiv \sum_{{x_1, x_2, \ldots}} \p(x_1, x_2, \ldots) \ln \frac{\p(x_1, x_2, \ldots)}{\r(x_1, x_2, \ldots)} \prod_i \frac{\r(x_i)}{\p(x_i)}
\label{eq:15}  
\\
& = \DDf{\p(X_{1,2,\ldots})}{ \r(X_{1,2,\ldots})} - \sum_i \DDf{\p(X_i)}{ \r(X_i)}
\label{eq:15a}	
\end{align}
When the context is clear I sometimes abbreviate this as $\IIDDf{\p(X_{1,2,\ldots})}{\r(X_{1,2,\ldots})}$.

Multi-divergence measures how much of the divergence between $\p$ and $\r$ arises from the correlations among the variables $X_1,X_2,\ldots$, rather than the marginal distributions of each variable considered separately. See App.~\ref{sec:multidivappendix} for a discussion of the elementary properties of multi-divergence and
its relation to conventional multi-information, e.g., as a measure of the ``joint information''
among a set of more than two random variables. 

Finally, the difference between conventional multi-information and multi-divergence
is another new information-theoretic quantity, called \textbf{cross multi-information}:
\begin{align*}
&\IIKKf{\p(X_{1}; X_2;\ldots)}{\r(X_{1}; X_2;\ldots)} \\
&\quad \equiv  \II(\p(X_{1,2,\ldots})) - \IIDDf{\p(X_{1,2,\ldots})}{\r(X_{1,2,\ldots})} \\
&\quad = \bigg[ \sum_i \KKf{\p(X_i)}{\r(X_i)} \bigg] - \KKf{\p(X_{1,2,\ldots})}{\r(X_{1,2,\ldots})}
\end{align*}
(When the context is clear, I abbreviate cross multi-information as $\IIKKf{\p(X_{1,2,\ldots})}{\r(X_{1,2,\ldots})}$.)

Adopting an information-theoretic perspective,
$\IIKKf{\p(X_{1,2,\ldots})}{\r(X_{1,2,\ldots})}$ is the reduction in expected codeword 
length if we use one type of coding rather than the other. Under the first, less efficient
coding scheme, a codeword is created by 
concatenating codewords produced separately for each component of $x_i$
using distinct codebooks that are separately optimized for each of those components,
where all of these codebooks are optimized for $\r$ while events
are actually generated by sampling $\p$. Under the second, more efficient 
coding scheme, a codeword is created by using a single
codebook that is optimized for joint events $(x_1,x_2,\dots)$. Thus, $\IIKKf{\p(X_{1,2,\ldots})}{\r(X_{1,2,\ldots})} = 0$ if $\r$ is
a product distribution.

As shorthand, I often write $ \II(\p)$, $\IIDDf{\p}{\r}$, and $\IIKKf{\p}{\r}$ when the set of random variables $X_{1,2,\ldots}$ is clear from context. In the usual way, if $\r(x)$ and $\p(x)$ both equal $0$ for some $x$'s, then I implicitly redefine $\DDf{\p}{\r}$ and $\IIDDf{\p}{\r}$ to be
the limit as those probabilities go to zero. 


\section{Computational Machines}
\label{sec:comp_machines}

The term ``computer'' means many different things in the literature. To avoid this imprecision, here
I will instead use the term \textbf{computational machine}, 
defining it to mean one of the members of the Chomsky hierarchy~\cite{hopcroft2000jd}.

In this section I introduce the particular
computational machines whose entropy dynamics I will consider in this paper. 
These machines are introduced in order of increasing computational power; the set of
computations that straight-line circuits can perform are a strict subset of the set of
computations that finite automata can perform; these in turn are a strict subset of the set of computations that
transducers can perform; and these in turn
are a strict subset of the set of computations that a Turing machine can perform.\footnote{The 
reader should be warned that there are some statements in 
the recent nonequilibrium statistical physics literature to the effect that
transducers are as computationally powerful as Turing machines. These are mistaken.}

However, I will also consider some digital systems that ``process information'',
but occur as subsystems of some computational machines, and are 
less computationally powerful than any of them. 
As an example, the information processing system that has perhaps been studied
more than other one in the physics literature is simple bit erasure, which
is not a member of the Chomsky hierarchy. Another example is a single gate in a circuit,
e.g., a XOR gate. I will sometimes refer to these kinds of extremely simple information processing systems
as \textbf{sub-computers}, and use the term \textbf{(computational) device} to refer
to either sub-computers or full-fledged computational machines. I
will sometimes make the further distinction that a device considered as an abstract
mathematical transformation (as in computer science theory) is a ``logical device'', while a 
device considered as a physical system (as in entropy dynamics) is a ``physical
device''.

\subsection{Bayes nets}
\label{sec:bayes_net}

Although Bayes nets are not normally viewed as computational machines,
they provide a natural framework for investigating several kinds of computational machines,
in particular straight-line circuits (considered in the next subsection).
Accordingly, I summarize what Bayes nets are in this subsection.

Formally, a \textbf{Bayes net} $\CC$ is a tuple $(V, E, F, X)$~\cite{kofr09}. The pair 
$(V, E)$ specifies the vertices and edges of a directed acyclic graph (DAG). 
$X$ is a Cartesian product $\prod_v X_v$, where each $X_v$ is the space of the variable associated with node $v$. $F$ is a set of conditional distributions, one for
each non-root node $v$ of the DAG, mapping the joint value of the (variables at the) parents
of $v$, $x_{\pa(v)}$, to the value of (the variable at) $v$, $x_v$.

An \textbf{input} node is a root node, and an \textbf{output} node is a leaf node. 
I assume that no output node is also an input node.
%
I indicate the set of all nodes that are 
the parents of node $v$ as $\pa(v)$.
For any Bayes net $\CC$, I write the conditional distribution implemented at its node $v$
(specified in $F$) as $\map^\CC_v( x_v \vert x_{\pa(v)})$.
When $\CC$ is implicit, I sometimes shorten this to $\map_v$. Note that this
conditional distribution reduces to a prior distribution
$\map_v(x_v)$ whenever $v$ is a root node, 

In terms of  
this notation, the conditional distribution implemented by the overall Bayes net is 
\[
\map^\CC( x \vert x_{\Cin}) = \prod_{v \in V} \map^\CC_v(x_v \vert x_{\pa(v)}) 
\]
Similarly, the combination of the distribution $\p_{\Cin}(x_{\Cin})$ and 
the conditional distributions at the non-root nodes of the Bayes net specifies
a joint distribution over $x \in X$, the joint space of all nodes in the Bayes net, 
given by
\[
\p(x) = \p_{\Cin}(x_{\Cin})\map^\CC( x \vert x_{\Cin})
\]
This joint distribution in turn
specifies a conditional distribution of the joint state of the output nodes
of the Bayes net, given the joint state of the input nodes. 
As an important example, when the
distributions in $F$ are all single-valued functions, this conditional distribution
reduces to a single-valued input-output function implemented by the Bayes net as a whole.
In general, there are an infinite number of Bayes nets that implement the same
conditional distribution of outputs given inputs.

For each node $v$ in the Bayes net, I 
write the distribution formed by marginalizing $\p(x)$ down to $X_v$ 
as  $\p_{v}(x_{v})$. I also write the marginal distribution over the joint state of the parents of any node $v$ in the Bayes net as  $\p_{\pa(v)}(x_{\pa(v)})$.
I refer to these two distributions as the distribution $\p_{\Cin}$ \textbf{propagated} 
to $v$ and to $\pa(v)$, respectively.

\subsection{Straight-line circuits}
\label{sec:circuit_theory}

A \textbf{(straight-line) circuit} $\CC$ is a tuple $(V, E, F, X)$ that can be
viewed as a special type of Bayes net~\cite{kofr09,ito2013information,ito_information_2015}. 
Intuitively, the DAG of the Bayes net, $(V, E)$, is the {wiring diagram} of the circuit.
In conventional circuit theory, the all conditional distributions
in the set $F$ are single-valued functions. 

Note that following the convention in the Bayes nets literature, with this definition
of a circuit we orient edges in the direction
of logical implication, i.e., in the direction of information flow.
So the inputs to the circuit are the roots of the
associated DAG, and the outputs are the leaves. The reader should
be warned that this is the \textit{opposite} of the convention in computer science theory.
When there is no
risk of confusion, I simply refer to a circuit $\CC$, with all references to
$V, E, F$ or  $X$ implicitly assumed as the associated elements defining $\CC$.\footnote{A
more general type of circuit than the one considered here allows branching conditions
at the nodes and allows loops. Such circuits cannot be represented as a
Bayes net. To make clear what kind of circuit is being considered,
sometimes the branch-free, loop-free type of circuit is called a ``straight-line''
circuit~\cite{savage1998models}.} Just like in Bayes nets, we refer to the maximal
number of nodes on any path going from a root node to a leaf node of a circuit as the \textbf{depth}
of that circuit. (In other conventions, the depth is the number of links along that path,
which is one less than the number of nodes on it.)

I use the term \textbf{gate} to refer to any non-root node in a circuit.
In the special case where 
all non-output nodes in a circuit have outdegree 1 and there is a single output node, the  
circuit is called a (circuit) \textbf{formula}. 
In the context of formulas, I use $v_\Cout$ to indicate the single element of $V_\Cout$.
In the context of circuits, I use the term \textbf{all-at-once (AO) circuit} to mean
a circuit with a single gate, and
write $AO(\CC)$ to mean an AO circuit that implements the same conditional distribution
$\map^\CC$ as $\CC$. 

Finally, I sometimes
use the terms ``gate'', ``circuit'', etc., to refer to physical systems with physical states,
and sometimes to refer to the associated abstract mathematical conditional distributions in a DAG.
The intended meaning should be clear from context -- when I need to be explicit,
I will refer to \textbf{physical circuits} and \textbf{computational circuits}.

Straight line circuits are an example of \textbf{non-uniform} computers. These are
computers that can only work with inputs of some fixed
length. (In the case of a circuit, that length is specified
by the number of root nodes in the circuit's DAG.) One can use a single circuit to compute the output
for any input in a set of inputs, so long as all those inputs have the same length. If on the other hand 
one wishes to consider using circuits to compute the output for any input in a set that contains
inputs of all possible lengths, then one must use a \textbf{circuit family},
i.e., an infinite set of circuits $\{C_i : i \in \Z^+ \}$, where each circuit $C_i$ has
$i$ root nodes.

In contrast to non-uniform computers, \textbf{uniform} computers are machines that can work with arbitrary length
inputs.\footnote{The reader should be warned that computer scientists also consider ``uniform circuit
families'', which is something that is related but different.} 
In general, the number of iterations a particular uniform computational machine requires
to produce an output is not pre-fixed, in contrast to the case with any particular nonuniform computational machine. 
Indeed, for some inputs, a uniform computational machine may never
finish computing. The rest of this section introduces some of 
the more prominent uniform computational machines.

\subsection{Finite Automata}
\label{sec:sfa_def}

One important class of (uniform) computational machines are the finite automata.
There are several different, very similar definitions of finite
automata, some of which overlap with common definitions of
``finite state machines''. To fix the discussion, here I adopt the following definition:

\begin{definition}
\label{def:fa}
A \textbf{finite automaton} (FA) is a 5-tuple $(R, \Lambda, r^\varnothing, r^A, \rho)$ where:
\begin{enumerate}
\item $R$ is a finite set of \textbf{computational states};
\item $\Lambda$ is a finite (input) \textbf{alphabet};
\item $r^\varnothing \in R$ is the \textbf{start state};
\item $r^A \in R$ is the \textbf{accept state}; and
\item $\rho : R \times \Lambda \rightarrow R$ is the \textbf{update function},
mapping a current input symbol and the current computational state to a next
computational state.
\end{enumerate}
\end{definition}

A finite string of successive
input symbols, i.e., an element of $\Lambda^*$,
is sometimes called an \textbf{(input) word}, written as $\vl$. 
To operate a finite automaton on a particular input word, one
begins with the automaton in its start state, and feeds that state
together with the first symbol in the input word into the update function,
to produce a new computational state. Then one feeds in the next symbol
in the input word (if any), to produce a next computational state, and so on.
I will sometimes say that the \textbf{head} is in some state $r \in R$,
rather than say that the computational state of the automaton is $r$.

Often one is interested in whether the head is in state $r^A$
after the last symbol from the input word is processed. If
that is the case, one says that the automaton \textbf{accepts} that input word. In this
way any given automaton uniquely specifies a \textbf{language} of all input
words that that automaton accepts, which is called 
a \textbf{regular} language. As an example, any finite language (consisting
of a finite set of words) is a regular language. On the other hand,
the set of all palindromes over $\Lambda$, 
to give a simple example, is \textit{not} a regular language.

Importantly, any particular FA can process input words of arbitrary length.
This means that
one cannot model a given FA as some specific (and therefore fixed width) circuit, in general. 
The FA will have properties that are not captured by that circuit.
In this sense, individual FAs are computationally more powerful than individual circuits.
(This does not mean that individual FAs are more powerful than entire circuit \textit{families}
however; see the discussion in \cref{sec:TMs_def} of how circuit families can be even
more powerful than Turing machines.)

Finite automata play an important role in many different fields, including electrical
engineering, linguistics, computer science, philosophy, biology, mathematics, and logic. 
In computer science specifically, they are widely used in the
design of hardware digital systems, of compilers, of network protocols, and 
in the study of computation and languages more broadly.


In all these applications of FAs, the automaton is viewed as a system that maps an
input \textit{word} to an output \textit{computational state}. This motivates the
following alternative definition of an FA:

\begin{definition}
A \textbf{word-based} finite automaton is a 6-tuple 
$(R, \Lambda, r^\varnothing, r^A, \rho^*, \tau)$ where:
\begin{enumerate}
\item $R$ is a finite set of \textbf{computational states};
\item $\Lambda$ is a finite (input) \textbf{alphabet};
\item $r^\varnothing \in R$ is the \textbf{start state};
\item $r^A \in R$ is the \textbf{accept state}; 
\item $\tau \in \Z^+$ is the \textbf{counter}; and
\item $\hat{\rho} : R \times \Lambda^* \times \Z^+ \rightarrow R$ 
is the \textbf{(computational state) update function},
mapping a current input word, current counter pointing to one of that word's symbols,
and current computational state, to a next
computational state.
\label{def:word_based_sfa}
\end{enumerate}
\end{definition}
\noindent When I need to distinguish FAs as defined in Def.~\ref{def:fa} from
word-based FAs, I will refer to the former as \textbf{symbol-based} FAs.
 
To map a symbol-based finite automaton (Def.~\ref{def:fa})
with update function $\rho$
into an equivalent word-based update function $\hat{\rho}$, we set 
\eq{
\hat{\rho}(r, \lambda^*, \tau) = \rho(r, \lambda^*_\tau)
\label{eq:word_based_sfa}
} 
At the first iteration of a word-based FA, not only is the computational
state set to the start state, but the counter has the value $0$. 
In addition, at the end of each iteration $\hn$
of the FA, after its computational state is updated by $\hat{\rho}$,
the counter is incremented, i.e., $\tau_\hn \rightarrow \tau_{\hn}+1$.
From now on I will implicitly move back and forth between the two definitions 
of an FA as is convenient.

To allow us to analyze a physical system that implements the running
of an FA on many successive input words, we need a way to signal to the system
when one input word ends and then another one begins. Accordingly, without loss of generality we assume
that $\Lambda$ contains a special \textbf{blank} state, written $b$, that delimits
the end of a word. I write $\Lambda_-$ for $\Lambda \setminus \{b\}$, so that
words are elements of $\Lambda_-^*$.

In a \textbf{stochastic} finite automaton (sometimes called a ``probabilistic
automaton''), the single-valued function $\rho$ is replaced by a conditional distribution. 
In order to use notation that covers all iterations $i$
of a stochastic finite automaton, I write this \textbf{update distribution} 
as $\Map(r_{i+1} \mid r_i, \lambda_i)$. The analogous extension of the word-based
definition of finite automata into a word-based definition of a stochastic
finite automata is immediate. For simplicity, from now on I will simply refer to ``finite
automaton'', using the acronym ``FA'', to refer to a finite automaton that
is either stochastic or deterministic.

Typically in the literature there is a set of multiple accept states --- called
``terminal states'', or ``accepting states'' --- not just one. Sometimes
there are also multiple start states.
%

%
%
%

\subsection{Transducers - Moore machines and Mealy machines}
\label{sec:transducer_def}
 
In the literature the definition of FA is sometimes extended 
so that in each transition from one computational state to the next an output symbol is 
generated.  Such systems are also called ``{transducers}'' in the computer
science community. 

\begin{definition}
\label{def:trans}
A \textbf{transducer} is a 6-tuple $(R, \Lambda, \Gamma, r^\varnothing, 
x^A, \rho)$ such that:

\begin{enumerate}
\item $R$ is a finite set, the set of \textbf{computational states};
\item $\Lambda$ is a finite set, called the \textbf{input alphabet};
\item $\Gamma$ is a finite set, called the \textbf{output alphabet};
\item $r^\varnothing \in R$ is the \textbf{start} state;
\item $r^A \in R$ is the \textbf{accept state};
\item $\rho: R \times {\Lambda} \rightarrow R \times {\Gamma}$
is the \textbf{update rule}.
\end{enumerate}
\end{definition}
\noindent
Sometimes the computational states of a transducer are referred to as
the states of its \textbf{head}. I refer to the (semifinite) string of symbols that have yet
to be processed by a transducer at some
moment as the \textbf{input (data) stream} at that moment. I refer to the string of symbols
that have already been produced by the information ratchet at some
moment as the \textbf{output (data) stream} at that moment.

To operate a transducer on a particular input data stream, one
begins with the machine in its start state, and feeds that state
together with the first symbol in the input stream into the update function,
to produce a new computational state, and a new output symbol. Then one feeds in the next symbol
in the input stream (if any), to produce a next computational state
and associated output symbol, and so on.

In a \textbf{stochastic} transducer, the single-valued function $\rho$ is replaced by
a conditional distribution. In order to use notation that covers all iterations $i$
of the transducer, I write this \textbf{update distribution} as
$\Map({\gamma}_{i+1}, r_{i+1} \mid r_{i},{\lambda}_i)$. 
%
Stochastic transducers are used in fields ranging from linguistics to
natural language processing (in particular machine translation) to
machine learning more broadly.
From now on I implicitly mean ``stochastic transducer'' when I use the
term ``transducer''.\footnote{The reader should be warned that
some of the literature refers to both FAs and transducers
as ``finite state machines'', using the term ``acceptor'' or ``recognizer''
to refer to the system defined in Def.~\ref{def:fa}. Similarly, the word
``transducer'' is sometimes used loosely in the physics community, to
apply to a specific system that transforms one variable --- often
energy --- into another variable, or even just into another form.}

As with FAs, typically in the literature 
transducers are defined to have a set of multiple accept states, not just one. Sometimes
there are also multiple start states.
Similarly, in some of the literature
the transition function allows the transducer to receive
the empty string as an input and/or produce the empty string as
an output. 

A \textbf{Moore machine} is a transducer where the output ${\gamma}$ is
determined purely by the current state of the transducer, $r$. In contrast,
a transducer in which the output depends on both the current state $x$ and 
the current input ${\lambda}$ is called a \textbf{Mealy machine}.

As a final comment, an interesting variant of the transducers defined
in Def.~\ref{def:trans} arises if we remove the requirement that there be
accept states (and maybe even remove the requirement of start states). 
In this variant, rather than feeding an infinite
sequence of input words into the system, each of which results
in its own output word, one feeds in a single input word, which is
infinitely long, producing a single (infinitely long) output word. 
This variant is used to define so-called ``automata groups'' or ``self-similar 
groups''~\cite{MR2162164}.

Somewhat confusingly, although the computational properties of
this variant of transducers differs in crucial ways from those defined
in Def.~\ref{def:trans}, this variant is also called ``transducers''
in the literature on ``computational mechanics'', a branch of hidden Markov model 
theory~\cite{barnett2015computational}. Fortunately, 
this same variant have also been given a \textit{different} name, \textbf{information ratchets}, 
in work analyzing their statistical physics properties~\cite{mandal2012work}. 
Accordingly, here I adopt that term for this variant of (computer science)
transducers.

\subsection{Turing machines}
\label{sec:TMs_def}

Perhaps the most famous class of computational machines are Turing 
machines~\cite{hopcroft2000jd,arora2009computational,savage1998models}.
One reason for their fame is that it seems one can model any computational machine that
is constructable by humans as a Turing machine. A bit more formally, the \textit{Church-Turing
thesis} states that, ``A function on the natural numbers is computable by a human 
being following an algorithm, ignoring resource limitations, if and only if 
it is computable by a Turing machine.'' 
The ``physical Church-Turing thesis'' modifies
that to hypothesize that the set of functions computable with Turing machines includes 
all functions that are computable using mechanical algorithmic procedures admissible by 
the laws of
physics~\cite{arrighi2012physical,piccinini2011physical,pour1982noncomputability,moore1990unpredictability}.

In part due to this thesis,
Turing machines form one of the keystones of the entire field
of computer science theory, and in particular of computational complexity~\cite{moore2011nature}.
For example, the famous Clay prize question of whether $\cs{P} = \cs{NP}$ --- widely considered one of the deepest and most profound open questions in mathematics ---
concerns the properties of Turing machines. As another example, the theory
of Turing machines is intimately related to deep results on the limitations of mathematics,
like G{\"o}del's incompleteness theorems, and seems to have broader
implications for other parts of philosophy
as well~\cite{aaronson2013philosophers}. Indeed, it has been argued that the foundations
of physics may be restricted by some of the properties of Turing machines~\cite{barrow2011godel,aaro05}.

Along these lines, some authors have suggested that the foundations of statistical
physics should be modified to account for the properties of Turing machines, e.g.,
by adding terms to the definition of entropy. After all, given the
Church-Turing thesis, one might argue that the probability distributions at the heart of
statistical physics are distributions ``stored in the
mind'' of the human being analyzing a given statistical physical system (i.e., 
of a human being running a particular algorithm to compute a property of a given
system). Accordingly, so goes the argument, the costs
of generating, storing, and transforming the minimal specifications of
the distributions concerning a statistical physics system should be included 
in one's thermodynamic analysis of those changes in the distribution of states of the system.  See~\cite{caves1990entropy,caves1993information,zure89b}. 

There are many different definitions of Turing machines that are ``computationally equivalent''
to one another. This means that
any computation that can be done with one type of Turing machine can be done with the other.
It also means that the ``scaling function'' of one type of Turing machine, 
mapping the size of a computation to the minimal amount of resources needed to 
perform that computation by that type of Turing machine, is at most a polynomial
function of the scaling function of any other type of Turing machine. (See for example the
relation between the scaling functions of single-tape and multi-tape
Turing machines~\cite{arora2009computational}.) The following definition will be useful
for our purposes, even though it is more complicated than strictly needed:

\begin{definition}
A \textbf{Turing machine} (TM) is a 7-tuple $(R,\Lambda ,b,v,r^\varnothing,r^A,\rho)$ where:

%

\begin{enumerate}
\item $R$ is a finite set of \textbf{computational states};
\item $\Lambda$ is a finite \textbf{alphabet} containing at least three symbols;
\item $b \in \Lambda$ is a special \textbf{blank} symbol;
\item $v \in \Z$ is a \textbf{pointer};
\item $r^\varnothing \in R$ is the \textbf{start state};
\item $r^A \in R$ is the \textbf{accept state}; and
\item $\rho : R \times \Z \times \Lambda^\infty \rightarrow 
R \times \Z \times \Lambda^\infty$ is the \textbf{update function}.
It is required that for all triples $(r, v, T)$, that if we write
$(r', v', T') = \rho(r, v, T)$, then $v'$ does not differ by more than $1$
from $v$, and the vector $T'$ is identical to the vectors $T$ for all components
with the possible exception of the component with index $v$;\footnote{Technically 
the update function only needs to be defined on the ``finitary'' subset of $R \times \Z 
\times \Lambda^\infty$, namely, those elements of $R \times \Z 
\times \Lambda^\infty$ for which the tape contents has a non-blank value in only finitely many positions.}
\end{enumerate}
\label{def:tm}
\end{definition}

$r^A$ is often called the ``halt state'' of the TM rather than the accept state. 
(In some alternative, computationally equivalent definitions of TMs, 
there is a set of multiple accept states rather than
a single accept state, but for simplicity I do not
consider them here.) $\rho$ is sometimes called the ``transition function'' of the TM.
We sometimes refer to $R$ as the states of the ``head'' of the TM,
and refer to the third argument of $\rho$ as a \textbf{tape}, writing a
value of the tape (i.e., semi-infinite string of elements of the alphabet) as $T$.
The set of triples that are possible arguments to the update function 
of a given TM are sometimes called the set of \textbf{instantaneous descriptions}
(IDs) of the TM. (These are sometimes instead referred to as ``configurations''.)
Note that as an alternative to Def.~\ref{def:tm}, we
could define the update function of any TM as a map over an associated space of IDs.

Any TM $(R,\Sigma ,b,v,r^\varnothing, r^A, \rho)$ starts with $r = r^\varnothing$, the counter
set to a specific initial value (e.g, $0$), and with $T$
consisting of a finite contiguous set of non-blank symbols, with
all other symbols equal to $b$. The TM operates by iteratively
applying $\rho$, until the computational state falls in $r^A$, at
which time it stops, i.e., any ID with the head in the halt state is a
fixed point of $\rho$.

If running a TM on a given initial state of the tape results in the TM eventually halting,
the largest blank-delimited string that contains the position of the pointer 
when the TM halts is called the TM's \textbf{output}. The initial
state of $T$ (excluding the blanks) is sometimes called the associated 
\textbf{input}, or \textbf{program}. (However,
the reader should be warned that the term ``program'' has been used by some physicists to
mean specifically the shortest input to a TM that results in it computing
a given output.) We also say that the TM \textbf{computes} an output
from an input.  In general, there will be inputs for which the TM never halts. 
The set of all those inputs to a TM that cause it to eventually
halt is called its \textbf{halting set}. 

We say that a total function $f$ from $(\Lambda \setminus \{b\})^*$ to itself
is \textbf{recursive} if there is a TM with input alphabet $\Lambda$ such that for all 
$x \in (\Lambda \setminus \{b\})^*$, the TM computes $f(x)$. If $f$ is instead a partial function,
then we say it is \textbf{partial recursive} if there is a TM with input alphabet $\Lambda$
that computes $f(x)$ for all $x$ for which $f(x)$ is defined. (Sometimes we simply refer 
to a function that is either recursive or partial recursive as ``computable''.) Famously,
Turing showed that there are total functions that are not recursive. In light of the Church-Turing thesis,
this result is arguably one of the deepest philosophical truths concerning fundamental 
limitations on human capabilities ever discovered.

As mentioned, there are many variants of the definition of TMs provided above. In one
particularly popular variant the single tape in \cref{def:tm}
is replaced by multiple tapes. Typically one of
those tapes contains the input, one contains the TM's output (if and) when the TM
halts, and there are one or more intermediate ``work tapes'' that are
in essence used as scratch pads. The advantage of using this more complicated
variant of TMs is that it is often easier to prove theorems for such machines
than for single-tape TMs. However, there is no difference in
their computational power. More precisely, one can transform any single-tape TM
into an equivalent multi-tape TM (i.e., one that computes the same partial function),
as well as vice-versa~\cite{arora2009computational,livi08,sipser2006introduction}.

Returning to the TM variant defined in \cref{def:tm},
a \textbf{universal Turing machine} (UTM), $M$, is one that can be used
to emulate any other TM. More precisely, a UTM $M$ has the property that
for any other TM $M'$, there is an invertible map $f$ from the set of possible
states of the tape of $M'$ into the set of possible states of the tape of $M$, such
that if we:
\begin{enumerate}
\item  apply $f$ to an input string $\sigma'$ of $M'$ to fix an input string $\sigma$
of $M$; 
\item run $M$ on $\sigma$ until it halts; 
\item apply $f^{-1}$ to the resultant output of $M$;
\end{enumerate} 
then we get exactly the output computed by $M'$ if it is run directly on $\sigma'$.

An important theorem of computer science is that there exists universal TMs.
Intuitively, this just means that there exists programming languages which are ``universal'',
in that we can use them to implement any desired program in any other language, after
appropriate translation of that program from that other
language. This universality leads to a very important concept:

\begin{definition}
The \textbf{Kolmogorov complexity} of a UTM $M$ to compute a string $\sigma \in \Lambda^*$
is the length of the shortest input string $s$ such that $M$ computes $\sigma$ from $s$.
\end{definition}
\noindent 
Intuitively, (output) strings that have low Kolmogorov complexity for some specific UTM $M$ are those with 
short, simple programs in the language of $M$. For example, in all common (universal) programming
languages (e.g., \textit{C, Python, Java}, etc.), 
the first $\hn$ digits of $\pi$ have low Kolmogorov complexity, since those
digits can be generated using a relatively short program.  
Strings that have high (Kolmogorov) complexity are sometimes referred to as
``incompressible''. These strings have no patterns in them that can be generated by
a simple program. As a result,
it is often argued that the expression ``random string'' should only be used for strings that 
are incompressible.

Suppose we have two strings $s^1$ and $s^2$ where $s^1$ is a proper prefix of $s^2$. 
If we run the TM on $s^1$, it can detect when it gets to the end of its input, by
noting that the following symbol on the tape is a blank. Therefore, it can
behave differently after having reached the end of $s^1$ from how it behaves
when it reaches the end of the first $\ell(s^1)$ bits in $s^2$. As a result,
it may be that both of those input strings are in its halting set, but result
in different outputs.

A \textbf{prefix (free) TM} is one in which this can never happen:
there is no string in its halting set that
is a proper prefix of another string in its halting 
set.\footnote{It is not trivial to construct prefix single-tape TMs directly.
For that reason it is common to use prefix three-tape TMs, in
which there is a separate input tape that can only be read from, output tape that
can only be written to, and work tape that can be both read from and written
to. To ensure that the TM is prefix, we require that the head cannot ever back up on
the input tape to reread earlier input bits, nor can it ever back up on the output
tape, to overwrite earlier output bits. To construct a single-tape prefix TM,
we can start with some such three-tape prefix TM and transform
it into an equivalent single-tape prefix
TM, using any of the conventional techniques for transforming between
single-tape and multi-tape TMs.
}
%

are related to one another the same way that various kinds of Shannon entropy
are related to one another. For example, loosely speaking, the
conditional Kolmogorov complexity of string $s$ conditioned on string $s'$,
written as $K(s \mid s')$, is the length of the shortest string x such
that if the TM starts with an input string given by the concatenation
$xs'$, then it computes $s$ and halts. If we restrict attention
to prefix-free TMs, then for all strings $x, y \in \Lambda^*$, we have~\cite{livi08}
\eq{
K(x, y) \le K(x) + K(x \mid y) + O(1) \le K(x) + K(y) + O(1)
\label{eq:kolmogorov_conditional}
}
(where ``$O(1)$'' means a term that is independent of both $x$ and $y$).
Indeed, in a certain technical sense, the expected value of $K(x)$ under any
distribution $P(x \in \Lambda^*)$ equals the Shannon entropy of $P$. (See~\cite{livi08}.)

The \textbf{coin-flipping prior} of a prefix TM $M$ is the probability distribution 
over the strings in $M$'s halting set generated by IID ``tossing a coin'' 
to generate those strings, in a Bernoulli process, and then normalizing.\footnote{Kraft's 
inequality guarantees that since the set of strings in the halting set is a prefix-free
set, the sum over all its elements of their probabilities cannot exceed $1$, and
so it can be normalized. However, in general that normalization constant
is uncomputable, as discussed below. Also, in many contexts we can actually
assign arbitrary non-zero probabilities to the strings outside the halting set,
so long as the overall distribution is still normalizable. See~\cite{livi08}.} 
So any string $\sigma$ in the halting set
has probability $2^{-|\sigma|} / \Omega$ under the coin-flipping prior, where
$\Omega$ is the normalization constant for the TM in question.

The coin-flipping prior provides a simple Bayesian interpretation of Kolmogorov
complexity: Under that
prior, the Kolmogorov complexity of any string $\sigma$ for any prefix TM $M$ is just
(the log of) the maximum, over all strings 
$\sigma'$ in the halting
set of $M$, of the joint probability that $\sigma'$ is the \textit{input} to $M$,
and 
the \textit{output} of that TM is $\sigma$. 
%

The normalization constant $\Omega$ for any fixed prefix UTM, sometimes called ``Chaitin's Omega'',
has some extraordinary properties. For example, the successive digits of $\Omega$
provide the answers to \textit{all} well-posed mathematical problems. So if we
knew Chaitin's Omega for some particular prefix UTM, we could answer every problem in mathematics.
Alas, the value of $\Omega$ for any prefix UTM $M$ 
cannot be computed by any TM (either $M$ or some other one).
So under the Church-Turing thesis, we cannot calculate $\Omega$. 
(See also~\cite{baez2012algorithmic} for a discussion of a
``statistical physics'' interpretation of $\Omega$ that results if we view the coin-flipping prior
as a Boltzmann distribution for an appropriate Hamiltonian, 
so that $\Omega$ plays the role of a partition function.)

It is now conventional to analyze Kolmogorov complexity using prefix UTMs, with the coin-flipping
prior, since this removes some undesirable technical properties that Kolmogorov complexity has
for more general TMs and priors. Reflecting this, all analyses in the physics
community that concern TMs assume prefix UTMs. 
(See~\cite{livi08} for a discussion of the extraordinary properties of such UTMs.)

Interestingly, for all their computational power, there are some surprising ways
in which TMs are \textit{weaker} than the other computational machines introduced above.
For example, there are an infinite number of TMs that are more powerful than any given circuit, i.e., 
given any circuit $C$, there are an infinite number of TMs that compute the same function as $C$.
Indeed, any single UTM is more powerful than \textit{every} circuit in this sense. On the other hand,
it turns out that there are circuit \textit{families} that are more powerful than any single TM.
In particular, there are circuit families that can solve the halting problem~\cite{arora2009computational}.
 
I end this subsection with some terminological comments and definitions that will
be useful below. It is conventional when dealing with Turing machines to implicitly
assume some invertible map $R(.)$ from $\Z$ into $\Lambda^*$. Given
such a map $R(.)$, we can exploit it to implicitly assume an additional invertible
map taking $\mathbb{Q}$ into $\Lambda$, e.g., by uniquely expressing any rational
number as one product of primes, $a$, divided by a product of different primes, $b$;
invertibly mapping those two products of primes into the single integer $2^a 3^b$; and then evaluating
$R(2^a 3^b)$.
Using these definitions, we say that a real number $z$ is \textbf{computable}
iff there is a recursive function $f$ mapping rational numbers to rational
numbers such that for all rational-valued accuracies $\epsilon > 0$,
$|f(\epsilon) - z| < \epsilon$. We define computable functions from $\mathbb{Q} \rightarrow \R$
similarly.

\section{Entropy dynamics}
\label{sec:entropy_dynamics}

This section reviews those aspects of stochastic thermodynamics that are necessary
to analyze the dynamics of various types of entropy during the
evolution of computational machines. As illustrated with
examples, the familiar quantities at the heart of thermodynamics
(e.g., heat, thermodynamic entropy, work) arise in special cases of this analysis.

In the first subsection, I review the conventional decomposition
of the entropy flow (EF) out of a physical system into the change in
entropy of that system plus the (irreversible) entropy creation (EP) produced
as that system evolves~\cite{van2015ensemble,seifert2012stochastic}. 
To fix details, I will concentrate on the total amount of EF, EP and entropy change that arise
over a time-interval $[0, 1]$.\footnote{In this paper
I will not specify units of time, and often implicitly change
them. For example, when analyzing the entropy dynamics of a given circuit,
sometimes the time interval $[0, 1]$ will refer to the time to 
run the entire circuit, and the attendant entropic costs.
However at other times $[0, 1]$ will refer to the time to run a single gate within
that circuit, and the entropic costs of running just that gate. In addition,
for computational machines that take more than one iteration to run, I will 
usually just refer
to a ``time interval $[0, 1]$'' without specifying which iteration of the machine 
that interval corresponds to.
Always the context will make the meaning clear.}

In the second subsection, I review recent results~\cite{wolpert2018exact} 
that specify how the EP generated by the evolution of some system depends on the initial distribution of states of the system.
These recent results allow us to evaluate how the EF of an arbitrary system, whose dynamics
implements some conditional distribution $\map$ of final states
given initial states, depends on the initial distribution of states of the system
that evolves according to $\map$. 
(As elaborated in subsequent sections, this dependence is one of the central features determining the
entropic costs of running any computational machine.)

I end this section with some general cautions about translating a computer science definition
of a computational machine into a physics definition of a system that implements that machine.

\subsection{Entropy flow, entropy production, and Landauer cost}
\label{sec:gen_land}

To make explicit connection with thermodynamics,
consider a physical system with countable state space $X$ that evolves over time 
interval $t\in [0,1]$ while in contact with one or more thermal reservoirs, while possibly also undergoing driving by 
one or more work reservoirs.\footnote{In statistical physics, a ``reservoir'' $R$ in contact 
with a system $S$ is loosely 
taken to mean an infinitely large system that interacts with $S$ on
time scales infinitely faster than the explicitly modeled dynamical evolution
of the state of $S$. For example, a ``particle reservoir'' exchanges particles with the
system, a ``thermal reservoir'' exchanges heat, and a ``work reservoir'' is an external system that
changes the energy spectrum of the system $S$.} In this chapter I focus on the scenario where 
the system dynamics over the time interval
is governed by a continuous-time Markov chain (CTMC). However many of the results 
presented below are more general.

Let $W_{x;x'}(t)$ be the rate matrix of the CTMC.
So the probability that the system is in state $x$ at time $t$ evolves according
to the linear, time-dependent equation
\eq{
\frac{d}{dt} p_x(t) = \sum_{x'} W_{x;x'}(t)p_{x'}(t)
\label{eq:rate_matix}
}
which I can write in vector form as $\dot{p}(t) = W(t) p(t)$. 
I just write ``$W$''
to refer to the entire time-history of the rate matrix.
$W$ and $p(0)$ jointly fix the conditional distribution of the system's
state at $t=1$ given its state at $t=0$, which I write as $\map(x_1 \mid x_0)$.
Note that in general no finite rate matrix can implement a map $\map$
that is a single-valued function. However, we can always implement such a function to
any desired finite accuracy by appropriate construction of the rate matrix~\cite{owen_number_2018}.
Accordingly, throughout this paper I will refer to CTMCs implementing
a single-valued function, when what I really mean is that it implements
that function up to any desired finite accuracy.

As shorthand, I sometimes abbreviate $x(0)$ as $x$, and sometimes abbreviate the initial distribution
$p(0)$ as $p$. (So for example, $\map p$ is the ending distribution over
states.) I will also sometimes abbreviate $p(1)$ as $p'$, and $x(1)$ as $x'$; the
context should always make the meaning clear. 

Next, define the {\textbf{entropy flow (rate)}}
at time $t$ as
\eq{
\sum_{x';x''} W_{x';x''}(t) p_{x''}(t) 
    \ln \bigg[\frac{W_{x';x''}}{W_{x'';x'}}\bigg]
\label{eq:EF_rate_def}
}
Physically, this corresponds to an entropy flow rate out of the system, into
reservoirs it is coupled to. 

In order to define an associated total amount of entropy flow 
during a non-infinitesimal time interval (EF), define $\bm{x} = (N, \vec{x}, \vec{\tau})$ 
as a trajectory of $N+1$ successive states $\vec{x} = (x(0), x(1), \dots, x(N))$, 
along with times $\vec{\tau} = (\tau_0 = 0, \tau_1, \tau_2, \dots, \tau_{N-1})$ of the
associated state transitions, where $\tau_{N-1} \le 1$ (the time of the end of the process),
$x(0)$ is the beginning, $t=0$ state of the system, and $x(N)$ is the ending, $t=1$ state of the system.
Then under the dynamics of \cref{eq:rate_matix}, the probability of any particular trajectory
is~\cite{esposito2010three,seifert2012stochastic,esposito2010threedetailed}
\begin{align}
\label{eq:trajweight}
p(\bm{x} \vert x(0)) =  \left(\prod_{i=1}^{N-1} S_{\tau_{i-1}}^{\tau_i}(x(i-1)) 
        W_{x(i), x(i-1)}(\tau_i) \right) S_{\tau_{N-1}}^1(x_{N})
\end{align}
where $S_{\tau}^{\tau'}(x) = e^{\int_{\tau}^{\tau'} W_{x, x}(t) dt}$ is the ``survival probability'' of remaining in state $x$ throughout the interval $t\in[\tau,\tau']$.
The total EF out of the system during the interval can be written as an integral
weighted by these probabilities:
\eq{
\WWW_\PP(p_0) &= \int  p_0(x_0) p(\bm{x} \vert x(0)) 
\sum_{i=1}^{N-1} W_{x(i), x(i-1)}(\tau_i)
\ln \frac{W_{x(i), x(i-1)}(\tau_i)}{W_{x(i-1), x(i)}(\tau_i)} \; D\bm{x} 
\label{eq:ctmc-ef}
}
%
(Note that I use the convention that EF reflects 
total entropy flow \textit{out} of the
system, whereas much of the literature defines EF as the entropy flow \textit{into} the 
system.) 

EF will be the central concern in the analysis below.
By plugging in the evolution equation for a CTMC, we can decompose 
EF as the sum of two terms. The first is just the change in entropy of the system during
the time interval. The second, is the \textbf{(total) entropy production} (EP)
in the system during the process~\cite{esposito2011second,seifert2012stochastic,van2015stochastic}.
I write EP as $\EP{} (p)$. It is the integral over the interval of the 
instantaneous EP rate,
\eq{
\sum_{x';x''} W_{x';x''}(t) p_{x''}(t)
   \ln \bigg[\frac{W_{x';x''}p_{x''}(t)}{W_{x'';x'}p_{x'}(t)}\bigg] 
\label{eq:ep_def}
}

I will use the expressions ``EF incurred by running a process'', 
``EF to run a process'', or
``EF generated by a process'' interchangeably, and similarly for EP.\footnote{Confusingly,
sometimes in the literature the term ``dissipation'' is used to refer to EP,
and sometimes it is used to refer to EF. Similarly, sometimes EP is instead referred to
as ``irreversible EP'', to contrast it with any change in the entropy of the
system that arises due to entropy flow.}
EF can be positive or negative. However, for any CTMC, the EP rate given in~\erf{eq:ep_def}
is non-negative~\cite{esposito2011second,seifert2012stochastic}, and therefore
so is the EP generated by the process. So
\ba
\WWW(p_0) &=& \EP{}(p_0) + S(p_0) - S(\map p_0) 
\label{eq:epv2}  \\
  &\ge& S(p_0) - S(\map p_0)
\label{eq:landauer_one_gate}
\ea
where throughout this paper, 
$\map$ refers to the conditional distribution of the state of the system at $t = 1$
given its state at $t = 0$, which is implicitly fixed by $W(t)$.

Total entropy flow across a time interval can be written as
a linear function of the initial distribution: 
\eq{
\WWW_\PP(p_0) = \sum_{x_0} \mathcal{F}(x_0) p_0(x_0)
\label{eq:ef_linear}
}
for a function $\mathcal{F}(x)$ that depends on the entire
function $W_{x;x'}(t)$ for all $t \in [0, 1)$, and so
is related to the discrete time dynamics of the entire process, $\map(x_1 \mid x_0)$. (See~\erf{eq:ctmc-ef}.) However, the \textit{minimal}
entropy flow for a fixed transition matrix $\map$, occurring when EP is zero, is the drop in entropy
from $S(p_0)$ to $S(\map P_0)$. This is not a linear
function of the initial distribution $p_0$. In addition, the entropy
production --- the difference between
actual entropy flow and minimal entropy flow --- is not a linear
function of $p_0$. (So the two nonlinearities ``cancel out'' when they are added, to give
a linearity.) These nonlinearities are the basis of much of the richness of statistical
physics, and in particular of its relation with information theory.

There are no temperatures in any of this analysis. Indeed, in this very
general setting, temperatures need not even be defined.
However, suppose that
there exists an ``energy function'' $E: X \rightarrow \R$ such that 
for all $x, x' \in X$,
\eq{
\ln \bigg[\frac{W_{x;x'}}{W_{x';x}}\bigg] = \frac{E(x') - E(x)}{T}
}
Then we can interpret the dynamics of the system at $t$
as though the system were coupled to a single heat bath
with a well-defined temperature $T$, and the energy function
obeyed ``detailed balance'' (DB) with that heat bath.
(We also say that such a rate matrix is ``reversible''~\cite{zia2007probability}.) 
If this is the case for all $t \in [0, 1]$, then  EF can be written as~\cite{esposito2010three}
\ba
\WWW = k_B T^{-1} Q
\label{eq:efheat}
\ea
where $k_B$ is Boltzmann constant, and $Q$ is the expected amount of heat transfered from the system into bath $\nu$ during the course of the process. In addition, if DB holds for a system at 
time $t$ for a heat bath with temperature $T$, then we say
that the system is \textbf{at (thermal) equilibrium} at that time if 
\eq{
p_t(x) \propto e^{-E(x) / k_B T}
}
This is just the familiar canonical ensemble from equilibrium statistical
physics~\cite{sethna2006statistical}. At thermal equilibrium, the Shannon entropy analyzed
in this paper is identical to thermodynamic entropy.

\begin{example}
Consider the special case of an \emph{isothermal} process, meaning there is a single heat bath  with time-invariant temperature $T$ (although possibly one or more work reservoirs
and particle reservoirs). Suppose that the process
transforms an initial distribution $p$ and Hamiltonian $H$ 
into a final distribution $p'$ and Hamiltonian $H'$. 
%

In this scenario, EF equals $(k_B T)^{-1}$ times the total heat flow into the bath.
EP, on the other hand, equals $(k_B T)^{-1}$ times the \emph{dissipated work} of the process, which is the work done on the system over and above the minimal work required by any isothermal process that performs the transformation $(p,H)\mapsto (p',H')$~\cite{parrondo2015thermodynamics}. 
So by~\erf{eq:landauer_one_gate} and energy conservation, the minimal work is 
the change in the expected energy of the system plus ($k_B T$ times)
the drop in Shannon entropy of the system.
This is just the change in the \emph{nonequilibrium free energy} of the
system from the beginning to the end of the process~\cite{deffner2013information,parrondo2015thermodynamics,hasegawa2010generalization}.
\label{ex:single_bath}
\end{example}

There are many different physical phenomena that can result in nonzero EP. One
broad class of such phenomena arises if we take an ``inclusive'' approach,
modeling the dynamics of the system and bath together: 

\begin{example}
Continuing with the special case of an isothermal process,
suppose that the heat bath never
produces any entropy by itself, i.e., that the change in the entropy of the
bath equals the EF from the system into the bath. 
Then the change in the sum, \{marginal entropy of the system\}
$+$ \{marginal entropy of the heat bath\} must equal
the EF from the system to the bath plus the change in the marginal entropy of the system
by itself. By~\erf{eq:epv2} though, this is just the EP of the system.

On the other hand, Liouville's theorem tells us that the \textit{joint} entropy of the 
system and the bath is constant. Combining establishes that EP of the system 
equals the change in the difference between the joint entropy and
the sum of the marginal entropies, i.e., EP equals the change in the
mutual information between the system and the bath.

To illustrate this, suppose we start with system and bath statistically independent. So
the mutual information between them
originally equals zero. Since mutual information cannot be negative, the change 
of that mutual information during the process is non-negative. This confirms that
EP is non-negative, for this particular case where we start with no statistical dependence
between the system and the bath. See~\cite{esposito2010entropy}.
\label{ex:3a}
\end{example}

Variants of \erf{eq:landauer_one_gate} are sometimes referred to in the literature
as the \textbf{generalized Landauer's bound}.
To motivate this name, suppose that there is a single heat bath, at
temperature $T$, and that the system has two possible states, $X=\{0,1\}$. Suppose
further that the initial distribution $p(x)$ is uniform over these two states, and that the
conditional distribution $\map$ implements the function $\{0,1\} \mapsto 0$, i.e., it is a 2-to-1 `bit-erasure' map. So by \erf{eq:efheat} and the non-negativity of EP, the minimal heat flow {out}
of the system accompanying any process that performs that bit erasure is 
$k_B T (\ln [2] - \ln 1 ) = k_B T \ln[2]$, in accord with the bound proposed by Landauer. 

Note though that in contrast to the bound proposed
by Landauer, the {generalized Landauer's bound} holds for systems with 
an arbitrary number of states, an arbitrary initial distribution over their states, and
an arbitrary conditional distribution $\map$. Most strikingly, the generalized Landauer bound
holds even if the system is coupled to multiple thermal reservoirs, all at
different temperatures, e.g., in a 
steady state heat engine~\cite{esposito2009thermoelectric,pietzonka2018universal}
(see Ex.~\ref{ex:multiple_baths} below). In such a case $\ktlntwo$
is not defined. Indeed, the generalized Landauer bound
holds even if the system does not obey detailed balance
with any of the one or more reservoirs it's coupled to.

Motivated by the generalized Landauer's bound, we define the \textbf{(unconstrained)
Landauer cost} as the minimal EF required to compute $\map$ on initial distribution
 $p$ using \textit{any} process, with no constraints:
\eq{
\WW(p, \map) := \SSS(p) - \SSS(\map p) \,.
\label{eq:land_cost_def}
}
With this definition we can write
\ba
\WWW(p) = \WW(p,\map) + \EP{\PP}(p)
\label{eq:efcombined}
\ea

\begin{example}
Landauer's bound is often stated in terms of the minimal amount of \emph{work} 
that must be done in order to perform a given computation, rather than the \emph{heat} that must be generated. 
This is appropriate for physical processes that both
begin and end with a constant, state-independent energy function.
For such processes,
there cannot be any change in expected energy between the beginning and end of the process. 
Moreover, by the first Law of thermodynamics, 
\[
\Delta E = W - Q(p) \,
\]
where $\Delta E$ is the change in expected energy from the beginning and end 
of the process, $W$ is work incurred by the process, and as before,
$Q(p)$ is the expected amount of heat that leaves the system and enters the bath.  
Since $\Delta E=0$, $W = Q$. So the bounds in \cref{ex:single_bath} 
on the minimal heat that must flow out of the system also give the
minimal work that must be done on the system.
\label{ex:3b}
\end{example}

Any process which achieves $\EP{}=0$ (i.e., the generalized Landauer's bound) for some 
particular initial distribution $p$ is said to be 
\textbf{thermodynamically reversible} when run on that distribution. (For simplicity,
I use this terminology even if there are no heat baths connected to the system, so that
we cannot interpret entropic costs as thermodynamic quantities.) 
In the special case that the system is coupled to a single heat
bath and obeys DB, for its dynamics over the interval $[0 1]$ to be thermodynamically
reversible the system must be at equilibrium with that heat bath at all 
$t \in [0, 1]$.\footnote{This way of 
characterizing thermodynamic reversibility has been central to statistical physics since it was invented in the 19th century.}

A necessary condition
for a CTMC to be thermodynamically reversible when run on some $q_0$ is that if 
we run it forward on that initial distribution $q_0$ to produce $q_1$, 
and then ``run the process backward'', by
changing the signs of all momenta and reversing the time-sequence of any driving
by work reservoirs, we return to $q_0$.
(See~\cite{jarzynski_equalities_2011,van2015ensemble,sagawa2014thermodynamic,ouldridge_thermo_comp_book_2018}.) Moreover, it has recently been proven that for any $\map$ and initial distribution $q_0$, we can
always design a CTMC that implements $\map$ on any initial distribution, and in addition is
thermodynamically reversible if the initial distribution is $q_0$.
(See~\cite{owen_number_2018} 
for a proof based on stochastic thermodynamics,
~\cite{turgut_relations_2009,maroney2009generalizing,faist2012quantitative,wolpert2016free,wolpert2016correction,wolpert_arxiv_beyond_bit_erasure_2015} for earlier, less detailed
analyses based on general nonequilibrium statistical physics, and \cref{sec:cyclic} and \cref{sec:hidden}
below for general discussion.) 

\begin{example}
%
Suppose we design a CTMC to implement bit erasure and to be
thermodynamically reversible if run on some initial distribution $q_0$~\cite{esposito2010finite}.
So if we run the bit erasure process backwards from the ending (delta function) distribution,
we have to arrive back at the initial distribution $q_0$.
This means that if we run that bit-erasure
process on any initial distribution $p_0 \ne q_0$ and then run it
backwards, we would not arrive back at $p_0$ (we arrive at $q_0$ instead). 
This proves that the bit-erasure process cannot
be thermodynamically reversible when run on any such $p_0 \ne q_0$. This phenomenon
is analyzed in the next subsection, and operations like bit erasure are discussed more broadly
in \cref{sec:bit_erasure}.
\end{example}

\subsection{Mismatch cost and residual EP}

\label{sec:motivation}

Computational machines are built of multiple interconnected computational devices.
A crucial concern in calculating the entropic costs of running such a computational machine
is how the costs incurred by running any of its component devices, implementing some distribution $\map$,
depends on the distribution over the inputs to that device, $p_0$.

For a fixed $\map$, we can write Landauer cost of any process that implements $\map$
as a single-valued function of the initial distribution
$p_0$; no properties of the rate matrix $W$ matter for calculating Landauer cost, beyond the fact that
that matrix implements $\map$. However, even if we fix $\map$, we cannot write EP as a single-valued
function of $p_0$, because EP \textit{does} depend on the details of how $W$ implements $\map$. 
(Intuitively, it is the EP, not the Landauer cost,
that reflects the ``nitty gritty details'' of the the dynamics of the rate matrix
implementing the computation.) In this subsection
I review recent results establishing precisely how $W$ determines the dependence of EP on $p_0$.

It has long been known how the entropy production \textit{rate},
at a single moment $t$, jointly depends on the rate matrix $W(t)$ and on the 
distribution over states $p_t$. (In fact, those dependencies are given by the expression
in~\erf{eq:ep_def}.) On the other hand, 
until recently nothing was known about how the EP of a discrete time process, evolving over
an extended time interval, depends on the initial distribution over states. 
Initial progress was made in~\cite{kolchinsky2016dependence}, in which the dependence
of EP on the initial distribution was derived for the special case where
$\pi(x_1 \mid x_0)$ is nonzero for all $x_0, x_1$. However, this
restriction on the form of $\pi$ is violated in deterministic computations. 

Motivated by this difficulty,~\cite{wolpert2018exact} extended the earlier work
in~\cite{kolchinsky2016dependence}, to give
the full dependence of EP on the initial distribution for arbitrary $\pi$.
That extended analysis shows that EP can always be written as a sum of two terms. Each of 
those terms depends on $p_0$, as well as on the ``nitty gritty details'' of
the process, embodied in $W(t)$.

The first of those EP terms depends on $p_0$ linearly. By appropriately constructing
the nitty gritty details of the system (e.g., by having the system implementing $\pi$ run
a quasi-static process), it is possible to have this first term equal
zero identically, for all $p_0$. The second of the EP terms instead is
given by a drop in the KL divergence between $p$ and a distribution $q$ that
is specified by the nitty gritty details, during the time interval
$t \in [0, 1]$. For nontrivial distributions $\map$, this term can\textit{not}
be made to equal zero for any distribution $p_0$ that differs from $q_0$.
This is unavoidable EP incurred in running the system, which arises whenever one changes the input distribution
to a different distribution from the optimal one, without modifying the device itself.

To review these recent results,  
recall the definition of islands $c$ and associated distributions $\Delta_c$
from \cref{sec:notation}. Next make the following definition:
\begin{definition}
\label{def:prior}
For any conditional distribution $\map$ implemented by a CTMC,
and any
island $c \in L(\map)$, the associated \textbf{prior} is
\eq{
    q^c \in  \argmin_{\rr : \supp(r) \in \Delta_c} \EP{\PP}(\rr) \nonumber
}
We write the associated lower bound on EP as
\eq{
    \EPmin{\PP}(c) &:= \min_{\rr : \supp(r) \in \Delta_c} \EP{\PP}(\rr)    \nonumber
}
\end{definition}
It will simplify the exposition to introduce an
arbitrary distribution over islands, $q(c)$, and define
\eq{
q^\PP(x) := \sum_{c \in L(\map)} q(c) q^c(x) \nonumber
}
In~\cite{wolpert2018exact} it is shown that
\eq{
\EP{\PP}(p) = \DDf{p}{q^\PP} - \DDf{\map p}{\map q^\PP}  + \sum_{\mathclap{c \in L(\map)}} p(c) \EPmin{\PP}(c) 
\label{eq:27}
}
where $p(c) = \sum_{x \in c} p(x)$ for all $c$.
(Due to the definition of islands, while the choice of distribution $q(c)$ affects
the precise distribution $q$ inside the two KL divergences, it has no effect on their
difference; see~\cite{wolpert2018exact}.)

The drop of KL divergences on the RHS of~\erf{eq:27} is called the
the \textbf{mismatch cost} of running the CTMC on the initial distribution $p$, and is written
as $\EF_\PP(p)$.\footnote{In~\cite{kolchinsky2016dependence}, due to a Bayesian interpretation of $q$,
the mismatch cost is instead called the ``dissipation due to incorrect priors''.} 
Given the priors $q^c$, both of these KL divergences
depend only on $p$ and on $\map$; no attributes of $W$ beyond those
that implicitly set the priors $q^c$ matter for mismatch cost. 
By the data-processing inequality for KL divergence,
mismatch cost is non-negative. It equals zero if $p = q$ or if $\pi$ is a measure-preserving
map, i.e., a permutation of the elements of $X$.

The remaining sum on the RHS of~\erf{eq:27} is called
the \textbf{residual EP} of the CTMC. It is a linear function of $p(c)$,
without any information theoretic character. In addition, it has
no explicit dependence on $\map$. It is (the $p(c)$-weighted average of) the 
minimal EP within each island. $\EPmin{\PP}(c) \ge 0$ for all $c$,
and residual EP equals zero only if the process is thermodynamically reversible. 
I will refer to $\EPmin{\PP}(c)$ as the \textbf{residual EP (parameter) vector} of the
process. The ``nitty-gritty'' physics details of 
how the process operates is captured by the residual EP  vector together with the
priors.

Combining \erf{eq:27} with the definitions of EF and of cross entropy establishes the following
set of equivalent ways of expressing the EF:
\begin{proposition}
\label{prop:EF_formula}
The total EF incurred in running a process  that 
applies map $\map$ to an initial distribution $p$ is
\begin{align*}
\WWW_\proc(p) &= \WW(p,\map) + \EF_\PP(p) + \sum_{c \in L(\map)} p(c) \EPmin{\PP}(c) \\
&= [\KKf{p}{q} - \KKf{\map p}{\map q}]  + \sum_c p(c) \EPmin{\PP}(c)
\end{align*}
\label{prop:central_equation_entropy_dynamics}
\end{proposition}
\noindent
Unlike the generalized Landauer's bound, which is an inequality, 
Prop.~\ref{prop:EF_formula} is exact. It holds for both macroscopic and
microscopic systems, whether they are computational devices or not.

I will use the term \textbf{entropic cost} to broadly
refer to entropy flow, entropy production, mismatch cost,
residual entropy, or Landauer cost. Note that the entropic
cost of any computational device is only properly defined if we
have fixed the distribution over possible inputs of the device.

It is important to realize that we can\textit{not} ignore the residual EP when calculating EF
of real-world computational devices. 
In particular, in real-world computers --- even real-world
quantum computers --- a sizable portion of the heat generation
occurs in the wires connecting the devices inside the computer
(often a majority of the heat generation, in fact). However,
wires are designed to simply copy their inputs to their outputs, which is a logically invertible map.
As a result, the Landauer cost of running a wire is zero (to within the accuracy
of the wire's implementing the copy operation with zero error), no matter
what the initial distribution over states of the wire $p_0$ is. For the same reason,
the mismatch cost of any wire is zero. This means that the entire EF incurred by
running any wire is just the residual EP incurred by running that wire.
So in real-world wires, in which $\EPmin{\PP}(c)$ invariably varies with $c$ (i.e., in which
the heat generated by using the wire depends on whether it transmits a 0 or a 1), 
the dependence of EF on the initial distribution $p_0$ must be linear. 
In contrast, for the other devices in a computer (e.g., the digital gates
in the computer), both Landauer cost and mismatch cost can be quite large,
resulting in nonlinear dependencies on the initial distribution.


\begin{example}
It is common in the literature to decompose
the rate matrix into a sum of rate matrices of one or more \textbf{mechanisms} $v$:
\eq{
W_{x;x'}(t) &= \sum_\nu W^v_{x;x'} (t)
\label{eq:W_mult_mech}
}
In such cases one replaces the definitions of
the EF rate and EP rate in~\erf{eq:EF_rate_def},\eqref{eq:ep_def}, 
with the similar definitions,
\eq{
\sum_{x';x'',\nu} W^\nu_{x';x''}(t) p_{x''}(t) 
    \ln \bigg[\frac{W^\nu_{x';x''}}{W^\nu_{x'';x'}}\bigg]
\label{eq:EF_rate_def_multi_mechanisms}
}
and
\eq{
\sum_{x';x'',\nu} W^\nu_{x';x''}(t) p_{x''}(t)
   \ln \bigg[\frac{W^\nu_{x';x''}p_{x''}(t)}{W^\nu_{x'';x'}p_x'(t)}\bigg]
\label{eq:EP_rate_def_multi_mechanisms}
}
respectively. 

When there is more than one mechanism, since 
the log of  a sum is not the same as the sum of a log, these redefined
EF and EP rates differ from the analogous quantities given by plugging 
$\sum_\nu W^v_{x;x'} (t)$ 
into~\erf{eq:EF_rate_def},\eqref{eq:ep_def}. For example,
if we were to evaluate~\erf{eq:EF_rate_def} for this multiple-mechanism $W(t)$,
we would get 
\eq{
\sum_{x';x'',\nu} W^\nu_{x';x''}(t) p_{x''}(t) 
    \ln \bigg[\frac{\sum_{\nu'} W^{\nu'}_{x';x''}}{\sum_{\nu''} W^{\nu''}_{x'';x'}}\bigg]
}
which differs from the expression in~\erf{eq:EF_rate_def_multi_mechanisms}.

Nonetheless, all the results presented above apply
just as well with these redefinitions of EF and EP. In particular, under
these redefinitions the time-derivative
of the entropy still equals the difference between the EP rate and the EF rate, total
EP is still non-negative, and total EF is still a linear function of
the initial distribution. Moreover, that linearity of EF means that
with this redefinition we can still write (total) EP as a sum of the mismatch cost, 
defined in terms of a prior, and a residual EP that is a linear function of the initial
distribution.

By themselves, neither the pair of definitions in~\erf{eq:EF_rate_def},\eqref{eq:ep_def}
nor the pair in~\erf{eq:EF_rate_def_multi_mechanisms},\eqref{eq:EP_rate_def_multi_mechanisms} 
is ``right'' or ``wrong''. Rather, the primary basis for choosing between them
arises when we try to apply the resulting mathematics to analyze 
specific thermodynamic scenarios. The development 
starting from~\erf{eq:EF_rate_def},\eqref{eq:ep_def}, for a single mechanism, can
be interpreted as giving heat flow rates and work rates for the thermodynamic 
scenario of a single heat bath coupled to the system. (See Ex.~\ref{ex:3a} and~\ref{ex:3b} above.) 
However, in many thermodynamic scenarios there are multiple heat baths coupled to the system. 
The standard approach for analyzing these scenarios is to identify each heat bath with a separate
mechanism, so that there is a separate temperature for each mechanism,
$T^\nu$. Typically one then assumes \textbf{local detailed balance} (LDB), meaning that
separately for each mechanism $\nu$, the associated matrix $W^\nu(t)$ 
obeys detailed balance for the
(shared) Hamiltonian $H(t)$ and resultant ($\nu$-specific) Boltzmann distribution defined
in terms of the temperature $T^\nu$, i.e., for all $\nu, x, x', t$,
\eq{
\frac{W^\nu_{x;x'}(t)}{W^\nu_{x';x}(t)} &= e^{[H_{x'}(t) -H_{x}(t)] / T^\nu}
\label{eq:ldb}
}

This allows us to identify
the EF rate in~\erf{eq:EF_rate_def_multi_mechanisms} as the 
rate of heat flow to all of the baths. So the EP rate in~\erf{eq:EP_rate_def_multi_mechanisms} is the rate of 
irreversible gain in entropy that remains after accounting for that EF rate and for the change
in entropy of the system.
See~\cite{van2015ensemble,esposito2010three,seifert2012stochastic}.
\label{ex:multiple_baths}
\end{example}

It is important to emphasize that
some of the analysis above assumes that there are no constraints on how
the physical system can implement $\map$. In particular, the Landauer cost given in \cref{eq:land_cost_def} and \cref{prop:EF_formula} is the
{unconstrained} minimal amount of EF necessary to implement the conditional distribution $\map$
on any physical system, when there are no restrictions on
the rate matrix underlying the dynamics of that system. However, in practice there will always be \textit{some}
constraints on what rate matrices the engineer of a system can use to implement
a desired logical state dynamics. In particular, the architectures of 
the computational machines defined in \cref{sec:comp_machines} constrain which
variables in a system implementing those machines
are allowed to be directly coupled with one another by the rate matrix.
Such constraints can substantially increase the minimal feasible EF, as illustrated
by the following example.


\begin{example}
Suppose our computational machine's state space is two bits, $x^1$ and $x^2$, and that
the function $f(x)$ erases both of those bits. Let $p_0(x)$ be the initial
distribution over joint states of the two bits. (As a practical matter,
$p_0$ would be determined by the preferences of the users of the system,
e.g., as given by the frequency counts over a long time interval in which they
repeatedly use the system.)  In this scenario, the unconstrained Landauer cost is  
\eq{
S(p_0(X)) - S(p_1(X)) &= S(p_0(X)) \nonumber \\
    &= S(p_0(X^1)) + S(p_0(X^2 \mid X^1))
}

Now modify this scenario by supposing that we are constrained to 
implement the parallel bit erasure with
two subsystems acting independently of one another, one subsystem acting
on the first bit and one subsystem acting on the second bit. This changes the Landauer
cost to 
\eq{
S(p_0(X^1)) - S(p_1(X^1)) + S(p_0(X^2)) - S(p_1(X^2) &= S(p_0(X^1)) + S(p_0(X^2)) 
}
The gain in Landauer cost due to the constraint is $S(p_0(X^2)) - S(p_0(X^2 \mid X^1))$.
This is just the mutual information between the two bits under the initial
distribution $p_0$, which in general is nonzero. 

To understand the implications of this phenomenon, suppose that the parallel bit erasing subsystems are thermodynamically
reversible \textit{when considered by themselves}. It is still the
case that if they are run in parallel as two subsystems
of an overall system, and if
their initial states are statistically correlated, then that overall system is not
thermodynamically reversible. Indeed, if we start with $p_0$, implement the
parallel bit erasure using two thermodynamically reversible bit-erasers, and then run that process 
in reverse, we end up with the distribution $p_0(x^1)p_0(x^1)$ rather than $p_0(x^1, x^2)$. 
\label{ex:double_bit_erasure}
\end{example}

This phenomenon is a key aspect of the thermodynamics of computation, coupling
the global structure of a computational machine that implements some function $\map$
to the minimal EF one could achieve with that machine by optimizing all of its
components. It is the focus of \cref{sec:subsystem} below.

It is important to emphasize that all of the results in this section
for mismatch costs and residual EP hold for processes that
implement general stochastic matrices
as well as those that implement single-valued functions, 
if one uses the appropriate generalization of islands. 

Despite the importance of EP to the entropic costs of real world systems,
after \cref{sec:circuit_entropy_dynamics} I
will assume that the residual EP of any island of any device I
consider equals zero. The resultant analysis gives the least possible
EF, which would arise if (as is almost
always the case in the real world) we don't design the prior of a device
to match the actual distribution of its input states, and so must account for its possible
mismatch cost. In addition, this assumption
simplifies the calculations, since we no longer have to consider 
the islands of the processes.

%
%

\section{Logical versus thermodynamic reversibility}
\label{sec:bit_erasure}

As mentioned in \cref{sec:entropy_dynamics}, 
it is now understood that for any initial distribution over states, and any conditional distribution
over those states, it is possible to design a process that implements that
conditional distribution and in addition is thermodynamically reversible if run
on that specified initial distribution.
In particular, this is true if the conditional distribution is a single-valued,
logically irreversible map.
As a concrete example,~\cite{diana2013finite} shows how
to build a dynamic process over the state of a (binary) quantum dot that implements
(an arbitrarily accurate approximation of) bit erasure, while having arbitrarily small total EP,
if the process is run on a given (but arbitrary) initial distribution over states. 

Ultimately, logical and thermodynamic reversibility are independent for the simple
reason that they concern different properties of physically evolving systems.
For a physical process to be logically reversible means two things. First, if the 
process is run up to time $t=1$ after
starting in some state $x_0$ at $t = 0$, it always executes some (continuous time) trajectory
$x_{0 \ldots 1}$ that ends at some specific state $x_1$ with probability 1, i.e., it is
deterministic. Second, any ending state $x_1$ that arises from some initial
state $x_0$ only arises for that initial state, i.e., the map is invertible
over the states of the system.

However, as discussed at the end of \cref{sec:gen_land},
thermodynamic reversibility involves changes in marginal distributions,
not changes in states. Moreover, for a process to be 
thermodynamically {reversible} does \textit{not} mean
it implements an {invertible} map over the space of all distributions.
Indeed, if a process implements a single-valued, non-invertible function
over states,
then 
in general the process will map multiple initial distributions
to the same final distribution. As an example, bit erasure is
a non-invertible map over the associated unit simplex, and so maps any
initial distribution to the same ending, delta function distribution.


Time-reversing a process might recover the initial distribution even if the dynamical
system is very noisy. For example, evolving a bit in a process that 
(quasi-statically) takes an initially
uniform distribution into a uniform distribution, but is so noisy
that it loses all information about the precise initial state by the time it
ends, is thermodynamically reversible but logically irreversible. Conversely,
one can easily build a system that implements the identity map 
while producing an arbitrarily large amount of EP, e.g., if one has
sufficiently high energy barriers between the states of the system. 

In sum, the fact that a given process obeys 
one kind of reversibility (or not), by itself, has no implications about whether it obeys the other kind of reversibility.\footnote{Of course, this does not mean that
there are no thermodynamic
distinctions between logically irreversible and logically irreversible
processes. To give a trivial example, the Landauer cost of a logical 
reversible process is zero, no matter what the initial distribution is, which is
not true of logically irreversible processes in general.} See~\cite{sagawa2014thermodynamic,hasegawa2010generalization,esposito2010finite,parrondo2015thermodynamics}
for more on this topic.

%

As a historical comment, it is worth noting that
modern nonequilibrium statistical physics wasn't developed when 
Landauer, Bennett, and co-workers did their pioneering work on the thermodynamics
of computation.
As a result, they had to couch their insights concerning a fundamentally
\textit{non}-equilibrium process --- computation --- in terms of
equilibrium statistical physics. 
%
Unfortunately, this led to confusion in the early work
on the thermodynamics of computation, even in some of the 
most important and path-breaking of that work (e.g., in~\cite{livi08,zure89a},
although the confusion seems to be absent in~\cite{bennett1993thermodynamics}).
%
This confusion resulted in significant controversy,
which lasted for over three decades in
the philosophy of science community
and even longer in the physics and computer science communities~\cite{maroney2009generalizing,sagawa2009minimal,dillenschneider2010comment,ouldridge_thermo_comp_book_2018}. 

Only with the recent breakthroughs in nonequilibrium statistical
physics do we have a fully formal framework for understanding the issues
that Landauer and Bennett focused on. In particular, only now do
we have the tools to analyze the relationship between
the thermodynamics of a logically irreversible computer and the thermodynamics of a
logically reversible computer that computes the same input-output function in 
a fully rigorous manner.
(See \cref{sec:fredkin,sec:bennett,sec:zurek} below for some preliminary
work related to this topic.)

\section{Conventions for calculating entropic costs in computational machines}
\label{sec:accounting_section}

The definitions of computational machines given in \cref{sec:comp_machines} are  
standard ones found in computer science textbooks. However, these definitions do
not specify enough detail of how to physically
implement those machines to uniquely fix the associated entropy dynamics.
(A related point was made in~\cite{barato2014unifying}.) 
In fact, in general there are an infinite number of different (continuous time)
Markov chains that implement the same (discrete time) dynamics of a given computational device.
Those different CTMCs will result in different entropic dynamics, 
in general. 

One example of these phenomena arise with circuits. Suppose that
we require that the dependency structure of the rate matrices in a CTMC 
we use to implement some specific circuit reflects the dependency structure
of the wiring diagram of that circuit. Concretely, this means that the 
dynamics of every physical variables in a given gate in the circuit
is not allowed to depend on any information in the circuit that resides outside of the
gate and the
inputs to that gate, even if the gate's inputs are statistically correlated with such
information in the state of other variables. When there is such a constraint
it is possible that long-range mutual information between the physical variables in different gates
gets lost as the circuit performs its computation. 
That results in nonzero irreversible entropy production in running the circuit. (This phenomenon is
analyzed in depth in~\cite{wolpert2018exact}, and summarized in \cref{sec:circuit_entropy_dynamics}
below; see also~\cite{Boyd:2018aa}.) In contrast, if there are no such constraints on how
the CTMC implements the circuit, allowing each gate to exploit the
state of any variable in the circuit, then we can achieve optimal thermodynamic efficiency,
with zero irreversible entropy production. However, the computer science definition
of a circuit does not specify whether there are such constraints on the CTMC that
can be used to to implement the circuit.

Much of the material in \cref{sec:circuit_entropy_dynamics} through \cref{sec:TMs_us}
below consists of filling in these physical details that are absent from the computer science
definitions of computational machines, and then analyzing the consequences for the entropic
costs of physical implementations of those
machines. In addition, \cref{sec:hidden} describes some of the surprising
aspects of the relationship between a given discrete time dynamical system and
the set of CTMCs that can implement that system. 

First though, in the rest of the current
section I present some limitations I will impose on how we are allowed
to ``fill in the details'' in going from a computer science definition of a computational
machine to a CTMC that implements it. Then
in \cref{sec:overwrite_inputs} I illustrate CTMCs that adhere to
these limitations and that implement some
very simple information processing systems (sub-computers, in essence),
before moving on to consider full-blown computational machines.

\subsection{Conventions that are often explicit}
\label{sec:explicit}

The goal in this paper is to focus as much as possible on the entropic costs in a physical process
that implements a computation that are
due to the computation being implemented, rather than due to details of the underlying physical system. Accordingly, as is standard in the literature, I 
assume that the Hamiltonian at both the beginning and the end of each iteration of a computational device is uniform across all states.

In addition, the processes discussed in this paper are all time-\textit{in}homogeneous,
i.e., the rate matrices can vary with time.
In practice this is typically accompanied by a time-dependence of the Hamiltonian function
of the system, $H_t(x)$, providing it arbitrary freedom at all times $t \in (0, 1)$.
(The dependence of the Hamiltonian on time is
sometimes referred to as a ``{protocol}'' in the literature.) Physically,
that time-dependence will arise if the Hamiltonian actually has the form
$H(x, \lambda(t))$, where $\lambda(t)$ is a physical variable outside of the system of
interest, whose state varies in time. However, the thermodynamics
of the evolution of $\lambda(t)$ is not considered in most of the
literature on stochastic thermodynamics. (See~\cite{deffner2013information,kawai_dissipation:_2007,still2012thermodynamics}
for some exceptions.) 

In the interests of circumscribing the 
scope of this paper, I adopt this same convention here, and simply
allow the Hamiltonian to depend explicitly on time, without considering
the thermodynamics of external systems that cause the Hamiltonian to do that.
As is also conventional in much of the literature,
from now on I choose units so that $k_BT = 1$, except where explicitly
stated otherwise.


\subsection{Conventions that are usually implicit}
\label{sec:accounting}

In addition to the explicit conventions described in \cref{sec:explicit}, there are
many other conventions that are often made implicitly, and which vary from paper to paper.
In particular, much of the confusion in the literature concerning the entropic costs of 
computers can be traced to researchers using different implicit
``accounting conventions'', for how to measure the entropic costs of an iteration
of a computational device. In general, there is no right or wrong choice for
such a convention. However, it is imperative that the convention one adopts be made
explicit. Moreover, some conventions are easier to motivate than others. 

To present the convention I will follow in this paper, first I fix some terminology.
Computational machines involve connected sub-computers, e.g., circuits involve
connected gates. Define an \textbf{iteration} of such a machine to be the
process of it simultaneously updating the state of all its sub-computers.
So each sub-computer runs the same logical map in each iteration.
Define a \textbf{run} of a machine as the entire sequence of iterations it
must run in order to complete a computation.
The number of iterations in each run is not pre-fixed for some computational machines
(and in fact might not be well-defined for certain inputs), e.g., for typical
FAs or TMs. We cannot choose units of time so that such a machine
always finishes its computation in the same interval, $[0, 1)$. So I implicitly
choose units so that each iteration of the machine occurs in a time interval of the form
$[t, t+1)$ for $t \in \Z^+$.

No computer operates in isolation. At a minimum, it must be coupled to an external
system to generate its inputs, and to a (possibly different) 
external system to record or act upon its outputs.
For simplicity assume that this coupling occurs through
an explicitly designated set of input variables and output variables
of the device, $x^{IN}$ and $x^{OUT}$, respectively. In many real-world
systems there is some overlap between $x^{IN}$ and
$x^{OUT}$ (indeed, they may even be identical). However, for simplicity,
in this paper I restrict attention to devices in which $x^{IN}$ and $x^{OUT}$ have zero overlap. 

In thermodynamics, it is conventional to focus on complete cycles of
physical systems, in which the system ends in the same state that it
started (i.e., to focus on processes that ultimately map an initial distribution
over states of the system to an identical ending distribution). 
This convention can greatly clarify the analysis.
For example, arguably the crucial breakthrough in the current understanding of
why Maxwell's demon does not violate the second law occurred when 
a full cycle of the joint gas-demon system was considered,
in which the demon's memory gets reinitialized after each iteration,
before that memory is used to record the 
next observation of the state of the gas~\cite{benn82}.

To formalize this convention in the current context, 
I define a \textbf{cyclic device} as one that has two properties.
First, all of the device's variables --- input, output, or internal (in its
sub-computers, if it has any) --- are reinitialized by the end of each run of the device,
before a next input is read in and the next run of the device begins. 
Second, the rate matrix for the device is periodic with period $1$,
the time it takes for it to run each iteration.
This ensures that if there are sub-computers in the device, that each reruns its 
logical update function in every iteration, with the same entropic cost functions of its 
initial distribution in each iteration.

In the real world, many computational devices are not cyclic. For example, often
variables are not reinitialized after the device is used; instead, they get overwritten
with new values the next time the device is used. However, not requiring that 
the devices run a complete cycle complicates the
analysis, which is (partly) why it is conventional in thermodynamics to
analyze the behavior of systems that go through a complete cycle, returning to 
the state that they started from. For the same reason, from now on I assume
that any device being discussed is a cyclic device, unless explicitly stated otherwise.

However, simply requiring that the device goes through a complete
cycle doesn't specify what portion of the entropic costs generated as
it completes that cycle are ascribed to the
device, and what portion are instead ascribed to the external systems the device interacts with
during that cycle.
To motivate a convention for how to ascribe those costs, first note that
in practice we will almost always want to have an external copy of the ending
value of $x^{OUT}$, the variable in the device that contains its output,
that persists after that variable gets reinitialized. Such a copy of the ending value of $x^{OUT}$ 
will then be used by other devices, e.g., as input to a subsequent computation, or as a signal to a controller,
or as data to be stored for later use. Importantly, we can exploit that external copy of the 
value of the output variable
to reinitialize that output variable, before the beginning of the next run of the device. 

%
I use the term \textbf{answer-reinitialization} to refer to such a process
that makes a copy of the ending state of $x^{OUT}$ in an offboard system and
reinitializes $x^{OUT}$ as it does so. I will ascribe the entropic costs of 
answer-reinitialization, if any, to the external system rather than the device
itself.\footnote{Such costs may be zero. For example,
under the presumption that such a copy is stored in a variable that was
itself in a well-specified initialized state (e.g., if that copy of the device's output is stored
in the initialized input variable of a downstream device), there is zero net Landauer cost
in the joint system during answer-reinitialization. See~\cite{parrondo2015thermodynamics}
and the discussion in \cref{sec:fredkin} below.} 

In contrast to the case with outputs, in many real world computational devices
some offboard system, perhaps even the brain of a human user of the computational
device, generates new inputs for the device. Those then need to be ``read into the input of the 
device''. This can be done with zero EF --- if $x^{IN}$ 
has been initialized to some standard value, with probability $1$, every time it gets a
new value.\footnote{If $x^{IN}$ is in its initialized state with probability $1$ whenever a next input is copied into it,
the Landauer cost of that copy operation is zero. Similarly, since the
copy operation is logically invertible,  the mismatch cost is zero.
Given the default assumption in this paper that residual
entropy costs are zero, this means that EF equals zero, as claimed.
See also~\cite{parrondo2015thermodynamics}.} 
I refer to this as the \textbf{input-reinitialization} of the device. I ascribe the
costs of input-reinitialization to the device itself. 

These two conventions, concerning how we ascribe the costs of reinitializing the inputs and the outputs,
fit together to ensure that we do not get the ``wrong'' answer for the entropic
costs of many simple computations. For example, 
suppose we want to run some device multiple times that receives inputs IID distributed
according to some distribution $p_0(x^{IN})$ and uses those inputs to produce outputs
that are distributed according to some distribution $p_1(x^{OUT})$. To accord
with common use of the term ``Landauer's bound'' (e.g., to describe the Landauer cost of
bit erasure), it would be good if the Landauer
cost that we ascribe to running the device is $S(p_0(X^{in})) - S(p_1(X^{out}))$.
Now under our convention concerning the entropic costs of the input-reinitialization,
that reinitialization adds a term $S(p_0(X^{in})) - 0$ to the Landauer cost
ascribed to the device. In addition to reinitializing its
input though, the device generates the output (after which an
external system copies the output offboard and then performs the answer-reinitialization,
at no cost to the device).
That computation results in a contribution $0-S(p_1(X^{out}))$ to the Landauer cost
(since by hypothesis, with probability $1$, $x^{OUT}$ was in its initialized state before receiving
the results of the computation). Summing these two contributions
gives a total Landauer cost of $S(p_0(X^{in})) - S(p_1(X^{out}))$ to ascribe
to the device, exactly as desired.

In addition to these two conventions, I ascribe the costs of 
\textit{generating} a new input for each successive computation
to the external system that does so.\footnote{Such 
choices not to include certain costs in the analysis are similar 
to the common choice in the stochastic 
thermodynamics literature \textit{not} to include the thermodynamics of the (usually
un-modeled) process that drives the time-variation of a CTMC, as
discussed at the end of \cref{sec:gen_land}.} 
I refer to this full set of requirements for how to ascribe the entropic costs of running
a cyclic device as the \textbf{standard accounting} convention.\footnote{Of course, 
in some circumstances one might want to use some non-standard accounting. 
However, for simplicity no such alternative accounting conventions
are considered in this paper.}

\begin{example}
Consider iterating a cyclic device $K$ times, but only answer-reinitializing
and input-reinitializing when all $K$ iterations are complete. As an example,
this is what happens to the RAM in a modern synchronous computer if you run a 
``computation'' that requires $K$ clock cycles, where you only copy
the state of the RAM  to a disk drive at the end of those $K$
cycles, after which you reinitialize the RAM, ready to receive the
input for a next computation.

Summing the Landauer costs of the $K$ iterations, almost all terms cancel,
leaving the entropy of the initial distribution over states
minus the entropy of the ending distribution over states. That difference is exactly what one would expect the minimal EF
to be, if we considered the entire sequence of $K$ iterations as a single-iteration process.
So as far as Landauer cost is concerned, the number of iterations
used to transform the initial distribution into the final distribution
is irrelevant; only the overall transformation between those distributions matters.

On the other hand, each iteration will contribute a strictly positive residual EP 
in general, with no such cancellations. Moreover, in current real-world systems, the prior
distribution over states of the system will be the same at the beginning of each iteration. 
(For example, that is the case in modern synchronous computers.) On the other hand, the actual
distribution over states will change from one iteration to the next. As a result,
each iteration will contribute a strictly positive mismatch cost in general,
without cancellation of terms.
So in contrast to the case with the Landauer cost, the number of iterations
used to transform the initial distribution into the final distribution will
have a big effect on total EP, if we don't update the prior distribution
from one iteration to the next.

However, suppose we \textit{do} update the prior in each iteration, by propagating
it forward one iteration, i.e., by applying the
conditional distribution of that iteration to the prior. 
(See \cref{sec:bayes_net}.) In this case, 
there will be cancellations just like those with Landauer costs.
Specifically, the sum of the mismatch costs over $K$ iterations will just
equal the KL divergence between the actual and prior distributions before
the first iteration begins, minus that divergence after the $K$'th iteration ends.
Since cross entropy is the sum of entropy and KL divergence, the same property
also holds for cross entropy, i.e., we get the same cancelling of terms for the
total EF, if the residual EP is zero.
\label{ex:cancelling_terms}
\end{example}

As a final comment, typically
in current real-world computers there is no copy of the original input into a device
that is maintained until the input-reinitialization process of that device.\footnote{A 
simple practical reason for this is that many real-world computational machines comprise
a huge number of computational devices (e.g., real-world circuits comprise
millions of gates), and the memory requirements for storing copies of all the inputs to all
those devices would be exorbitant. Moreover, if we were to rerun the computational
machine, then either we would have to erase all those copies of the original inputs
to make room for the new inputs -- at high
entropic cost, in general -- or use yet more storage for the new inputs. However,
see \cref{sec:fredkin} below.}
Suppose though that a copy of the initial state of $x^{IN}$ \textit{were} to persist 
until the input-reinitialization, e.g., in some external, ``offboard'' system.
Then the Landauer cost of that input-reinitialization could be reduced to zero, simply by using
that offboard copy of $x^{IN}$ to change the state of the input variable of the device
(which contains the same value $x^{IN}$) back to its initialized state.
(This point is also emphasized in~\cite{ouldridge_thermo_comp_book_2018}.)
As a result, under standard accounting, the Landauer cost of the full cycle
would be $0 - S(p_1(X^{OUT}))$, i.e., it would be \textit{negative}.
Moreover, if  a copy of the input were to persist, mismatch cost would be zero, whether or not
the computation were noisy.\footnote{To see this, use the chain rule for KL divergence twice
to expand the mismatch cost for the case where the input persists as
\eq{
& D\bigg[p_0(X^{in},X^{out}) \;\Vert\; q_0(X^{in}, X^{out})\bigg] - D\bigg[\pi(X^{out}|X^{in})p_0(X^{in})
    \;\Vert\; \pi(X^{out}|X^{in})q_0(X^{in})\bigg] \nonumber \\
& \qquad= D\bigg[p_0(X^{in} \;\mid\; X^{out}) \;\Vert\; q_0(X^{in} \mid X^{out})\bigg] 
    +  D\bigg[p_0(X^{out}) \;\Vert\; q_0(X^{out})\bigg] \nonumber \\
&\qquad\qquad\qquad\qquad \qquad\qquad\qquad\qquad- D\bigg[\pi(X^{out}|X^{in})p_0(X^{in})
    \;\Vert\; \pi(X^{out}|X^{in})q_0(X^{in})\bigg] \nonumber \\
& \qquad=  D\bigg[p_0(X^{in}) \;\Vert\;  q_0(X^{in})\bigg] - 
    \bigg(D\bigg[\pi(X^{out}|X^{in})\;\Vert\; \pi(X^{out}|X^{in})\bigg] + 
        D\bigg[ p_0(X^{in}) \;\Vert\;  q_0(X^{in})\bigg]\bigg) \nonumber \\
& \qquad=  D\bigg[p_0(X^{in}) \;\Vert\;  q_0(X^{in})\bigg] - D\bigg[ p_0(X^{in}) \;\Vert\;  q_0(X^{in})\bigg] 
            \nonumber \\ & \qquad = 0  \nonumber}
}

On the other hand, there will invariably be entropic costs involved in making the offboard copy of the input in the first place.
To properly analyze whether the entropic benefits of storing a copy of the
input in an offboard system outweigh the costs of doing so requires us to expand 
the scope of our analysis of entropic costs to include
the offboard system. This is done below in \cref{sec:fredkin}.

\newpage

For the 
rest of this paper, to keep the analysis focused on the costs of running
the device without consideration of the costs of any associated offboard systems,
I do not allow any copy of the input to the device to persist after the run finishes. More generally, I do not allow any coupling of the device with the external universe 
in a run except during
the initial process that reads in the inputs for the run or the ending 
process that copies offboard the values of the output variables of the device 
(with the entropic costs 
of those two processes not ascribed to the device). In addition, I require
that the ending value of $x^{OUT}$ of any device implementing some computational machine
only contains the output specified in the computer science definition of
the machine, as in \cref{sec:comp_machines}.


\subsection{Accumulating entropic costs until an event occurs}
\label{sec:variable_duration}

How should we measure the entropic costs incurred by a computational machine during a run
that takes more than a single iteration? 
%
If the number of iterations that the machine runs is a constant, never varying
from run to run, there is no difficulty due to having multiple iterations per
run. An example of such a \textbf{fixed-duration} (or ``fixed length'') machine
is a circuit of depth $\KK$. In each iteration $1 \le \hn < \KK$ 
the states of the gates in level $\hn$ of the circuit are used
to set the states of the gates in level $\hn + 1$, and so the
total number of iterations the circuit runs is $\KK-1$. To calculate the entropic costs of a run of the
circuit for some initial distribution over input nodes $p_0$, 
we just add up the entropic costs accrued by running all $\KK-1$ 
iterations of the circuit, starting from that distribution $p_0$. (See \cref{sec:circuit_entropy_dynamics}.)

However, the situation is more complicated if the number of iterations is
a random variable, e.g., as occurs with most FAs or TMs whose input is
generated by a random process.
This complexity can be illustrated with some recent papers in the stochastic dynamics literature
which analyze the entropic costs incurred by running a system until a first passage time occurs.
(N.b., the systems analyzed in these papers need not be computational machines.)
As an example,~\cite{neri2017statistics} analyzes the probability density function 
as a function of time $t$ for the event, \{the total EP generated by a given CTMC
first reaches some pre-specified value $\epsilon$  at $t$\}.  
As another, related example,~\cite{gingrich2017fundamental} derives formulas concerning the 
probability density function of the event, \{the
first time $t$ after a process starts that a net current reaches a given 
threshold value\}. The formulas derived in this paper involve the total amount of EP 
incurred up to $t$.\footnote{In~\cite{gingrich2017fundamental}, this is referred to as
``time fluctuations in a fixed-current ensemble''.
} 

A scientist experimentally
testing the predictions made in these papers would need to observe the system continually, to see exactly
when the event of interest occurs.\footnote{In
fact, in many scenarios the scientist is continually observing many other attributes of the system,
in addition to the bit of whether the event of interest has occurred. For example, in~\cite{gingrich2017fundamental}, the
total level of net current due to transitions that has occurred up to $t$ must
be observed, and continually updated, just to determine that bit of whether the event of interest 
has occurred.} 
Such observation by an external system
(namely the scientist) incurs entropic costs in general. Arguably, to
do the thermodynamic analysis properly, one should incorporate those costs incurred by
observing the system. (Cf., the importance in analyzing Maxwell's demon
of including the thermodynamics of its observing the system it wants to extract work from.)
Indeed, because the total time the system runs is variable, arguably one should also incorporate the entropic
costs that are incurred \textit{after} the event occurs,
unless one has yet another system that will ``turn off'' the 
primary system and observation apparatus when the event
occurs, so that they no longer generate entropic costs.

If we do not model the entropic costs of the external system observing when the
event of interest occurs, and / or do not incorporate entropic costs incurred
by the system after that event occurs, we can come to misleading conclusions. 
This is illustrated in the following example, involving a machine
that implements bit erasure, but takes either one or two iterations to do so, with the
number of iterations determined randomly.

\begin{example}
Suppose our system has four states, $\{a, b, c\}$. It starts at $t=0$ with
$P(a) = P(b) = 1/2$. The ``end of the computation'' is signaled by having
the state equal $c$. Without loss of generality we assume that $c$ is a fixed point
of the dynamics at all times.

Suppose that the process iterates a map that takes the state $a \rightarrow b$
and the state $b \rightarrow d$. 
So the full, two-iteration process implements a two-to-one
map, sending both $a$ and $b$ to $c$. In this sense, the map
implements bit erasure.

The probability that the computation ends after
the first iteration is $1/2$. The total drop in entropy that a scientist observing the
system would record by then is $\ln[2] - \ln[2] = 0$, since the state of
the system is equally uncertain at $t=0$ and $t=1$. Under standard accounting, 
this would be the Landauer cost if the computation ended at $t=1$.

The probability that the computation instead ends after the second iteration is also $1/2$. However,
the total entropy drop if the computation ends then is just the normal bit-erasure 
Landauer cost, $\ln[2]$. So {\rm{if we were to stop accumulating drops in entropy
when the computation ends}}, then the expected Landauer cost would
be $(1/2) \cdot 0 + (1/2) \ln [2] = \ln [2] / 2$. This is half the Landauer cost
for bit erasure in a single iteration, the value given by ``Landauer's bound''. 
On the other hand, 
if we calculated the total drop in entropy during the full interval from $t=0$ to $t=2$,
ignoring whether the computation has ended early, under standard
accounting we would again get a value
that equals the Landauer bound for bit erasure, $\ln[2]$.

So it might seem that we can avoid Landauer's bound, cutting the drop in entropy
by half, on average, so long as we introduced stochasticity concerning when exactly the
bit erasure completes.
\end{example} 

These kinds of issues concerning the entropic costs of runs that last for a random number of iterations 
arise when interpreting some of the early work on the thermodynamics
of Turing machines, in which entropic costs are only accumulated
up to the time that the TM halts, and no consideration is given to
the entropic costs of an external observation apparatus watching to see exactly when
the TM halts. (See the discussion in \cref{sec:TMs_us}.)

In the next section I review earlier work that (implicitly) used standard accounting
to analyze the entropic costs of any device that in one iteration implements
an arbitrary transformation of its entire state space, without any constraints
on how it implements that transformation. 
Then in the rest of this paper I consider
full computational machines, all of which run for multiple iterations,
and have constraints of one type or another
on how they process their inputs, and therefore have a more complicated
relationship between the computation they perform and the entropic costs they
incur as they do so. Except in \cref{sec:fredkin} and (to a degree)
in \cref{sec:strasberg}, in those sections I restrict attention to the entropic costs of cyclic devices,
under standard accounting.

\section{Entropy dynamics of unconstrained devices}
\label{sec:overwrite_inputs}

Extending the definition of an AO circuit, define an \textbf{all at once} (AO) device that
 implements $\map$
in some fixed time interval as some device
such that for some initial distribution $p_0$, the EF generated
by running the device on $p_0$ is $S(p_0) - S(\map p_0)$. As illustrated
below, much of the richness in the thermodynamics of computation
arises when we consider non-AO devices. However, to illustrate the
concepts introduced above, in this section
I begin by reviewing the thermodynamics of AO cyclic devices under standard
accounting. 

\erf{eq:landauer_one_gate} gives the theoretical minimal EF that could be produced
by any process that applies a conditional
distribution $\pi(x_1 \mid x_0)$, relating the state of the system at time $1$ to
its earlier state at time $0$, to an initial distribution $p_0$. 
However, that equation doesn't say whether the bound it provides can be achieved,
and if so how.

For a process to achieve that bound means that it
is thermodynamically reversible for the initial distribution $p_0$. 
A large body of literature has developed showing that for certain
maps $\map$, if we are given any
$p_0$, it is possible to construct an associated quasi-static process 
that both implements $\map$ and is arbitrarily close to thermodynamic reversibility if run on 
$p_0$~\cite{parrondo2015thermodynamics,hasegawa2010generalization,takara_generalization_2010}.
Many of these papers were motivated by considering bit erasure. 
However, bit erasure has the unusual property that 
the distribution over outputs $x_1$ is independent of the precise input $x_0 \in X$.
Reflecting this, even when they considered other $\map$'s besides bit erasure,
these papers implicitly restricted themselves to consideration of
maps $\map(x_1 \mid x_0)$ where the distribution over outputs is independent of the 
precise input $x_0$.

This restriction means the analyses in these papers do not directly apply 
to the most common maps performed by gates in real computers,
in which the output distribution is \textit{dependent} on the
initial state, e.g., the logical AND map, the bit flip, etc.
(See~\cite{maroney2009generalizing} for an early discussion touching on this limitation.)
Indeed, in quasi-static processes in which the system is in equilibrium with the heat
bath at all times, 
\textit{all information about the initial state is lost by the end of the process}, being
transferred to the heat bath. (This can be shown explicitly using stochastic
thermodynamics; see \cref{sec:hidden}.) Hence, no such process over $X$
can implement a map $\map$ in which the output distribution depends on the input state.

However, no matter what $\map$ and $p_0$ are, it is now
known how to construct a cyclic device that implements $\map$,
and whose EF equals the Landauer cost in the special case where the prior of the device
equals $p_0$ --- if one uses 
a process defined over a space that is strictly larger than
$X$~\cite{turgut_relations_2009,maroney2009generalizing,faist2012quantitative,wolpert2016free,wolpert2016correction,wolpert_arxiv_beyond_bit_erasure_2015,owen_number_2018,wolpert_minimal_2017}.
The key idea is to expand the space of the system beyond
the original space $X$, used to define $\map$ and $p_0$, into a 
``partially hidden'' space $X\times X'$,
where a specific state in $X'$ (here written as $\bot$) is identified as the ``initialized
state'' of $X'$. The analyses in those papers show how to design an associated ``partially hidden'' 
cyclic device, with a rate matrix $\mathscr{W}$ operating over $X \times X'$, such that 
any initial value $(x_0, \bot) \in X \times X'$ gets mapped to an ending distribution
$(\map \delta(x, .), \bot)$. 
So one can read off the dynamics of the ``visible'' device $\map$ evolving over
$X$ from the dynamics of the ``visible states'', $X$ of the 
expanded device with rate matrix $\mathscr{W}$
evolving over $X\times X'$.

In these analyses, the prior of the device operating over $X'\times X$ 
is assumed to be of the form $q_0(x)\delta(x', \bot)$ for some arbitrary
distribution $q_0(x)$. It is 
shown in~\cite{wolpert2016free,wolpert2016correction,wolpert_arxiv_beyond_bit_erasure_2015} that if
$q_0 = p_0$, then this hidden device implements the map over $\map$ over $X$
and incurs zero EP, achieving the Landauer bound for applying $\map$ to $p_0$.
In addition, it is explicitly shown there that if $q_0 \ne p_0$, then
the resultant mismatch cost of the hidden device over $X \times X'$ 
is exactly the mismatch cost
for the original device implementing the distribution $\map$, given by
evaluating ~\erf{eq:27} for 
the initial distribution $p_0(x)$ over the visible space $X$, the prior $q_0(x)$
over $X$, and the conditional distribution $\map$.

In more detail, as it is formulated in~\cite{wolpert2016free,wolpert2016correction,wolpert_landauer_2016a},
this 
partially hidden device works as follows. First, the device stores a copy of $x_0$
in $X'$. So $|X'| = |X|$. (As described in~\cite{parrondo2015thermodynamics}, since 
$X'$ starts in its initialized state with probability 1, this
copy operation is logically reversible, and so has both zero Landauer cost and
zero mismatch cost.) Then, without changing the value of this copy, 
that copy is used to guide
a quasi-static process that evolves the 
distribution $p_0$ over the visible states, to implement $\map$. After this is done
the value $x' \in X$ is reinitialized to $\bot$.
This reinitialization is guided by the ending value $x_1$, and is based on the prior
$q_0(x)$, which must govern the value of the copy of $x_0$ stored at this point
in $X'$. It is this reinitialization step that results in the mismatch 
cost. After all this is done,
as required by standard accounting, a copy of $x_1$ is copied to an offboard
system, and $X$ is reinitialized, in preparation for the next input.~\footnote{This
type of construction using a hidden space $X'$ was rediscovered 
in ~\cite{garner2017thermodynamics} and App. A of~\cite{Boyd:2018aa}.}


In addition, in many real-world physical systems, there is not a single, fixed
actual distribution $p_0$ of the initial states of a system, i.e., 
there is stochastic uncertainty about the distribution
of the ``inputs'' to the device. So have a distribution over
such distributions, $P(p_0)$. For example, in a real-world digital computer, a given
user implicitly specifies some distribution $p_0$ of inputs to the computer.
However a different user would have some different distribution $p'_0$ over
inputs to the computer. In this case $P$ is the probability distribution over
which user is using the computer. As another example, a cell operating in some
particular environment will be subject to a distribution $p_0$ over
physical characteristics of that environment, e.g., over macromolecules
it may be detecting in that environment. In this case a given environment
specifies a distribution $p_0$, and $P$ is the distribution over environments.
~\cite{wolpert_arxiv_beyond_bit_erasure_2015} analyzes the implications of such
distributions over initial distributions $p_0$ for the entropic costs of the
device, showing that the Landauer cost increases by a Jensen-Shannon divergence
between $P(p_0)$ and the associated distributions $p_0$. (This 
Jensen-Shannon divergence is called ``entropic
variance'' in~\cite{wolpert_arxiv_beyond_bit_erasure_2015}.)
Finally,~\cite{wolpert_arxiv_beyond_bit_erasure_2015,wolpert2016free,wolpert2016correction} 
also shows how
to extend this analysis to the case where the set of visible states $X$ 
is a coarse-graining over some some underlying space. 

This early analysis showing how to use an extra, hidden space $X'$
to implement an arbitrary conditional distribution $\map$
raises many interesting research questions. For example,
since hidden state spaces can be expensive 
to build into physical devices in the real world, this early analysis raises
the question of what the fundamental limits are on the minimal $X'$ that is required in
order to implement a desired $\map$. Some results concerning this question and related ones
are summarized in \cref{sec:hidden}. In particular, 
it turns out that augmenting the space $X$ with additional hidden states
is {necessary} to implement nontrivial $\map$, \textit{even without introducing
considerations of  thermodynamical reversibility}. Surprisingly, as
discussed in \cref{sec:hidden}, it is impossible for \textit{any} CTMC to
implement any non-trivial map $\map$.

\section{Effects on entropic cost of constraining the rate matrix}
\label{sec:subsystem}

All modern physical systems that implement computational machines use sub-computers that
can only access a subset of the variables in the overall machine. This means
that there are constraints on how the overall machine can operate. These constraints are a major distinction between computational machines and the systems analyzed in \cref{sec:overwrite_inputs}. 
In this section I review some recent results relating the precise constraint on what
variables a given sub-computer inside a computational machine is allowed to access,
and the consequences of that constraint for the entropic costs of
running the machine. 

As notational shorthand, I assume that any
conditional distribution governing the dynamics of a device operating within an
overall computational machine that is written in the form $\map(y \mid x)$
is implemented with a CTMC and Hamiltonian that do not involve any of the variables in the overall 
computational machine other than $x$ and $y$. As an example, suppose we
have an overall system (computational machine) with three subsystems (devices), $A, B, Z$,
having states $x, y$ and $z$, respectively. Suppose that the device
$B$ gets its input $x$ from the output of device $A$
and uses that to compute its output $y$ during some specific iteration $n$ of
the overall system. Suppose further that to reflect engineering constraints
in how we can build the device, 
we want to make sure that we do not model the system with a CTMC and Hamiltonian 
that ever couple the state of $B$'s variables to any external variables besides
the state of $A$. Then I write the update distribution
for device $B$ as $\map(y \mid x)$. On the other hand, suppose that $B$ implements the
same logical function taking $x \rightarrow y$, 
but that the Hamiltonian governing the system during iteration $n$ is allowed
to couple the variables in $B$ to the values of some external variable $z$ in the overall system
that does not lie in $A$.\footnote{As discussed below, such 
coupling would often allow the entropic costs of running $B$ to be
reduced, even if
the evolution of $y$ is independent of $z$. Loosely speaking, this is the case 
if the initial value of $y$ is statistically coupled with the value $z$. 
In such a case, the physical process that updates the state $y$ can
be more thermodynamically efficient if it can exploit the information about $y$
that is contained in $z$.}
In this case I write the conditional distribution governing $B$'s computation
as $\map(y \mid x, z)$, even though the distribution over values $y$
is independent of the precise value of $z$. The advantage of this
shorthand it that is allows us to use the form of the update distribution $\map$ to specify
constraints on which physical variables are allowed to be directly physically
coupled to which other ones, e.g., through an interaction Hamiltonian.

Under this notation, any {physical system} implementing $\map(y \mid x)$ 
is subject to extra constraints on what rate matrix CTMC it can be implemented with, compared
to a physical system implementing $\map(y \mid x, z)$.
However, typically a physical process obeying such constraints on the rate matrix CTMC and Hamiltonian
cannot achieve the minimal EF given in~\erf{eq:landauer_one_gate} 
(which requires a constraint-free system). This will result in unavoidable EP.

\subsection{Subsystem processes}

To make these general considerations precise, I now
introduce a type of process which will play a central role in 
the rest of this paper:
\begin{definition}
Suppose that the 
rate matrix of a CTMC defined over $X = X_A \times X_B$
can be expressed as a sum over mechanisms $\nu$ of associated
rate matrices, each of which is of the form 
\begin{align*}
W^v_{x_A,x_B; x'_A, x'_B}(t) = W^{\nu}_{x_A;x'_A}(t) \delta(b', b)
 +  W^{\nu}_{x_B;x'_B} \delta(a', a)
\end{align*}
for appropriate rate matrices $W^{\nu}_{x_A;x'_A}(t), W^{\nu}_{x_B;x'_B}(t)$.
Then the CTMC is a \textbf{subsystem process} over $X_A \times X_B$.
\label{def:subsystem_proc}
\end{definition}
\noindent (See~\cite{wolpert2018exact} for a more general definition, applicable 
even if we do not assume that the system evolves according to a CTMC.)
As an example, a physical system can evolve as a subsystem process if the 
reservoirs it is attached to are infinitely large (so that $A$ and $B$ cannot couple
indirectly through the reservoirs), 
and if at all times $t\in[0,1]$ the energy function decouples subsystems
$A$ and $B$; i.e., the Hamiltonian obeys
\[
H_{t}(x_{A},x_{B})=H_{t}^{A}(x_{A})+H_{t}^{B}(x_{B})
\]
for all $t\in[0,1]$.

All the computational machines considered in this paper operate with synchronous
rather than asynchronous updating, in which each device inside an overall computational
machine runs in a ``modular'' fashion, not reacting to interrupts
from outside of itself while it runs. Physically, this means that 
once the variables in any such device are set to their initial values,
the device is run by a process in which those variables evolve independently of the rest of
the variables in the overall computational machine. So it is run by a subsystem process.
As an example, any process that runs a gate in a circuit is a subsystem process.

The following result in proven in App. \ref{sec:subsystem_proc_proof}:

\begin{proposition}
\label{prop:subsystem_proc_props} 
For any subsystem process over a state space $X_A \times X_B$,
\begin{enumerate}
\item Subsystems $A$ and $B$ evolve independently over that interval, 
i.e., the conditional distribution of the state evolution is of the form
\begin{align*}
\map^{A,B}(a_1, b_1 \vert a_0, b_0) = \map^A(a_1\vert a_0)\map^B(b_1\vert b_0) 
\end{align*}
\label{enu:fracsystem1}
\item The EF of the joint system during the process can be written as
\begin{align*}
\WWW_{\PP}(p^{A,B}_0)=\WWW_{A}(p^{A}_0) + \WWW_{B}(p^{B}_0)
\end{align*}
where  $\WWW_{A}(p_A)$ is the EF that would be incurred 
for a CTMC over state space $X_A$ with rate matrices 
$W^{\nu}_{x_A;x'_A}(t)$
(i.e., if subsystem $B$ did not exist at all), and similarly for $\WWW_{B}(p_B)$.
\label{enu:fracsystem2}
\end{enumerate}
\end{proposition}
\noindent
In light of \cref{prop:subsystem_proc_props}(\ref{enu:fracsystem2}), 
I sometimes refer to $\WWW_{A}(p^{A}_0)$ as the \textbf{subsystem EF} of subsystem $A$, and
similarly for $\WWW_{B}(p^{B}_0)$. I also sometimes refer to $\WWW_{\PP}(p^{A,B}_0)$
as the \textbf{system-wide EF}. 

Note that by the Second Law of Thermodynamics~\cite{seifert2012stochastic}, 
$\WWW_{A}(p^A_0) \ge 
S(p^A_0) - S(p^A_1)$, where equality arises for appropriate (typically quasi-static)
rate matrices $W^{\nu}_{x_A;x'_A}(t)$. (Similar considerations hold for $\WWW_{B}(p_0^B)$.)
Accordingly, I refer to $S(p^A_0) - S(p^A_1)$
as the \textbf{subsystem Landauer cost} of subsystem $A$, and
and write it as $\subWW(p^A_0,\map^A)$, 
or just $\subWW(p^A_0)$ for short.
(Again, similar considerations hold for $S(p^B_0) - S(p^B_1)$.) 

Along the same lines, I write
the minimal EF of any process that implements $\map(a_1, b_1 \vert a_0, b_0)$ as
\eq{
\subWW(p^{A,B}_0,\map^{A,B}) = S(p_0^{A,B}) -  S(\map p_0^{A,B}) 
}
I refer to this as the \textbf{system-wide Landauer cost}, or sometimes as 
the AO Landauer cost, since it's the Landauer cost of an AO device that
implements $\map$.

While the EF of a full system undergoing a subsystem process is
additive over the EFs of its subsystems, in general the {Landauer cost} of the
full system undergoing a subsystem process
is \textit{not} additive over the Landauer costs of its subsystems. More precisely,
in general the sum of the subsystem Landauer costs, 
$S(p^A_0) - S(p^A_1) + S(p^B_0) - S(p^B_1)$, differs
from the system-wide Landauer cost, $S(p^{A,B}_0) - S(p^{A,B}_1)$. This
reflects the fact that the system-wide Landauer cost is
the minimal EF of implementing $\map^{A, B}$ using \textit{any} process, 
without imposing the constraint that the process must 
be a subsystem process. 

I will refer to the difference
${\mathcal{L}}(p^A_0,\pi^A) + {\mathcal{L}}(p^B_0,\pi^B) 
        - {\mathcal{L}}(p^{A,B}_0,\pi^{A,B})$
as the \textbf{Landauer loss} of the subsystem process.
It equals the change in mutual information
between subsystems $A$ and $B$ during the process:
\begin{align}
& \subWW(p^A_0,\map^A) + \subWW(p^B_0,\map^B) - \subWW(p_0^{A,B},\map^{A,B}) \nonumber \\
& \qquad = S(p^A_0) + S(p^B_0) - S(p_0^{A,B}) \nonumber \\
&\qquad\qquad\qquad - \bigg[S(\map^A p^A_0) + S(\map^B p^B_0) 
        - S(\map^{A,B} p_0^{A,B})\bigg] \nonumber \\
& \qquad =I_{p^{A,B}}(A ; B) - I_{\map p^{A,B}}(A ; B) 
\label{eq:drop_in_mutual}
\end{align}
A simple example of Landauer loss was presented in \cref{ex:double_bit_erasure}. 
The concept of Landauer loss is extended to involve more than
two random variables in \cref{sec:circuit_entropy_dynamics}, in order to
analyze the entropic costs of circuits. In particular,
when there are more than two random variables,
the mutual information terms in \cref{eq:drop_in_mutual} get replaced by
multi-information terms.

Now that we have the appropriate definitions for distinguishing the EFs and Landauer
costs of the subsystems and the entire system, we can do the same for their EPs.
Define the \textbf{system-wide EP} of a 
subsystem process, $\sigma(p^{A,B}_0)$, as the difference between the 
system-wide EF of that process and the system-wide Landauer cost of the process. 
Then by combining the additivity of EF, the nonadditivity
of Landauer cost, and Eq.~\eqref{eq:efcombined}, we see that
the system-wide {EP} of a subsystem process is not additive over the EPs of its subsystems. 
To make this more precise, define the \textbf{subsystem EP} of subsystem $A$ as 
\begin{align}
\subEP{A}(p^{A}) = \WWW(p^A)  - \subWW(p^A,\map^A)
\label{eq:C3}
\end{align}
where for simplicity I have dropped the subscript on $p^A$ that indicates it is
an initial distribution. (I make analogous definitions for subsystem $B$. )
Since $\WWW(p^A)$ is linear in $p^A$,
we can apply the decomposition of EP in \cref{sec:motivation} to 
subsystem processes and write
\begin{align}
&\subEP{A}(p^{A}) \;=\; \sum_{\mathclap{c\in L(\map^{A})}} p^{A}(c)  \nonumber \\
& \;\;\; \times \Big(\DDf{p^{A}({c})}{ q^{A}({c})} -\DDf{\map^{A}p^{A}({c})}{\map^{A}p^{A}({c})} + \subEPmin{A}(c)\Big)
\label{eq:subsystem_EP_decomp}
\end{align}
where in analogy to the development in \cref{sec:entropy_dynamics}, 
the \textbf{subsystem prior} $\qcPP_A$ is any distribution
\[
q^{A}({c}) \in  \argmin_{\rr \in \Delta_c} \subEP{A}(\rr)
\]
Again following the development in \cref{sec:entropy_dynamics},
it can be useful to re-express $\subEP{A}(p^{A})$ as the sum of two terms.
The first is the \textbf{subsystem mismatch cost},
\[
\subEF{A}(p^{A})=\DDf{p^{A}}{ q_{A}}-\DDf{\map^{A}p^{A}}{\map^{A}q_{A}}
\]
(where $q_A(x_A) = \sum_c q_A(c) \qcPP(x_A)$ and as before
the choice of the probabilities $q_A(c)$ is arbitrary). The second is the
\textbf{subsystem residual EP}, 
\ba
\sum_{{c\in L(\map^{A})}} p^{A}(c) \subEPmin{A}(c)
\label{eq:subep}
\ea
(I make an analogous decomposition for subsystem $B$.) 
Note that by \cref{eq:C3},
$\subEP{A}(p^A) \ge 0$ for all $p^A$. 
In particular, this must be true if $p^A(c) = q^A(c)$ for all $c \in L(\map^A)$.
Plugging that into \cref{eq:subsystem_EP_decomp}, we see that
subsystem residual EP is non-negative.

Finally, expand
\begin{align}
\sigma(p) &= S(\map p) - S(p) + \WWW(p)  \nonumber \\
   &= S(\map p) - S(p) + \WWW(p^A)  + \WWW(p^B)  \nonumber \\
   &= S(\map p) - S(p)  \nonumber \\
    & \qquad + \sigma_A(p^A) + \subWW(p^A,\map^A) + \sigma_B(p^B) + \subWW(p^B,\map^B)
\label{eq:full_EP}
\end{align}
where I have used \cref{eq:C3}. Eq.~\eqref{eq:full_EP} formalizes the
statement above, that system-wide EP is not additive over its subsystems. 
Eq.~\eqref{eq:full_EP} also can be used to establish the following:

\begin{corollary}
Given any subsystem process that implements $\map$ on initial distribution $p^{A,B}$,
\begin{enumerate}
\item The minimal system-wide EP equals the Landauer loss.
\item The minimal system-wide EP equals the system-wide mismatch cost.
\item If the process achieves the minimal EP and the priors of the
two subsystems are unique, then those priors are the marginals of $p^{A, B}$,
i.e., $q^A = p^A$, $q^B = p^B$.
\end{enumerate}
\end{corollary}
\begin{proof}
Fix the initial distribution $p$ and the conditional distribution $\map$.
Eq.~\eqref{eq:full_EP}
tells us that the minimal EP generated by any system that applies
$\map$ to $p$ while using a subsystem process occurs when the two subsystem 
EPs achieve their minimal values, i.e., when $\sigma_A(p^A) = \sigma_B(p^B) = 0$.
(For example, this could arise if the subsystems evolve quasi-statically \cite{esposito2012stochastic}.)
So the minimal system-wide EP is
$\subWW(p^A,\map^A) + \subWW(p^B,\map^B)  - \subWW(p,\map)$. This establishes the
first claim.

Next, the system achieves the minimal EP iff the two subsystem EPs
equal zero. In turn, they equal zero iff the subsystem residual EPs of both
subsystems equals $0$. This means that the residual EP of the
entire system is zero. That in turn implies that the system-wide EP must equal
system-wide mismatch cost, establishing the second claim.

Finally, again note that in order to achieve the minimal system-wide EP
for initial distribution $p^{A,B}$, the subsystem
EPs must both be zero. This means that their mismatch costs must both be zero
when the full system is run on distribution $p^{A,B}$.
Therefore a prior for subsystem $A$ is given by $p^A$, and similarly
for subsystem $B$. 
\end{proof}

\subsection{Solitary processes}
\label{sec:solitary}

An important special type of subsystem process is one in which
$W^\nu_{x_B;x'_B}(t) = 0$ for all $\nu, x_B, x'_B \ne x_B$ and $t \in [0, 1]$.
Any such subsystem process is called a \textbf{solitary process over $A$}.
I will sometimes refer to the value $x_A(0)$ in a solitary process as the
\textbf{input} to that process, with $x_A(1)$ being its \textbf{output}.
By~\cref{prop:subsystem_proc_props}, in a solitary process $\WWW_B(p_0^B)$
equals the EF that arises by evolving a system over $X_B$ with rate matrices
$W^\nu_{x_B;x'_B}(t)$. So in a solitary process over
subsystem $A$, $\WWW_B(p_0^B)$ equals zero identically, 
and therefore the EF of a solitary process equals the EF of subsystem $A$.
Similarly, since the state of subsystem $B$ cannot change in the subsystem
process, its subsystem Landauer cost is zero, and therefore its subsystem
EP is zero.

It is often easier to analyze sequences of solitary processes than sequences of
subsystem processes. Moreover, computational machines 
that are conventionally defined with some of their
devices running in parallel can almost always be redefined
to have those devices run serially,
according to some convenient topological order that respects the original
definition of the machine. We can then model the running of that
machine as a sequence of solitary 
processes.\footnote{As an example, suppose the computational machine
is a circuit, with the gates being the subsystems in question.
The associated topological order is specified
by the circuit's DAG, and running the gates serially according
to that order ensures that the circuit implements the same input-output map 
as does sequentially running various subsets
of the gates in parallel, as specified in the original DAG.} This allows
us to analyze the entropic costs of running the full system by decomposing
it into a sum of entropic costs of solitary processes.

Note though there are devices in real-world 
computational machines that are not governed by {solitary} processes. 
For example, suppose that the gates in a circuit other than some gate $g$ 
use dynamic (``active'') memory to maintain their values
while $g$ runs. Then the states of those gates are in a nonequilibrium 
steady state (NESS) while $g$ runs, i.e., the terms in the rate matrix that govern
their dynamics are not all zero. (Intuitively, the entropy flow in
such an NESS --- the work done on the system ---
is needed to maintain the NESS against the equilibrating thermodynamic
force arising from coupling with the heat bath(s).) In this case $g$ is run with a subsystem
process that is not a solitary process. In the interests of space, such situations
are not considered in this paper, or equivalently, it is assumed
that the entropic costs generated by maintaining NESSs are arbitrarily small
on the scale of the entropic costs of running the sub-computers that change
the distributions over the states of some variables. (See \cref{sec:future_work}).

Suppose we have a solitary process over a subsystem $A$ in which $A$ is
initially statistically coupled with some other subsystem $B$, and that
the subsystem EP of $A$ is zero. Suppose that that initial statistical coupling between
$A$ and $B$ is reduced as the process unfolds. Then by Eq.~\eqref{eq:full_EP}, the
system-wide EP, and therefore the Landauer loss, are nonzero, even though
the two subsystem EPs equal zero.

It might seem paradoxical that the Landauer loss in a solitary process can be nonzero.
After all, both the system-wide Landauer cost and the subsystem Landauer cost
involve processes over the same state space (namely, $X_A \times X_B$), and
implement the same conditional distribution $\Map$ over that space. How can
two processes, over the same space, implementing the same conditional distribution,
have different minimal EFs, especially if one of the subsystems doesn't even change 
its state under that conditional distribution?

The reason for the difference in minimal EFs is that system-wide Landauer cost is the minimal 
EF of any rate matrix that implements $\Map$, but
the subsystem Landauer cost is the minimal EF of any rate matrix that
implements $\Map$ 
\textit{while also satisfying the extra constraints defining a solitary process}.
Due to those constraints, while the first kind of rate matrix is allowed to exploit 
the values of $x_B$ as well as $x_A$ to
drive the dynamics of $x_A$, the second kind is not allowed to do that.
More precisely, in the process whose EF equals the system-wide Landauer cost, for all mechanisms
$v$, and all $t\in[0,1]$, 
\eq{
W^v_{x_A,x_B; x'_A, x'_B}(t) = K^v_{x_A; x'_A, x'_B}(t) \delta(x_B, x'_B)
}
for some functions $K^v_{x_A; x'_A, x'_B}(t)$ that vary with changes to $x'_B$.
This is \textit{not} a solitary process (indeed, it violates the definition of subsystem
processes). Yet it clearly leaves $x_B$ unchanged. Moreover,
\textit{for any given initial distribution} $p_0(x_A, x_B)$, we can design the 
functions $K^v_{x_A; x'_A, x'_B}(t)$ so that the CTMC not only implements $\Map$, but also
varies how the distribution at each $x_A$ value evolves in intermediate times $t \in (0, 1)$, 
based on the associated $x_B$ value, in such a way that zero 
system-wide EP is generated for the given initial distribution  
$p_0(x_A, x_B)$. It is this second capability that we exploit to reduce the EF below
the value that can be achieved with a solitary process.

This is illustrated in the following example.

\begin{example}
Consider a solitary process where $X = X_A \times X_B$ and both $X_A$ and
$X_B$ are binary. Suppose that under that solitary process $x_A$ undergoes 
bit erasure. So
\eq{
\Map(x^1_A, x^1_B \mid x^0_A, x^0_B) &= p(x^1_A \vert x^0_A) \delta(x^1_B, x^0_B) \nonumber \\
    &= \delta(x^1_A, 0)\delta(x^1_B, x^0_B) 
\label{eq:38first}
}
Suppose as well that initially, $x_A = x_B$ with probability 1, i.e., 
the two variables are initially perfectly correlated. So the Landauer loss, i.e., the drop in
mutual information, i.e., the minimal system-wide EP, is $\ln [2]$.

We can derive this value for the Landauer loss in more detail as follows.
First recall that EF is additive under a solitary process. Since $x_B$ does not change, this means that
the minimal EF of any solitary process that implements $\Map$ is the same as
the minimal EF of a process that implements the conditional distribution
$p(x^1_A \mid x^0_A)$ on the initial distribution $p(x_0^A)$. This
is $\ln [2]$. On the other hand, the entropy over the joint space is $\ln[2]$ both
under the initial distribution and under the final distribution. So the minimal
EF of a dynamics over the joint space is 0. Subtracting gives the value of the 
Landauer loss: $\ln [2]$.

To illustrate how $\Map$ can be implemented with zero EF, suppose that the system were 
connected to a single heat bath at a constant temperature, and obeyed LDB. Suppose as well that the Hamiltonian
over $X_A \times X_B$ evolved according to the
kind of ``quench then quasi-statically evolve'' process described in~\cite{parrondo2015thermodynamics,hasegawa2010generalization,takara_generalization_2010}. That
means that when the process begins, the Hamiltonian is instantaneously changed to
\eq{
H_0(x_A, x_B) &= -\ln[p(x_A(0), x_B(0))]   \nonumber \\
    &\propto -\ln\big[\delta(x_A(0), x_B(0)\big]   \nonumber \\
    &= 
        \begin{cases}  
            \ln[2] & \text{ if $x_A(0) = x_B(0)$} \\ 
            \infty & \text{ otherwise.} 
        \end{cases} 
}
(where ``$\infty$'' is shorthand for some arbitrarily large finite number).
This quench happens so fast that the distribution over the states
doesn't have time to change. So the system automatically starts at thermal equilibrium. 

After this starting quench, the Hamiltonian quasi-statically evolves to (be arbitrarily close to)
\eq{
H_1(x_A, x_B) &= -\ln\big[p(x_A(1), x_B(1))\big]  \nonumber \\
    &\propto -\ln\big[\delta(x_A(1), 0)\big]   \nonumber \\
    &= 
\begin{cases}  
\ln[2] & \text{ if $x_A(1) = 0$} \\ 
\infty & \text{ otherwise.} 
\end{cases} 
}
while keeping an infinite energy barrier between $(x_A = 0, x_B = 0)$ and $(x_A = 0, x_B = 1)$,
and an infinite energy barrier between $(x_A = 1, x_B = 0)$ and $(x_A = 1, x_B = 1)$.
(In other words, keeping the rate matrix entries zero for all transitions that would 
change $x_B$.) This evolution implements the desired map $\Map$.
Moreover, because the evolution is quasi-static, the system is always at thermal equilibrium, 
i.e., there is zero EP.\footnote{Note 
that if the initial distribution were not uniform, then this particular quench step would not
result in the system starting at thermal equilibrium, which would cause EP to be generated
as the system quasi-statically evolves. This illustrates the fact that any process
can be thermodynamically reversible for at most one specific initial distribution, in general,
due to mismatch cost. See \cref{sec:motivation}.}$^,$\footnote{Note that the same results 
would hold if the dynamics uniformly randomized
$X_B$ rather than preserved its value exactly.}

While $H_0(x_A, x_B)$ is symmetric under interchange of $x_A$ and $x_B$, 
$H_1(x_A, x_B)$ is not. Moreover, since the transformation
from $H_0$ into $H_1$ is done quasi-statically, the evolution of the Hamiltonian
is a continuous map (i.e., a homotopy). In addition, since the system is always
at thermal equilibrium, the rate matrix term for evolution from $(x_A = 1, x_B = 1)$ to $(x_A = 0, x_B = 1)$ must be nonzero whenever the relative values of the Hamiltonian
for those two joint states is changing, to allow probability mass to flow between
those two states. This means 
that for any rate matrix that obeys LDB for this changing Hamiltonian, 
there must be some $t \in (0, 1)$ at which the value of $K_{x_A; x'_A, x'_B}(t)$ for 
$x_A = 0, x'_A = x'_B = 1$ differs from its value for $x_A = 0, x'_A \ne x'_B = 0$. (Recall~\erf{eq:ldb}.) 
In contrast, the rate matrix of a solitary
process cannot ever vary depending on whether $x'_A = x'_B$. So a
solitary process cannot be implemented with this kind of zero EP 
quench-then-quasi-statically-evolve process.
\label{ex:stoch_matrix_land_loss}
\end{example}

\cite{wolpert2018exact} also considers the case where we have an overall system $M$
together with an associated set $\{A, A', \ldots\}$ 
of (potentially overlapping) subsystems of $M$. It is formally proven there
that if we run a sequence of consecutive solitary processes for those subsystems, 
one after the other, then the system-wide Landauer cost and system-wide EP
of the entire sequence are both additive over their values for running the separate
solitary processes over the subsystems $A, A', \ldots$, in order.

In light of this, define the \textbf{machine (subsystem) EF} (resp., machine Landauer cost,
machine EP, machine mismatch cost, machine 
residual EP) as the sum of the subsystem EFs (resp., subsystem Landauer costs,
subsystem EPs, subsystem mismatch costs, 
subsystem residual EPs) incurred by successively running each of the subsystems, $A, A', \ldots$. 
Define the \textbf{(machine) Landauer loss} as the difference between the system-wide
Landauer cost and the machine Landauer cost.
Since system-wide Landauer cost is additive, machine Landauer loss
equals the sum of the Landauer losses of each of the subsystems. It is shown
in~\cite{wolpert2018exact} that machine Landauer loss is the minimal EP of 
running the machine $M$, as one would expect.


\subsection{Related literature}

~\cite{Boyd:2018aa} contains some results that are closely
related to the analysis of solitary processes in this paper and in~\cite{wolpert2018exact}. 
In particular,~\cite{Boyd:2018aa} argues 
that the drop in mutual information between two independent
subsystems of a composite system is the minimal EP of that composite system during any
interval in which one of those subsystems does not change state.
That argument assumes that the
subsystem is coupled to a single heat bath and obeys LDB. It then uses 
general considerations of nonequilibrium statistical physics to make the case that 
``... \{such a\} subsystem matches the framework for an
open driven system described in~\cite{boyd2016identifying}, and so the entropy production 
\{is lower-bounded by the Landauer loss\}''.\footnote{~\cite{Boyd:2018aa}
uses the term ``modularity dissipation'' to mean what is called ``Landauer loss''
here and in~\cite{wolpert2018exact}.}

The analysis provided here (and in more detail in~\cite{wolpert2018exact})
goes beyond the scenarios considered in~\cite{Boyd:2018aa}, in that it
applies when there are multiple baths, and even when LDB does not hold.
So it applies if in addition to a thermal reservoir, there are
one or more chemical species reservoirs, as for example is often the case inside 
biological cells. Furthermore, the analysis here and in~\cite{wolpert2018exact}
considers mismatch cost and
residual EP, in addition to Landauer cost. These considerations are crucial, for
example, in analyzing the entropic costs of digital circuits (see \cref{sec:circuit_entropy_dynamics}.) In addition, the proofs  
here are explicit and complete (being formulated in terms of stochastic thermodynamics).

On the other hand,~\cite{Boyd:2018aa} extends the analysis 
to information ratchets, viewing them as a sequence of solitary processes, one for
each new symbol on an input tape that the information ratchet processes. 
(See \cref{sec:finite_state_ratchets}.)
Moreover, the related discussion in~\cite{Boyd2018thesis} emphasizes that subsystem 
processes are ubiquitous in Nature. In particular,
they occur throughout biological systems. Accordingly, 
biological systems have nonzero Landauer losses in general, which fact
alone prevents them from achieving the generalized Landauer bound.

~\cite{riechers_thermo_comp_book_2018} also contains work related  
to subsystem processes. Like~\cite{Boyd:2018aa}, the analysis
in~\cite{riechers_thermo_comp_book_2018} assumes a single heat bath, and LDB. 
However, the analysis in~\cite{riechers_thermo_comp_book_2018}
is simpler than both the analysis here
and the corresponding analysis in~\cite{Boyd:2018aa}. That is because rather than
consider the implications on a composite system's minimal EP of a constraint on how 
that system is allowed to operate (as in Def.~\ref{def:subsystem_proc}), the analysis
in~\cite{riechers_thermo_comp_book_2018} makes an assumption about the relationship 
between the prior of the physical process underlying the
composite system and the actual initial distribution of the states of the composite system. This allows~\cite{riechers_thermo_comp_book_2018}
to directly exploit the decomposition of EP into mismatch cost and residual entropy in order
to express the minimal EP as the drop in mutual information 
between the two subsystems.

\section{Entropy dynamics of straight-line circuits}
\label{sec:circuit_entropy_dynamics}

%
%

Suppose we fix some map $\map$ from a finite space of inputs to a finite space of outputs
that we want a circuit to implement, along with an associated distribution over inputs
to the circuit, $p_0$.
%
In general, the exact same gate $g$ run at different parts
of any circuit implementing $\map$ on $p_0$ will have different distributions
over its inputs, $X_g^{IN}$. As a result, the same
gate implemented with a solitary process
that is run at different parts of the circuit will incur different Landauer cost. 
This suggests that some of the circuits that implement $\map$
have greater (circuit) Landauer cost than the others, i.e., that some
have greater Landauer loss than others.
This in turn implies that some of those circuits  
have more EF than others. How precisely do the entropic costs of
a circuit depend on the distribution over its inputs and on
its wiring diagram, $(V, E, F, X)$?

To investigate this question, 
\cref{sec:cyclic} begins by introducing a broadly applicable model of
the variables in  straight-line circuits and how they evolve
as the circuit is run. Special care is taken so that
all of the details necessary to calculating entropic costs are made explicit. 
\cref{sec:circuit_simplifying_ass} then introduces some extra, less broadly
applicable modeling choices that simplify the analysis.
Next, \cref{sec:results_entropy_circuits} presents some
of the simpler results that have been derived concerning the entropic costs
of systems that are governed by this model.
These results involve multi-divergence and cross multi-information, which as mentioned in
\cref{sec:info_notation} appear to be novel to the literature. This section
ends with a high-level review of an analysis of a particular, fully specified real-world
circuit.

It is worth noting that 
a high level discussion of some of the issues discussed in this section 
can be found in a prescient and insightful  paper, written before the modern machinery
of nonequilibrium statistical physics was developed~\cite{gershenfeld1996signal}.

\subsection{How to model straight-line circuits}
\label{sec:cyclic}

Each gate $g$ in a circuit is modeled as a pair of random variables, with the
associated state space written as $X_g = X_g^{IN} \times X_g^{OUT}$.
The conditional distribution relating the ending
state of $X_g^{OUT}$ to the initial state of $X_g^{IN}$ is 
the distribution given by the circuit specification
(in the sense of \cref{sec:circuit_theory}), which I write as $\map_g(x_g^{OUT} \vert x_g^{IN})$. For simplicity, I restrict attention to Boolean formula
circuits, so that all gates have outdegree 1 (see \cref{sec:circuit_theory}).
Accordingly, wolog we can assume that the gates run sequentially, in some
convenient topological order respecting the circuit's wiring diagram.
This in turn means that we can model running each gate $g$ as
a solitary process that
updates the joint variable $X_g$ while leaving all other variables
in the circuit unchanged. ($X_g$ is ``$X_A$'' in the notation of \cref{sec:subsystem}.) 
For analyzing the entropic costs of running such sequences of solitary
processes in straight-line circuits, I will use the terms \textbf{circuit EF}, 
\textbf{circuit Landauer cost}, \textbf{circuit Landauer loss}, etc.,
to mean ``machine EF'', ``machine Landauer cost'', ``machine
Landauer loss'', etc. (See \cref{sec:solitary}.)

I assume that we can model the solitary process updating any
$X_g$ as running a cyclic AO device over the state space $X_g$, where we identify
the initial state of $X_g^{IN}$ when $g$ starts to run as the input 
of that AO device, and the ending state of $X_g^{OUT}$ when $g$ finishes its run as 
the output of that AO device (see \cref{sec:overwrite_inputs}).
Moreover, for simplicity I assume that standard accounting applies
not just to the entire circuit, but also to any single gate
(see \cref{sec:accounting_section}). 
So the initial value of $X_g^{OUT}$ when $g$ starts to run is 
some special initialized state (which I write as $0$), and similarly the
ending state of $X_g^{IN}$ when $g$ finishes is also some special initialized
state (which I also write as $0$).\footnote{This modeling assumption is violated
by many real-world circuits, in which there is no reinitialization of the
variables in a gate after it runs. In such circuits the entropic costs in
a run arise when the ``relic'' states of variables, set during the preceding
run of the circuit, get over-written. In general, to analyze the entropic
costs of such circuits requires analyzing sequences of multiple runs rather than
just a single run, which is why it is not pursued here.}

As briefly mentioned in \cref{sec:gen_land}, in real-world computers, a large
fraction of the total EF occurs in the interconnects (``wires'') between
the gates. To allow the analysis to include such EF in the wires, 
it will be useful to model wires themselves as gates, i.e., nodes in a Bayes net.
This means that the DAG $(V, E)$ I will use to represent any particular physical circuit
when calculating entropic costs 
is defined in terms of the DAG $(V', E')$ of the associated wiring diagram, but
differs from that DAG.

To make this formal, define a \textbf{wire gate} as any gate in a circuit
that has a single parent and implements the identity map. (So any wire gate $w$ has $\map_w(x_w \vert x_{\pa(w)}) = \delta(x_w, x_{\pa(w)})$; see \cref{sec:circuit_theory}.)
Suppose we are given a computational circuit whose DAG
is $(V', E')$. The DAG that represents the associated physical circuit is constructed
from $(V', E')$, in an iterative process. To begin, set $(V, E)$ to be a copy of $(V', E')$. 
So there is a map $\eta(.)$ between all nodes and edges in the computational circuit's DAG
and the corresponding members of the physical circuit's initial DAG, a map
which is initially a bijection. We grow this initial DAG $(V, E)$ by
iterating the following procedure over all edges in $E'$:
\begin{enumerate}
\item For each edge $e \in E'$ not yet considered which does not involve
a root node, insert a new node $v$ in the middle of the corresponding
edge in $E$, $e = \eta(e')$.
\item Replace that single edge $e$ with a pair of edges, one leading into $v$ from the
head node of the edge $e$, and one leading from $v$ to the tail node of $e$.
\end{enumerate}
Physically, each such new node introduced into $(V, E)$ in this procedure
represents a wire gate, one such wire gate for each edge in $E'$. 

By the end of this iterative procedure, when we have fully constructed
$(V, E)$, the edges in $E$
don't correspond to physical wires (unlike the edges in $E'$). Rather they
indicate physical identity: an edge $e \in E$ going out from a non-wire gate $g$ 
into a wire gate $w$ is simply an indication that $x_g^{OUT}$ is the same
physical variable as the corresponding component of
$x_w^{IN}$. Similarly, an edge $e \in E$ going into a non-wire gate $g$ 
from a wire gate $w$ is simply an indication that the (corresponding
component of) $x_g^{IN}$ is the same physical variable as $x_w^{OUT}$. 
Recall though that the solitary process that runs a non-wire gate $g$ modifies $x_g^{IN}$.
So that solitary process modifies $x_w^{OUT}$ for each wire $w$ leading
into $g$. Similarly, it modifies $x_{w'}^{IN}$ for each wire $w'$ leading
out of $g$.

This means that when a gate $v \in V$
completes its run, having reinitialized its input, it  has \textit{also} reinitialized
the corresponding output of its parent, regardless of whether $v$
is a wire gate or a non-wire gate. So $v$ plays the role
of an ``offboard'' system for the computational devices at its parent gates, reinitializing 
the outputs of those parents. (See \cref{sec:accounting_section}.)
As a result, when all gates $v$ in the (physical) circuit have finished,
all gates $v$ that are not outputs of the overall circuit are back in their
initialized state, with probability 1, as are all input nodes to the 
circuit. So the circuit as a whole is cyclic.

In principle, this model also allows us to 
simultaneously run multiple computations through a single circuit,
in staggered ``waves''. So long as there is
a gap of at least two gates separating each wave, 
both from that wave's predecessor and from its successor, we are guaranteed that
there is no interference between the processes at the gates that are run for different waves.

I will sometimes use the terms ``gate'', ``circuit'', etc., to refer to physical systems with physical states,
with the associated DAG $(V, E)$. Other times I will use those terms
to refer to the associated abstract mathematical conditional distributions in 
the DAG $(V', E')$. Such switching back and forth between $(V', E')$ and the associated
$(V, E)$ will usually be implicit. The intended meaning of the terms 
``gate'', ``circuit'', etc., will always be clear from context.

Note that since any (idealized) wire gate implements the identity function, no
matter what the distribution over the states of its parent it has both zero
Landauer cost and zero mismatch cost. So all of the dependence of the
EF of any real-world wire $w$ on the distribution of inputs $p_{pa(w)}$ 
to that wire arises in the (linear) dependence of the residual EP on $p_{pa(w)}$.  
In addition, since a wire gate $w$ implements the
identity function, its islands are just the separate values of $x_w$.
So the total EF generated by using a wire $w$ is a linear function of the
distribution over its inputs, i.e., it is a dot product of that distribution
with an associated vector $\EPmin{\PP_w}(.)$ giving the residual EP 
of that wire for each of its possible inputs.

From now on I  
adopt the shorthand that for any gate $g$, $p_g$ refers to the distribution of $p_g^{OUT}$ 
before it has been re-initialized, but after it has been run.  
Similarly, I write $p_{pa(g)}$ to refer to the distribution over the states
$x_{pa(g)}$ of (the output components of) the parents of $g$ at the beginning of 
a run of $g$.
%
So the actual
distribution over the initial state of $g$, just before it runs, is $p_{pa(g)}$.

\subsection{Simplifying assumptions for calculating minimal EF of a straight-line circuit}

\label{sec:circuit_simplifying_ass}

To simplify the analysis further, 
in the rest of this section I make two more assumptions. First, I assume that the residual EP terms
$\subEPmin{g}(c)$ equal $0$ for all gates $g$ and associated islands $c$.  
This means in particular that from now on I ignore contributions to the EF from wire gates.

Second, I assume that we can choose the prior of any (non-wire) gate $g$
to be anything we want.
To motivate one way of choosing the prior at each gate, 
suppose we assume the input distribution for the entire circuit 
$\CC$ is some 
specific distribution $q(x_{\Cin})$ (which I sometimes write as ``$q$'', for short).
Then \emph{if that assumption were correct}, we should set 
the prior distribution for each gate $g$ simply by running the entire circuit $\CC$ on 
inputs generated according to $q$ to determine the distribution
over the parents of $g$, i.e., by propagating $q$
from $V_{\Cin}$ to $\pa(g)$.
In this way any assumption of $q$ specifies what prior 
we should build into each gate $g$ in the circuit.
%
%
I call a set of priors at the gates that are set this way from a shared input distribution $q$
a set of \textbf{propagated priors}, and write them as $\{q_{pa(g)}\}$.

In the analysis below, I allow for the possibility that the circuit will be used in more than one environment,
e.g., with different users. Therefore I allow the priors to differ from the actual
distributions. So the (propagated) priors at the gates can differ from the actual
distributions at those gates.

It is important to note that both of the assumptions I am making are 
often violated in the real world. As mentioned above, in current computers
the EF in the wires is comparable to that in the (non-wire) gates. However,
that EF is \textit{all} residual EP. So by ignoring residual EP, we ignore
one of the major determiners of EF in current computer. Moreover, in many situations it will
be easiest to mass-manufacture the gates, so that while we can vary the 
characteristics of any physical gate $g$
that specifies the conditional distributions $\map_g$, all gates $g$
that implement the same conditional distribution $\map_g$ have the same prior, 
regardless of where they appear in a circuit. In this case the priors at 
the gates are not propagated priors, for any assumed distribution over inputs
to the circuit.
(See~\cite{wolpert2018exact} for an analysis of the more general case, where neither
of these two assumptions are made.)

\subsection{Entropy dynamics of straight-line circuits }
\label{sec:results_entropy_circuits}

Suppose we have some conditional distribution of outputs given
inputs, $\map$, that is implemented by some circuit, and an input distribution, $p$,
and some prior distribution over inputs, $q$.
Then paralleling the definition of circuit Landauer
loss, I define the \textbf{circuit mismatch loss} of running that circuit
on that input distribution with that prior input distribution, as the difference
between the mismatch cost of an AO device that implements $\map$ on $p$
with a prior $q$ over its inputs, and the circuit mismatch cost of a circuit
that also implements $\map$ on $p$, and has the prior at each gate $g$ set
to the associated propagated prior $q_{pa(g)}$.

The following notation is motivated by the 
fact that residual EP is taken to equal zero for all devices:

\begin{definition}
\label{def:omegas}
Let $\CC = (V, E, F, X)$ be a circuit, $p$ a distribution over its inputs and $q$ a prior over its inputs.
\begin{enumerate}

\item $\W_{AO(\CC)}(p, q)$ is the total EF used to run $AO(\CC)$
on actual distribution $p$ with prior distribution $q$.

\item $\W_\CC(p, q)$ is the total EF 
used to run $\CC$ on actual distribution $p$ with propagated prior distribution
$\qpag$ at each gate $g \in G$.
%

\item $\mathcal{E}_{AO(\CC)}(p, q) \equiv \W_{AO(\CC)}(p, q) - \W_{AO(\CC)}(p, p)$
is the total mismatch cost when running $AO(\CC)$
on actual distribution $p$ with prior distribution $q$.

\item $\mathcal{E}_{\CC}(p, q) \equiv \W_{\CC}(p, q) - \W_{\CC}(p, p)$
is the total mismatch cost when running $\CC$
on actual distribution $p$ with propagated prior distribution $\qpag$ at each gate 
$g \in \CC$.

%

\item The circuit EF loss is
\eq{
\Delta \W_\CC(p, q) := \W_{\CC}(p, q) - \W_{AO(\CC)}(p, q)
}

\item The circuit Landauer loss is
\eq{
\Delta \WW_\CC(p) &:= \W_{\CC}(p, p) - \W_{AO(\CC)}(p, p) \nonumber \\
    &= \Delta \W_\CC(p, p)   \nonumber
}

\item The circuit mismatch loss is
\eq{
\Delta \EF_\CC(p, q) &:= \EF_{\CC}(p, q) - \EF_{AO(\CC)}(p, q)  \nonumber \\
	&= \Delta \W_\CC(p, q) - \Delta \WW_\CC(p)  \nonumber
\label{eq:20a}	
}
\end{enumerate}
\end{definition}

\noindent In~\cite{wolpert2018exact} the following is proven:

\begin{proposition}
For any Boolean formula $\CC$, the Landauer circuit loss for input distribution $p$ is
\eqn{
\Delta \WW_\CC(p) &=  \II(p) - \sum_g \II(\ppag) 
}
\label{prop:mutuals}
\end{proposition}
\noindent 

Recall that \cref{eq:drop_in_mutual} gives the Landauer loss of a subsystem process,
as a drop in mutual information of the two subsystems as the process runs.
The expression for $\Delta \WW_\CC(p)$ in \cref{prop:mutuals} can be seen as an extension of
that result, to concern full circuits, comprising a sequence of multiple
subsystem processes.

In words, \cref{prop:mutuals} tells us 
that the difference between the Landauer cost of a formula
and that of an equivalent AO device is a sum of multi-informations.
One of those multi-informations is given directly by the input
distribution. However, the other ones depend on the wiring diagram of the circuit
in addition to the input distribution.
Intuitively, those other multi-informations reflect the fact
that the greater the correlation among the inputs to any given gate $g$, the less
information is contained in the joint distribution of those
inputs, and therefore the less information is
lost by running that gate. In turn, if every gate in a circuit actually does run without
losing much information, then the amount of extra (minimal) EF due to using
that circuit rather than an equivalent AO device is small. 

The term $\II(p)$ in Prop.~\ref{prop:mutuals} can
be seen as a ``normalization constant'', in the sense that for any pair of Boolean formulas $\CC$ and $\CC'$, 
both of which compute the same conditional distribution, the difference in their Landauer losses
is just the difference in their Landauer costs,
\begin{eqnarray}
\Delta \WW_\CC(p) - \Delta \WW_{\CC'}(p) &= 
    \sum_{g \in \CC'} \II(p_{\pa(g)}) -  \sum_{g \in \CC} \II(p_{\pa(g)})
\label{eq:45a}
\end{eqnarray}
In light of this, suppose we wish to design a circuit that implements a given 
Boolean formula $f$ and that has minimal possible
EF for some distribution $p^{IN}$ over the domain of $f$, say subject to the constraint
that we can only use gates in
some specified universal set. Then~\erf{eq:45a} means that we must
find the circuit $\CC$ made out of 
such gates that implements $f$ and that also minimizes the sum of the multi-informations at its gates,
$\sum_{g \in \CC} \II(p_{\pa(g)})$, for the given distribution $p^{IN}$.
This appears to be a nontrivial optimization, since making changes
in one part of the circuit can affect the distributions that are input to the gates in a 
different part of the circuit, and therefore affect the total Landauer cost of the gates in
that different part of the circuit.

It is proven in~\cite{wolpert2018exact} that circuit Landauer loss cannot
be negative. This provides a major advantage to using an AO device rather
than a circuit. Unfortunately, there are major engineering 
difficulties which prevent us from building AO devices to implement
the kinds of functions implemented in real-world circuits, due to the
huge state spaces of such real-world circuits.

The following is also proven in~\cite{wolpert2018exact}:
\begin{proposition}
For any formula $\CC$, the circuit mismatch loss for actual 
input distribution $p$ and prior input distribution $q$ is
\eqn{
\Delta \EF_\CC(p, q) &=  -\IIDDf{p}{q} +  \sum_g \IIDDf{\ppag}{\qpag}
}
\label{prop:22}
\end{proposition}	
\noindent (Compare to \cref{prop:mutuals}.)
Interestingly, mismatch loss can be negative. In fact, the sum of Landauer loss
and mismatch loss can be less than zero. This suggests that
in some situations we would actually use less EF if we run
a circuit rather than an equivalent AO device to implement
a given Boolean formula, if our assumption for the input distribution
$q$ differs from the actual input distribution, $p$. 
To see how this phenomenon might be exploited,
suppose that just like in robust optimization, we
do not presume that we know the actual distribution $p$, and so cannot
set $q = p$. However, we feel confident in saying that $p$ lies
within a ball of radius $\KK$ of our guess for the actual input
distribution, $\hat{p}$, i.e., that $||\hat{p}, p|| < \KK$ for some 
appropriate distance measure $||\cdot, \cdot||$. Then we might
want to set $q  = \hat{p}$, and choose between a circuit and an equivalent AO device
that both use that prior based
on which one would result in less total EF for any $p$ such that
$||q, p|| < \KK$.\footnote{There are some subtleties in applying this reasoning to real-world
circuits, which arise if we wish to ``compare apples with apples''.
See~\cite{wolpert2018exact}.}

~\cite{wolpert2018exact} contains other results not reproduced here, for
the case where gates can have nonzero residual EP. 
The analysis in~\cite{wolpert2018exact} also covers circuits with noisy gates.
In addition, the analysis in that paper cover circuits with
gates that have outdegree greater than 1. 

The thermodynamics of circuits is also considered in~\cite{riechers_thermo_comp_book_2018}.
Superficially, the analysis in that paper is similar to the analysis summarized here.
In particular, Prop.~\ref{prop:mutuals} above has a similar functional form
to Eq.\;26 in~\cite{riechers_thermo_comp_book_2018}, which also involves a multi-information
(though without using that term).
However, the analysis in~\cite{riechers_thermo_comp_book_2018} concerns a  different
kind of system from the circuits considered here. 
Viewed as a circuit, the system considered in~\cite{riechers_thermo_comp_book_2018}
is a set of $N$ disconnected gates, working
in parallel, never combining, but with statistical correlations among their inputs. 
Eq.\;26 in~\cite{riechers_thermo_comp_book_2018} 
concerns the mismatch cost that would arise for such a system
if we used propagated priors and took $q = p$.

\subsection{Entropy dynamics in a transition detector circuit}

The analysis in~\cite{anderson2013toward} concerns the entropic costs of
a ``transition detector''. This is a device that receives
a string of bits, one after the other, and iteratively determines for each new bit whether 
it is the same or different from the previous bit that the device received.
Although the authors describe the device they analyze as a finite state automaton, 
they only consider a single pass of its operation, in which it only decides
whether one second bit differs from the first bit.
This reduces their machine to a straight-line circuit with a two-bit input space.

One noteworthy contribution of~\cite{anderson2013toward} is the
level of detail in the circuit they analyze.
They describe the variables in this circuit as follows:

\begin{quote}
``The processor
utilizes two general purpose registers A and B, a four-function
arithmetic logic unit (ALU), a two-word internal ``scratchpad''
memory $M=M_0M_1$, that also functions as an output buffer, four
multiplexers for signal routing, and input and output buffers.
... \{In addition there are\} a 16-word instruction memory, a 4-bit program
clock (PC)--with a hardware provision for halting the clock
if and when the instruction at address PC(1111) is reached--
and control logic that decodes each instruction and generates all
control signals required to appropriately configure the data path,
perform register operations, adjust the program clock on jump
instructions, and enable memory access and I/O.''
\end{quote}
The authors then specify the control logic and associated data path of
their circuit in the same level of detail.

The authors compare an ``instruction-level analysis'' (ILA) of the entropic
cost of running the circuit to an ``architecture-level analysis'' (ALA) of those costs,
under the assumption that the input bits are generated IID.
It appears that their analysis concerns 
Landauer cost.\footnote{It is hard to be completely sure of the physical meaning
of the quantities the authors analyzed, since they perform their analysis
using a semi-formal ``referential approach'' they developed in earlier work that is not used by others
in the literature, and that is not formulated in terms of modern nonequilibrium statistical
physics. At a minimum though, no mismatch cost or residual entropy creation terms appear
in their analysis.} It also appears that their ILA is a way to calculate the Landauer cost
of an AO device that implements the transition detector. On the other hand,
the ALA appears to be a way of calculating the circuit Landauer cost. 

Under this interpretation of the analysis in~\cite{anderson2013toward}, 
the difference of the two entropic costs that they calculate is
the Landauer loss of a single pass through the particular circuit they define. In 
agreement with the results in \cref{sec:results_entropy_circuits}, 
they find that depending on 
the probability distribution of the input bits, the Landauer loss
can be positive, but is never negative.

The interested reader may also want to consult~\cite{ercan2013heat}, which
is another article, involving the same authors, that considers the thermodynamics of specific
straight-line circuits that are modeled in great detail.

\section{Entropy dynamics of logically reversible circuits}
\label{sec:fredkin}

An interesting body of research concerning ``reversible circuits'' has grown 
out of the early work by Landauer and Bennett, in isolation
from the recent breakthroughs in nonequilibrium statistical physics.
This research assumes that one is presented with a conventional circuit $C$ made of
logically \underline{{\textit{ir}}}reversible gates which implements some logically 
irreversible function $f$, and wants to construct a logically \textit{reversible} circuit, $C'$, 
that emulates $C$. The starting point for this research is
the observation that we can always create such an emulating circuit, by appropriately
wiring together a set of logically reversible gates (e.g., Fredkin gates) to
create a circuit $C'$ that maps any input bits $x^{IN} \in X^{IN}$ to a set of
output bits that contain both $f(x^{IN})$ and a copy of 
$x^{IN}$~\cite{fredkin1982conservative,drechsler2012reversible,perumalla2013introduction,frank2005introduction}.
Tautologically, the entropy of the distribution over the states of $C'$
after this map has completed is identical to the entropy of the initial distribution over states. 
So the Landauer cost is zero, it would appear. This has led 
to claims in the literature suggesting that by replacing a
conventional logically irreversible circuit with an equivalent logically reversible circuit,
we can reduce the ``thermodynamic cost'' of computing $f(x^{IN})$ to zero. 

This line of reasoning should be worrisome. As mentioned, we now know that
we can directly implement any logically irreversible map
$x^{IN} \rightarrow f(x^{IN})$ in a thermodynamically reversible manner. So by running
such a direct implementation of $f$ in reverse (which can be done thermodynamically
reversibly), we would extract heat from a heat bath. If we do that,
and then implement $f$ forward using a logically reversible circuit, 
we would return the system to its starting distribution, seemingly having
extracting heat from the heat bath, thereby violating the second law. This
suggests that any supposed thermodynamic benefits of finite time reversible deterministic computation
are exactly analogous to the supposed thermodynamic benefits of running a Maxwell's demon.
In both cases, yes, in just a single run of your system, you can have your 
thermodynamic miracle. But as soon
as you try to use the system more than once, to complete a full cycle, the
miracle vaporizes.

As it turns out, there \textit{are} some thermodynamic advantages to using a logically reversible
circuit rather than an equivalent logically irreversible circuit. However, there are also
some disadvantages to using logically reversible circuits. 
 In the following two subsections I
elaborate these relative advantages and disadvantages of using reversible circuits, 
in order to illustrate the results presented in the sections above.

Before doing that though, in the remainder of this subsection I present some needed
details concerning logically reversible circuits that are constructed out of logically reversible gates.
One of the properties of logically reversible gates that initially caused problems in
designing circuits out of them is that
running those gates typically produces ``garbage'' bits, to go with the bits that provide the output
of the conventional gate that they emulate. The problem is that these garbage bits
need to be reinitialized after the gate is used, so that the gate can be used again.
Recognizing this problem, \cite{fredkin1982conservative} shows how to
avoid the costs of reinitializing any garbage bits produced by using a reversible
gate in a reversible circuit $C'$, by extending $C'$ with yet more reversible gates (e.g.,
Fredkin gates). The result is an \textbf{extended circuit} that
takes as input a binary string of input data $x$, along with a binary string of  ``control signals'' $m \in M$,
whose role is to control the operation of the reversible gates in the circuit. The output
of the extended circuit is a binary string of the desired output for input $x^{IN}$, 
$x^{OUT} = f(x^{IN})$, together with a copy 
of $m$, and a copy of $x^{IN}$, which I will write as $x^{IN}_{copy}$. 
So in particular, none of the output garbage bits produced by the individual
gates in the original, unextended circuit of reversible gates still exists by the time
we get to the output bits of the extended circuit.\footnote{More precisely, in one 
popular form of reversible circuits, a map $f : X^{IN} \rightarrow X^{OUT}$ is
implemented in several steps. First, in a ``forward pass'', the circuit made out of reversible gates
sends $(x^{IN}, m, \vec{0}^{GARBAGE}, \vec{0}^{OUT}) \rightarrow (x^{IN}, m, m', f(x^{IN}))$, 
where $m'$ is the set of ``garbage bits'',
$\vec{0}^{OUT}$ is defined as the initialized state of the output bits, and
similarly for $\vec{0}^{GARBAGE}$. After completing this forward pass,
an offboard copy is made of $x^{OUT}$,
i.e., of $f(x^{IN})$. Then the original circuit is run ``in reverse'', sending
$(x^{IN}, m, m', f(x^{IN})) \rightarrow (x^{IN}, m, \vec{0}^{GARBAGE}, \vec{0}^{OUT})$. The end result is a process that transforms the input bit string $x^{IN}$ into the offboard copy
of $f(x^{IN})$, together with a copy of $x^{IN}$ (conventionally stored in the same physical variables that contained
the original version of $x^{IN}$), all while leaving the control bit string $m$ unchanged.}
 
While it removes the problem of erasing the garbage bits, this extension of the original
circuit with more gates does not come for free. In general it requires doubling
the total number of gates (i.e., the circuit's size), doubling the running time of the circuit (i.e., the
circuit's depth), and increasing the number of edges coming out of each gate, by up to a factor
of 3. (In special cases though, these extra cost can be reduced, sometimes substantially.)

\subsection{Reversible circuits compared to computationally equivalent all-at-once devices}


In general, there are many different ``basis sets'' of allowed gates we can use to
construct a conventional (logically irreversible) 
circuit that computes any given logically irreversible 
function $f$. Moreover, even once we fix a set of allowed gates, in general
there are an infinite number of logically 
irreversible circuits that implement $f$ using that set of gates. 
Due to all this flexibility, 
we need to clarify precisely what  ``logically irreversible circuit'' we wish to compare to any given extended circuit that implements the same function $f$ as that extended circuit. 

One extreme possibility is to compare the extended circuit to a single,
monolithic gate that computes the same function, and which distinguishes
input variables from output variables. In other words, we could
compare the extended circuit to a physical system that directly maps 
$(x^{IN}, \vec{0}^{OUT}) \rightarrow (x^{IN}, f(x^{IN}))$. However,
this map is logically reversible, just like the extended circuit, and
so not of interest for the comparison. 

A second possibility is to compare the extended circuit to an AO device with a state
space $X$ that directly maps $x \in X \rightarrow f(x) \in X$, without
distinguishing input variables and output variables. Such a map is \textit{not}
logically reversible, but (as mentioned above) can be implemented with
a thermodynamically reversible system, whatever the initial distribution over
$X$. 
If we implement $f$ with an AO device, then the minimal
EF we must expend to calculate $f$ is the drop in entropy of the distribution over $X$ as 
that distribution evolves according to $f$. This drop is
nonzero (assuming $f$ is not logically invertible). This would seem to mean that there
is an advantage to using the equivalent extended circuit rather than the AO device,
since the minimal EF with the extended circuit is zero.

However, we must be careful to compare apples to apples. The number of information-carrying bits
in the
extended circuit after it completes computing $f$ is $\log |X^{IN}| + \log |X^{OUT}| + \log |M|$.
The number of information-carrying bits in the AO device when it completes is 
just $\log |X^{OUT}|$. So strictly speaking, the two systems implement different functions, 
that have the same domains but different codomains.  

This means that the entropic costs of answer-reinitializing the two circuits
(i.e., reinitializing the codomain variables) will differ.
%
%
In general, the Landauer cost and mismatch cost of {answer-reinitialization}
of an extended circuit will be greater than
the corresponding answer-reinitialization costs of an equivalent AO device. This is for the 
simple reason that the answer-reinitialization of the extended circuit must reinitialize
the bits containing copies of $x$ and $m$, which do not even exist in the AO device. 

Phrased differently, if we allow the extended circuit to keep a copy of $x^{IN}$,
rather than erase it, then to compare apples to apples, we should also not impose all
the strictures of standard accounting to the AO device, and allow the AO device 
to also forego erasing $x^{IN}$. However, that would change the Landauer cost we ascribe to running
the AO device from $S(X^{IN}) - S(X^{OUT})$ to the \textit{negative} value $-S(X^{OUT})$.
(Recall from the discussion at the end of \cref{sec:accounting} that if any copies
of the input are allowed to persist after the computation ends, then we can even
get negative Landauer cost.) It is worth emphasizing that this importance of reinitializing the copy of $x^{IN}$
was recognized even in the early analyses based on the original formulation of
Landauer's bound; it is the primary motivation for one of 
the most sophisticated of those early analyses, which is discussed in detail in \cref{sec:zurek}.

To be more quantitative, first, for simplicity, assume that the initial distribution over the bits 
in the extended circuit that encode $m$ is a delta
function. (This would be the case if we do not want the physical circuit to implement 
a different computation from one run to the next,
so only one vector of control signals $m$ is allowed.) This means that
the ending distribution over those bits is also a delta function.
The Landauer cost of reinitializing
those bits is zero, and assuming that we perform the reinitialization using a prior
that equals the delta function over $m$, the mismatch cost is also
zero. So assuming the residual EP of reinitialization those
bits containing a copy of $m$ is zero, we can we can ignore those bits from now on.

To proceed further in our comparison of the entropic costs of the 
answer reinitialization of an AO device
with those of an equivalent extended circuit, we need to specify the detailed dynamics of the
answer-reinitialization process that is applied to the two devices.
Both the AO device and the equivalent extended circuit have a set of output bits
that contain $f(x^{IN})$ that need to be reinitialized, with some associated
entropic costs. In addition though, the extended circuit needs to reinitialize
its ending copy of $x^{IN}$, whereas there is no such requirement of the equivalent AO device.
To explore the consequences of this, 
I now consider several natural models of the answer-reinitialization:

$ $

\noindent 1) In one model, we require that the
answer-reinitialization of the circuit is performed within each output bit $g$ itself, 
separately from all other variables. Define $Fr(\CC)$ to mean an extended circuit 
that computes the same input-output function $f^\CC$
as a conventional circuit $\CC$, and define $AO(\CC)$ similarly. 
Assuming for simplicity that the residual entropy of reinitializing all output bits
is zero,
the EF for the answer-reinitialization of $Fr(\CC)$ using such a bit-by-bit process is
\eq{
\W_{\CC'}(p, q) =  \sum_{g \in V^{\Cout}} \KKf{p_g}{q_g}
}
where $V^{\Cout}$ indicates the set of all bits containing the final values 
of $x^{OUT}$ and $x^{IN}_{copy}$.

Using gate-by-gate answer-reinitialization,
the EF needed to erase the output bits containing $f^\CC(x^{IN})$ is the same
for both $AO(\CC)$ and $Fr(\CC)$. Therefore the additional Landauer cost incurred in answer-reinitialization
due to using  $Fr(\CC)$ rather than $AO(\CC)$ is
the Landauer cost of erasing the output bits in $Fr(\CC)$ that store $x^{IN}_{copy}$,
\eq{
\Delta \SSS_{Fr(\CC), \CC} (p) & \coloneqq  \sum_{v \in V_{IN}} \SSS(p_v)
\label{eq:52}
}
where I write ``$v \in V_{IN}$'' to mean the output bits that contain $x^{IN}_{copy}$,
and $p_v$ to mean the ending marginal distributions over those bits. Similarly,
the difference in mismatch cost is
\eq{
\Delta \DD_{Fr(\CC), \CC} (p, q) & \coloneqq  \sum_{v \in V_{IN}} \DDbase_v(p^v\Vert {q}^v)
}
where $q_v$ refers to a prior used to reinitialize the output bits in .$v \in V_{IN}$.

However, independent of issues of answer-reinitialization,
the Landauer cost of implementing a function using an AO
device that is optimized for an initial distribution $p_{\Cin}$ can be bounded as follows:
\eq{
\SSS(p_{\Cin}) - \SSS(f^\CC p_{\Cin}) &\le \SSS(p_{\Cin})  \nonumber \\
	& \le  \sum_{v \in V_{\Cin}} \SSS_v(p_v) \nonumber \\
	& = \Delta \SSS_{Fr(\CC), \CC}(p)
}
Combining this with \erf{eq:52} shows that under gate-by-gate answer-reinitialization, the \textit{total} 
Landauer cost of implementing a function using an AO device  --- including the costs of
reinitializing the gates containing the value $f^C(x^{IN})$ ---  is
upper-bounded by the \textit{extra} Landauer cost of implementing that same
function with an equivalent extended circuit, i.e., just that portion
of the cost that occurs in answer-reinitializing
the extra output bits of the extended circuit.
This disadvantage of using the extended circuit holds even if the 
equivalent AO device is logically irreversible. So as far as Landauer cost is concerned
there is no reason to consider using an extended circuit
to implement a logically irreversible computation with
this first type of answer-reinitialization.

On the other hand, in some situations, the mismatch
cost of running the AO device will be \emph{greater} 
than the mismatch cost of the answer-reinitialization of $x^{IN}_{copy}$ 
in the equivalent extended circuit. This illustrated in the following example:
\begin{example}
Suppose that the input to the circuit consists of two bits, $a$ and $b$,
where the actual distribution over those bits, $\p$, and prior distribution over
those bits, $q$, are:
\eqn{
\p(x_b) &= \delta(x_b, 0) \\
\p(x_a \mid x_b = 0) &= 1/2 \qquad  \forall x_a \\
q(x_b) &= \delta(x_b, 1) \\
q(x_a \mid x_b = 0) &= 1/2 \qquad  \forall x_a \\ 
q(x_a \mid x_b = 1) &= \delta(x_a, 0)
}

Suppose as well that $f^\CC$ is a many-to-one map. Then plugging in gives
\eq{
\DDf{p_{\Cin}}{q_{\Cin}} - \DDf{f^\CC p_{\Cin}}{f^\CC q_{\Cin}} &=    
    \DDf{p_{\Cin}}{q_{\Cin}}  \nonumber \\
& > \sum_{v \in V_\Cin} \DDf{p_v}{q_v} \nonumber 
}
This sum equals the mismatch cost of the answer-reinitialization of $x^{IN}_{copy}$, which
establishes the claim.
\end{example}

\noindent However, care should be taken in interpreting this result,
since there are subtleties in comparing mismatch costs between
circuits and AO devices, due to the need to compare apples to apples 
(see discussion of this point in~\cite{wolpert2018exact}).

$ $

\noindent 2) A second way we could answer-reinitialize an extended circuit involves
using a system that simultaneously accesses \textit{all} of the output bits
to reinitialize $x^{IN}_{copy}$, including the bits storing $f(x^{IN})$.

To analyze this approach, for simplicity assume there are no restrictions on
how this reinitializing system operates, i.e., that it is an AO device.
The Landauer cost of this type of answer-reinitialization
of $x^{IN}_{copy}$ is just $\SSS(p(X^{IN} \mid X^{OUT})) - \ln[1]$,
since this answer-reinitialization process is a many-to-one map over the state of $x^{IN}_{copy}$. 
Assuming $f^\CC$ is a deterministic map though, by Bayes' theorem
\eq{
\SSS(p(X^{IN} \mid X^{OUT})) &= \SSS(p(X^{IN})) - \SSS(p(X^{OUT}))
}
So in this type of answer-reinitialization, the extra Landauer cost of the 
answer-reinitialization in the extended
circuit that computes $f^\CC$ is identical to the total Landauer
cost of the AO device that computes the same function
$f^\CC$. 
On the other hand, in this type of answer-reinitialization process the mismatch cost of the extended circuit
may be either greater or smaller than that of the AO device, depending
on the associated priors.
%

%

$ $

\noindent 3) A third way we could answer-reinitialize $x^{IN}_{copy}$ in an extended circuit
arises if, after running the circuit, we happened upon a set of initialized external bits, just
lying around, as it were, ready to be exploited. In this case, after running the circuit, we could simply swap
those external bits with $x^{IN}_{copy}$, thereby answer-reinitializing the output bits at zero cost.

Arguably, this is more sleight-of-hand
than a real proposal for how to re-initialize the output bits. Even so, it's
worth pointing out that rather than use those initialized external bits to contain a copy of $x^{IN}_{copy}$,
we could have used them as an information battery, extracting up
to a maximum of $\ktlntwo$ from each one by thermalizing it. So the opportunity cost
in using those external bits to reinitialize the output bits of the extended circuit
rather than use them as a conventional battery is $|V_{IN}| \ktlntwo$. This is an upper
bound on the Landauer cost of implementing the desired computation using an AO device. So again,
as far as Landauer cost is concerned, there is no advantage to using an extended circuit
to implement a logically irreversible computation with this third type of answer-reinitialization.

%

$ $

There are other schemes for reinitializing $x^{IN}_{copy}$ of course. In particular,
see the discussion in \cref{sec:zurek} for a review of a particularly sophisticated
such scheme.

Summarizing, it is not clear that there is a way to implement a logically irreversible function
with an extended circuit built out of logically reversible gates that
reduces the Landauer cost below the Landauer cost of an equivalent AO device. 
The effect on the mismatch cost of using such a circuit rather than an AO device is 
more nuanced, varying with the priors, the actual distribution, etc.

\subsection{Reversible circuits compared to computationally equivalent irreversible circuits}

I now extend the analysis, from comparing the entropic costs of an extended circuit to those  
of a computationally equivalent {AO device}, to also compare to the costs of a
computationally equivalent conventional circuit, built with multiple logically irreversible gates. 
As illustrated below, the entropic costs of the answer-reinitialization of a conventional circuit 
(appropriately modeled) are the same as the entropic costs
of the answer-reinitialization of a computationally equivalent AO device.
So the analysis of the preceding subsection gives us the
relationship between the answer-reinitialization entropic costs of conventional
circuits and those of computationally equivalent extended circuits.
In particular, the minimal EF required to answer-reinitialize a conventional circuit is in general
lower than the minimal EF required to answer-reinitialize a computationally equivalent 
extended circuit. 

Accordingly, in this subsection
I focus instead on comparing the entropic costs of running conventional
circuits, before they undergo any answer-reinitialization,
with the entropic costs of running computationally equivalent extended circuits, before
\textit{they} undergo any answer-reinitialization.
While the full analysis of the entropic costs of running conventional circuits is rather
elaborate~\cite{wolpert2018exact}, some of the essential points can be illustrated with 
the following simple example.

Suppose we have a system that comprises two input bits and two output bits, 
with state space written as
$X = X^{IN}_1 \times X^{IN}_2 \times X^{OUT}_1 \times X^{OUT}_2$. Consider mapping
the input bits to the output bits by running the ``parallel  bit
erasure'' function. Suppose that while doing that we simultaneously reinitialize the
input bits $x^{IN}_1$ and $x^{IN}_2$, in preparation for the next run 
of the system on a new set of inputs.
So assuming both of the output bits are initialized before the process begins to 
the erased value $0$, the state space evolves according to the function 
$f : (x^{IN}_1, x^{IN}_2, 0, 0) \rightarrow (0, 0, 0, 0)$. (See \cref{ex:double_bit_erasure} for an
alternative way of doing parallel bit erasure, with a system that does not differentiate input
and output variables.)
 
Consider the following three systems that implement this $f$:
\begin{enumerate}
\item An AO device operating over
$X^{IN}_1 \times X^{IN}_2 \times X^{OUT}_1 \times X^{OUT}_2$ that directly implements $f$;
\label{enum:1}
\item A system that implements $f$ using two bit-erasure gates that are physically
isolated from one another, as
briefly described in~\cref{ex:double_bit_erasure}. Under this model
the system first uses one bit-erasure gate to send $(X^{IN}_1, X^{OUT}_1) = (x^{IN}_1, 0) \rightarrow (0, 0)$,
and then uses a second bit-erasure gate to apply the same 
map to the second pair of bits, $(X^{IN}_2, X^{OUT}_2)$. 

The requirement that the gates be physically isolated means that the rate matrix of the first gate
is only allowed to involve the pair of bits $(X^{IN}_1, X^{OUT}_1)$, i.e., it
is of the form $W_{x^{IN}_1, x^{OUT}_1;(x^{IN}_1)', (x^{OUT}_1)'}(t)$. So
the dynamics of $(x^{IN}_1, x^{OUT}_1)$ is independent of
the values of the other variables, $(x^{IN}_2, x^{OUT}_2)$. Similar restrictions apply to the
rate matrix of the second gate. (So in the language of \cref{sec:subsystem}, since
the two gates run sequentially, the each run a ``solitary process''.) 

\label{enum:2}

\item A system that uses two bit-erasure gates to implement $f$, just as in (\ref{enum:2}), but does not stagger
those two bit-erasure processes in time, instead running them simultaneously. Clearly this change cannot have any thermodynamic
consequences, on purely physical grounds.

\label{enum:2a}

\item A system that uses two bit-erasure gates to implement $f$, just as in (\ref{enum:2}), but does \textit{not}
require that those gates run in sequence and that
they be physically isolated. In other words, the rate matrix that
drives the first bit-erasure gate as it updates the variables $(x^{IN}_1, x^{OUT}_1)$ \textit{is} allowed to 
do so based on the values $(x^{IN}_2, x^{OUT}_2)$, and vice-versa. Formally,
this means that the joint rate matrix of the entire system is of the form
\eq{
& W_{x^{IN}_1, x^{OUT}_1;(x^{IN}_1)', (x^{OUT}_1)';(x^{IN}_2)', (x^{OUT}_2)'}(t) 
\;\delta(x^{IN}_2, (x^{IN}_2)') \; \delta(x^{OUT}_2, (x^{OUT}_2)')   \nonumber \\
&\qquad\qquad\qquad+ W_{x^{IN}_2, x^{OUT}_2;(x^{IN}_1)', (x^{OUT}_1)';(x^{IN}_2)', (x^{OUT}_2)'}(t) 
\;\delta(x^{IN}_1, (x^{IN}_1)') \; \delta(x^{OUT}_1, (x^{OUT}_1)') 
}
\label{enum:3}
\end{enumerate}

\noindent
Models (\ref{enum:2}) and (\ref{enum:3}) both represent a conventional circuit made out of two gates
that each implement logically-irreversible functions. However, they differ
in whether they only allow physical coupling among the variables in the circuit that
are logically needed for the circuit to compute the desired function (model (\ref{enum:2})), or
instead allow arbitrary coupling, e.g., to reduce entropic costs (model (\ref{enum:3})).
Moreover, the set of rate matrices allowed under model (\ref{enum:3}) 
is a strict superset of the set of rate matrices of all subsystem processes
that implement $f$, i.e., we have more freedom to reduce EF using
model (\ref{enum:3}) than we would with any subsystem process (see \cref{def:subsystem_proc}). On the other
hand, the set of allowed rate matrices under model (\ref{enum:1}) is a strict superset of the set of allowed rate
matrices under model (\ref{enum:3}).

To analyze the consequences of these differences, first consider the Landauer cost of model (1),
which we can expand as
\eq{
&S(p_0(X^{IN}_1, X^{IN}_2,
X^{OUT}_1, X^{OUT}_2)) - S(\hat{f}_{1,2} \; p_0(X^{IN}_1, X^{IN}_2, X^{OUT}_1, X^{OUT}_2)) \nonumber \\
&\qquad = S(p_0(X^{IN}_1, X^{IN}_2, X^{OUT}_1, X^{OUT}_2))  \nonumber \\
&\qquad = S(p_0(X^{IN}_1, X^{IN}_2))
}
where $\hat{f}_{1,2}$ is the conditional distribution implementing the parallel  bit erasure, 
so that $\hat{f}_{1,2} \, p_0(x)$ is the ending distribution, which is 
a delta function centered at $(0,0,0,0)$.

Next, assume that both of the gates in model (\ref{enum:2}) are thermodynamically reversible when considered
by themselves, isolated from the rest of the universe, i.e., that their subsystem EPs are both zero.
Then the minimal EF needed to run the first of those gates is 
\eq{S(p_0(X^{IN}_1) - S(\hat{f}_1 \; p_0(X^{IN}_1, X^{OUT}_1)) = S(p_0(X^{IN}_1))
}
Similarly, the minimal EF needed to run the second gate is $S(p_0(X^{IN}_2))$. The same calculation
must apply to the system in model (\ref{enum:2a})

Combining, we see that the difference between \{the minimal EF needed to
run a conventional circuit constructed as in model (\ref{enum:2})\} and \{the minimal EF
needed to run a computationally equivalent AO device (model (1))\} is
\eq{
S(p_0(X^{IN}_1) + S(p_0(X^{IN}_2) - S(p_0(X^{IN}_1, X^{IN}_2))
}
This is just the initial mutual information between $X^{IN}_1$ and $X^{IN}_2$.\footnote{One
could reach the same conclusion by using the fact that machine Landauer loss of a sequence
of solitary processes is additive over those process (see end of \cref{sec:solitary}), 
the fact that the Landauer loss of each solitary process is the drop in mutual information
between the two subsystems during that process (see \cref{eq:drop_in_mutual}),
and the fact that the ending entropy of a system that erases a bit is $0$.} So
the minimal EF needed to run model (\ref{enum:2}) will exceed the minimal EF needed to
run model (1) whenever $X^{IN}_1$ and $X^{IN}_2$
are statistically coupled under the initial distribution, $p_0$.

On the other hand, because of the increased flexibility in their rate matrices, it is
possible that the bit-erasure gates in model (\ref{enum:3}) each achieve zero EP 
\textit{even when considered as systems operating over the full set of four bits}. 
So each of those bit-erasure gates
is thermodynamically reversible even when considered in the context of the full system.
As a result, running the circuit defined in model (\ref{enum:3}) requires the same minimal
EF as running an AO device. (See also \cref{ex:stoch_matrix_land_loss}.) So in general, the minimal EF needed to
run to the conventional circuit defined in model (\ref{enum:3}) is less than the minimal EF needed to
run to the conventional circuit defined in model (\ref{enum:2}).

Summarizing, the minimal total EF (including both the EF needed to run the system and to 
reinitialize it at the end of the run) that is needed by the circuit
defined by model (\ref{enum:2}) exceeds
the minimal total EF needed by either the equivalent AO device (model (1))
or the equivalent conventional circuit defined by model (\ref{enum:3}). Those two differences
in those minimal EF's both equal the mutual information of the inputs bits under $p_0$.
In turn, the minimal EFs to run either model (1) or model (\ref{enum:3})
exceeds the minimal EF needed to run an equivalent extended circuit, by
$S(p_0(X^{IN}_1, X^{IN}_2))$. However, the minimal \textit{total} EF of
the models (1) and (\ref{enum:3}) will in general be no greater than the minimal
total EF of the extended circuit, and may be smaller (depending on the details of the answer-reinitialization process in the extended circuit).


On the other hand, as a purely practical matter, constructing a conventional circuit
as in (\ref{enum:3}) for circuits substantially larger than parallel bit-erasures may be quite challenging;
to do so requires that we
identify all sets of variables that are \textit{statistically} coupled,
at any stage of running the circuit,
and make sure that our gates are designed to \textit{physically} couple those variables.
There are no such difficulties with constructed an extended circuit.
Another advantage of an extended circuit is that {no matter what
the true distribution $p_0$ is}, an extended circuit has zero mismatch cost,
since there is no drop of KL divergence between $p_0$ and \textit{any} $q_0$
under a logically reversible dynamics. In contrast, all three models (1) - (\ref{enum:3}) 
can have nonzero mismatch cost, in general.

As yet another point of comparison, an extended circuit will often have far more wires
than an equivalent conventional circuit. And as mentioned above, the residual EP generated
in wires is one of the major sources of EF in modern digital gates. So even 
in a situation where a conventional circuit has nonzero mismatch cost, 
when the EF generated in the wires is taken into account, there may be no 
disadvantage to using that conventional circuit rather a computationally equivalent
extended circuit.

Clearly there is a rich relationship between the detailed wiring diagram of a conventional
logically irreversible circuit, the procedure for
answer-reinitializing the outputs of a computationally equivalent extended circuit, 
the distribution over the input bits of those circuits, and how the aggregate
entropic costs of those two circuits compare. Precisely delineating this relationship
is a topic for future research.

\section{Entropy dynamics of finite automata} 
\label{sec:sfa_costs}

\subsection{Entropy dynamics of FAs in a steady state}

There is some work in the literature that presents calculations related
to the thermodynamics of FAs. For example,~\cite{Ganesh2013} considers
the thermodynamics of systems that can be defined as
word-based deterministic FAs with no accepting states whose input symbols 
are generated in an IID manner, and which have no word-delimiting input symbol.
(Arguably this last property in particular distinguishes the systems they consider from
what computer scientists typically call ``finite
state automata''.) So the ``input word'' in the systems they consider
is actually a single, infinite string of symbols, and their systems undergo a
single, infinitely long, run. This allows them to avoid
including a counter variable in the model, 
despite their use of a word-based FA. Instead, the time index on the rate
matrix can directly encode which input symbol is being processed at any 
given time.

Rather than consider finite time behavior, they consider the asymptotic limit,
presuming the system reaches a (unique) steady state. In keeping with this, they do not 
require that the system be cyclic in any sense. 
Indeed, the successive input symbols in the input word are not reinitialized as the computation
proceeds, so if one \textit{were} to stop their system at a finite time, and try
to give it a new input string, entropic costs would be incurred which are not
considered in their model. Similarly, their model does not
designate some variable as being the output of the computer.
So they do not consider the issue of how such an
output might be copied out of the system at some finite time. Given this lack
of an output and their not reinitializing the input,
they do not use any convention akin to standard accounting to determine how to
ascribe entropic costs incurred when one run ends and another begins.

Their model does not involve solitary processes, but instead considers AO devices.
(Note that since they consider the asymptotic limit, this means that they
implicitly allow the interaction Hamiltonian between the computational
state of the FA and the input state to involve the arbitrarily large set of 
variables giving all earlier input symbols.) Moreover, since they focus 
on the steady state, the entropy of the computational state of the system doesn't 
change in an update step. Since they don't require the
input symbols to be reinitialized once the are processed, the joint entropy of the
(infinite) string of
input symbols does not change in an update step either. So the Landauer cost
in any single update step
is due to the loss of information between earlier inputs and the current computational state.
Given the constancy of the marginal entropies of those earlier symbols and of
the current computational state, this loss of information in an update step is exactly
the change in the mutual information between the joint state of the earlier input symbols
and the computational state in that update step. Finally, since they consider an
AO device, minimal EP is zero. So the minimal EF per update step
is just the Landauer cost, i.e., it equals this change in mutual information.

This is their primary result. They derive it in a quantum mechanical context, but 
the same analysis holds for classical systems undergoing Markovian dynamics. 
As a final comment, instead of viewing the systems 
considered in~\cite{Ganesh2013} as a version of FAs, those systems can be viewed as a 
variant of information ratchets, with the added
restriction that the ratchet is not allowed to change the state of an input
symbol after reading it. Accordingly, the reader may want to compare 
the analysis in~\cite{Ganesh2013} with the analyses in 
\cref{sec:thermo_mealy}.

\subsection{Entropy dynamics of FAs with thermal noise} 

Another paper related to the thermodynamics
of FAs is~\cite{chu2018thermodynamically}. In that paper the authors
consider deterministic finite automata that are subject to thermal noise
during their operation. 
The paper is careful to introduce the complete computer science definition of an FA,
including the fact that input words having finite lengths. 
However, they only consider a single iteration of
an FA. So they don't need to explicitly ensure that the model of an FA that they analyze 
is a cyclic device. Nor do they need to consider the problems that arise from the
fact that the duration of a run with an FA is a random variable. (Recall
\cref{sec:variable_duration}.)
They also do not consider the issues related to the entropic costs of
copying the output of the FA offboard and/or the entropic costs of copying
in a new input. 

Like the current paper,~\cite{chu2018thermodynamically} uses
stochastic thermodynamics to perform their analysis.
However, in contrast to most of the
stochastic thermodynamics literature, they stipulate that the underlying
rate matrix is time-homogeneous during each iteration of the system. As a result,
to get any dynamics, they assign different energy levels to each
of the states of the FA, relying on thermal relaxation to drive the dynamics.
(In contrast, as mentioned in \cref{sec:explicit}, 
the convention in the literature is to stipulate
that energy levels are identical at the beginning 
of each new iteration of an information-processing system, in order to focus 
on the information processing behavior of the system.) 

Although the relationship is not exact, it seems that in order
to implement arbitrary (deterministic) update functions in their model of
FAs,~\cite{chu2018thermodynamically} 
exploits the same kind of ``partially hidden'' state space construction
discussed in \cref{sec:overwrite_inputs,sec:hidden}.
As a cautionary 
comment, the reader should be aware that~\cite{chu2018thermodynamically} uses some
idiosyncratic terminology. For example, they refer to 
the thermal relaxation of a two-energy system down to the lower energy
state as a ``bit flip''.

\section{Entropy dynamics of information ratchets}
\label{sec:thermo_mealy}

Suppose that we know that inputs to an information ratchet
are generated according to a particular $N$'th
order Markov chain, and we know the prior distribution over the
first $N$ inputs to the ratchet. Even with this information,
we do not know the joint distribution over
strings of $N+1$ successive inputs to the ratchet --- that distribution will depend on how long
that Markov chain has been running. This means that to analyze the entropy dynamics of an
information ratchet with inputs generated according to an
$N$'th order Markov chain, we have to also specify how long it has been running.
This substantially complicates the analysis.
The problem is only compounded if we don't know $N$ -- or if in
fact the stochastic process generating the inputs is not a Markov
process of \textit{any} finite order.\footnote{For example, this could be the
case if the states of the information ratchet are actually coarse-grained
bins of some underlying physical fine-grained space. In this situation,
the dynamics over the coarse-grained bins -- over the states of the ratchet --
are given by a hidden Markov model (HMM).}

The natural way to analyze such scenarios
is to take the infinite time limit, assuming that the inputs are generated according
to a stationary process. In a series of papers~\cite{Boyd:2018aa,boyd2016identifying,boyd2017above,boyd2017correlation,boyd2017transient,boyd2017leveraging,mandal2012work}, Boyd, Mandal, Riechers, Crutchfield
and others have begun pursuing this line of research. The focus
in these papers has not been the behavior of 
an information ratchet with an arbitrary given update rule of its
computational states, running on an arbitrary given
input data stream. Instead, these papers primarily concern the case
where the update rule is optimized for the given 
input data stream.

Given the challenging nature of analyzing the thermodynamics of
information ratchets with HMM input data streams, to date these papers
have mostly focused on information ratchets that create output patterns
\textit{ab initio}, with no patterns in the input stream, 
or that completely destroy patterns in the input stream (producing an output 
stream with no patterns). A natural topic for future research would be the challenging
regime of intermediate cases, in which there are some patterns in the input
stream, and different patterns in the output stream.

The analysis in these papers has focused on discrete-time rather than
continuous-time models, e.g.,
discrete-time rather than continuous-time Markov chains. This
means that some of the machinery of stochastic thermodynamics cannot 
be applied. Moreover, there are
many subtle issues concerning what discrete-time systems can be be represented
by \textit{any} CTMC. It is (relatively) straight-forward
to address these issues when the system has a finite state space, e.g., if it
is a circuit. (See \cref{sec:overwrite_inputs,sec:hidden}.) However, the global
state space of an information ratchet is infinite, including all input sequences
of arbitrary length. In light of this,~\cite{still_thermo_comp_book_2018} 
considers some of the implications
for the earlier analysis on discrete-time information ratchets that arise if one
requires the information ratchet to be physically instantiated with a CTMC.

\subsection{Infinite-time limit of entropy dynamics of finite-state information ratchets 
with arbitrary non-IID inputs}
\label{sec:finite_state_ratchets}

In the literature on the thermodynamics of 
information ratchets, the term \textbf{global system} is sometimes used
to refer to the combination of the information ratchet, input data
stream, and output data stream. In addition, throughout throughout those papers
it is assumed that there is a single thermal reservoir coupled to the
global system, and that the global system's rate matrix always
obeys LDB for that reservoir, for some associated Hamiltonian.

One of the most important results to emerge to date concerning
the thermodynamics of information ratchets arose from considering 
the infinite time limit of information ratchets that have finite set of computational
states, $R$. Suppose that in that limit the
distribution of computational states of the ratchet reaches a stationary
state. Then in that
limit, the EF in the global system produced per unit iteration is bounded below
by the difference between the Kolmogorov-Sinai entropy of the
output data stream and the Kolmogorov-Sinai entropy of the
input data stream~\cite{boyd2016identifying}.\footnote{Typically, the
Kolmogorov-Sinai entropy reduces to the more familiar entropy rate~\cite{cover_elements_2012}.
}
The authors refer to this as the \textbf{Information processing second law} (IPSL).

The two entropy rates in the IPSL
each refer to changes in entropy of only a subset of the variables
in the global system, namely the input 
data stream and the output data stream, respectively. Moreover, the distribution
over the joint space of those two data streams cannot reach a fixed point, 
because the size of that space grows with each iteration. However, the rate of
change in (the entropy of) that distribution per iteration can reach a fixed point.
That is what the IPSL captures.

In general, for a given desired map from the input stream to the output stream,
the greater the number of states of the ratchet one can use to implement
that map, i.e., the larger $R$ is,
then the closer the minimal EF (per iteration, in the infinite-time limit) 
of those ratchets will come to matching the IPSL. Intuitively, the reason for
this is that the dynamics of the state of the ratchet is less constrained when $R$ is
larger, and therefore the ratchet can implement the desired map to the input stream 
with greater thermodynamic efficiency. Viewed differently, the greater $R$ is,
the better able the ratchet is to ``store in its current state a sufficient statistic'' 
concerning the most recent sequence
of inputs, and therefore the better able it is to anticipate what the next input will
be, and therefore act in a manner that is thermodynamically optimal for that input.

\subsection{Infinite time limit of Entropy dynamics of {infinite}-state information ratchets 
with arbitrary non-IID inputs}

The computational power of information
ratchets with a finite set of computational states $R$ is weaker than that of TMs (contrary to some informal claims
in the literature). More precisely, it is not the case that for any given TM
there is some information ratchet such that the (partial) function
computed by the TM equals the (partial) function computed by
the information ratchet (e.g., if we specify some special finite string
of output bits of the information ratchet to signal that the ratchet
has completed the computation). Only if one can map each ID of a TM to
a unique element of $R$
can we map the dynamics of that TM into the dynamics of an equivalent
information ratchet. However, that would require that $R$ be infinite,
in general.\footnote{
Indeed, we can view information ratchets as a variant of 
conventional prefix-free TMs, in which
there is a one-way tape that contains the input string as usual, 
a one-way tape that contains the output string as usual ---
but rather than an infinite work tape as in prefix-free TMs, in information
ratchets the work tape is finite.
}
%
%

This limited power of information ratchets with finite $R$ naturally leads to consideration of information ratchets that have infinite $R$. Analyzing the entropy dynamics of
such information ratchets presents some
significant technical challenges however. 

In addition, many of the nice properties
of finite $R$ ratchets no longer hold for infinite $R$ ratchets.
One example of this is that the IPSL no longer
applies with infinite $R$ (indeed, with an infinite state space, the information ratchet
may never reach a stationary state). Another example arises from the fact that
the thermodynamic benefit of expanding $R$ mentioned above relies on our ignoring the 
thermodynamic cost of initializing the state of the ratchet before it starts
to run. When we are considering the limit of iterating the ratchet an infinite number
of times starting from that single initialized state,
and there are only a finite number of computational states of the ratchet, the
ratio of this one-time initialization cost of the ratchet to the number of iterations becomes 
infinitesimal, and so
can be ignored. However, if $R$ is infinite, this ratio need not go to zero.
So the thermodynamic benefit of expanding $R$ may disappear once $R$ becomes infinite. 
%
%
This general point is 
emphasized in~\cite{Boyd2018thesis}, where it is pointed
out that Landauer's bound can \textit{appear} to be violated in the
asymptotic limit when the information ratchet has an infinite state space.
This apparent violation arises because if the state space
is infinite, then the initialized state essentially serves
as an infinite information battery.
%

\subsection{Finite time ``synchronization costs'' of finite information ratchets 
with arbitrary non-IID inputs}

At the opposite extreme from the asymptotic limit of running a ratchet for
an infinite number of iterations (where the IPSL
applies) is the regime of running the ratchet only a small number
of times, before the global system asymptotes.
In this regime, even if the \textit{input data stream} has reached its stationary
distribution, in general the EF of the ratchet exceeds the value given by the IPSL.
This is for the simple reason that the global system may not have reached
a stationary distribution by the time that the input data stream does.
Loosely speaking, it takes time for the dynamics of the 
state of the ratchet to ``synchronize'' with the dynamics
of the incoming data stream (and/or outgoing data stream, as the case may be).

A useful tool for analyzing this regime is the \textbf{implementation cost},
which is the mutual information at any given iteration 
between the state of the ratchet and the 
combination of the input data stream and the output data stream~\cite{boyd2017transient}.
An important result here is that for ratchets that are predictive of their 
previous outputs, the greater the number of states of the ratchet (i.e., the larger 
its ``memory''), the greater the transient dissipation~\cite{garner2017thermodynamics}.
However, there are other kinds of ratchets besides predictive ones. In particular,
``retrodictive ratchets'', in which the state of the ratchet has
minimal information about the previous outputs, but all the 
information that is shared between the past and future outputs,
can synchronize without paying
a penalty for a large state space.

\subsection{Necessary conditions to reach the bound of the IPSL}

%
%
The results reviewed so far in this section involve
bounds on (the rate of) Landauer cost required by an 
information ratchet that is optimized for the input stream and
the desired map of it into an output stream, e.g., as in the IPSL. 
In addition though, in~\cite{Boyd:2018aa}, results closely related to those reviewed
in \cref{sec:subsystem} are exploited, to derive properties that the information ratchet must
have in order to achieve those bounds. These confirm and refine
earlier results~\cite{boyd2017leveraging}, establishing that
the size of the ratchet's state space must match the size of the
``memory'' of the input data stream in order to achieve the IPSL bound.
(In~\cite{Boyd:2018aa,boyd2017leveraging}, a formalization of this
is referred to as the ``thermodynamic principle of requisite variety''.)

These analyses underscore an interesting fact. 
Information ratchets are not AO devices, and so cannot access the
entire global system. (In particular, they cannot access all elements of
the input data stream). However, it is still the case that if they have been running long enough
(to have passed the stage of transient synchronization costs), 
and have an update rule matched to the dynamics of the data streams, then they
face none of the usual thermodynamic inefficiencies one would expect of non-AO devices.
Intuitively, an information ratchet can do this because it \textit{does} ultimately
access all physical variables $z$ in the (infinitely long)
input that are relevant for the computation --- just not all at once.
To circumvent this problem of a delay in access to relevant variables, the ratchet
stores in its state
the information concerning each such successive variable $z$ that is relevant for
minimizing EP \textit{in subsequent iterations}, after the ratchet has completed
its interaction with $z$.

%

\section{Kolmogorov complexity and the entropy dynamics of Turing machines}
\label{sec:TMs_us}

\subsection{The entropy flow for implementing a TM}
\label{sec:TM_me_old}


Suppose we are given a physical system that implements a prefix-free, single-tape
UTM, whose state space $X$ is the set of IDs of that UTM. 
%
Suppose we are also
given a desired output string ${\sigma}$. That output string in turn specifies a set $I({\sigma})$
of all input strings ${i}$ that
result in the UTM producing ${\sigma}$ and then halting.\footnote{In general,
that set is infinite, since for any input string ${i}$ that 
causes the UTM to compute ${\sigma}$,
we can construct another input string that just loops an arbitrary
number of times, doing nothing, before it runs the computation
starting from ${i}$.} What distribution over the elements of $I(\sigma)$ results in the
smallest total EF by the time it halts? We can assume
without loss of generality that that optimal distribution over inputs is a delta function,
so an equivalent question is, what is the least amount
of EF that could be incurred by running a UTM to compute $\sigma$ with an appropriately 
chosen element of $I(\sigma)$ as the input?

An answer to this question was originally given
in~\cite{wolpert_arxiv_beyond_bit_erasure_2015}.
The analysis in that paper assumed  that the physical implementation of the UTM has no residual EP. 
In addition, it was assumed that the starting form of the prior over  IDs is a
delta function over the state of the head (centered on its prespecified initial state), times a
delta function over the position of the head on the tape (centered on its prespecified initial position), times
the usual (normalized) coin-flipping prior over the set of all input tape strings for
which the UTM halts. Abusing notation, we can write that coin-flipping prior over initial tape strings $x$ that lie
in the halting set as 
$q(x) = 2^{-\ell(x)} / \Omega $ where $\Omega$, the normalization constant, is sometimes called Chaitin's constant. 
(Note that due to Kraft's inequality, $\Omega \le 1$, and so $\ln \Omega \le 0$.)
%
In addition, for simplicity, that analysis restricted attention to UTMs such that if they halt
on some input, then when they do so
the string containing their output starts at tape position $1$, and their pointer 
is positioned over the last symbol in that output string. (So the symbol
just to the right of the pointer when the UTM halts is a blank.)

The analysis in~\cite{wolpert_arxiv_beyond_bit_erasure_2015} considered this scenario for the 
special case of
a conventional three-tape implementation of a prefix UTM rather than single a
tape implementation. In addition, to
avoid the problems that arise if we only sum entropic costs incurred
by running a system until a random event occurs (see \cref{sec:variable_duration}), the
analysis in~\cite{wolpert_arxiv_beyond_bit_erasure_2015} 
considered the case where we run the physical system that implements the UTM
for an infinite number of iterations, summing
entropic costs all along, even those costs that arise after the UTM stops changing its state,
because it has halted. 

Here, I will derive the same value of the minimal EF as in that paper, but in a simpler, more direct
manner. To do this I will consider the minimal EF to physically implement 
the entire \textit{partial function} $f_M$ of an arbitrary UTM $M$, and do so in just a single timestep
rather than by iterating the update function of $M$. In other words, I will consider
the minimal EF needed by any physical system which, in a single timestep,
sends the set of all initial bit strings in the halting set of $M$ into the associated ending bit string.
Other than this change, I will make the same assumptions that were made in~\cite{wolpert_arxiv_beyond_bit_erasure_2015},
described just above.

Since residual EP
is assumed to equal zero, the EF for our (delta-function) initial distribution $p_0$  is the sum of two terms. 
The first is the mismatch cost for $p_0$, given the prior $q_0$ over initial bit strings. The second
is the generalized Landauer cost, for the partial function $f_M$.
Since the ending distribution over the UTM's string is just a delta function
(centered on $\sigma$), and so is the initial distribution,
that generalized Landauer cost is zero. Therefore the minimal EF is just the 
minimal mismatch cost over all input strings in $I(\sigma)$,
\eq{
\inf_{\sigma' \in \supp(I(\sigma))}\left( \sum_x \delta(x, \sigma) \ln \left[ \dfrac{G(\sigma)\delta(x, \sigma)}{\delta(x, \sigma)}\right]
- \sum_x \delta(x, \sigma') \ln \left[ \dfrac{q(x)}{\delta(x, \sigma')} \right] \right)
\label{eq:66}
}
where
\eq{
G(\sigma) := \sum_x q_0(x) \delta(f(x), \sigma)
}
is the ``universal prior probability'' evaluated at $\sigma$, up to normalization constants~\cite{livi08}

Recall that by L'hopital's rule, $0 \ln 0 = 0$.
%
Therefore this minimal EF is
\eq{
G(\sigma) - \inf_{\sigma' \in \supp(I(\sigma))} \;\;
\sum_x \delta(x, \sigma') \ln \left[ \dfrac{q(x)}{\delta(x, \sigma')} \right] &\;\;=\;\; 
G(\sigma) - \inf_{\sigma' \in \supp(I(\sigma))} \;\; \ln[q(\sigma')] \nonumber \\
&\;\;=\;\; G(\sigma) + \inf_{\sigma' \in \supp(I(\sigma))} \;\; \left(\ell(\sigma') + \ln(\Omega)\right)
}
(up to the arbitrary additive constant $\ln(2)$.)
So the minimal EF required to compute $\sigma$ using the UTM is
\eq
{K(\sigma) + \log[q(I(\sigma))] + \log \Omega
\label{eq:kolmogorov_correction}
}
where $K(\sigma)$ is the Kolmogorov complexity of using the UTM $M$ to compute $\sigma$.
This sum can be viewed as a ``thermodynamic complexity'' of computing $\sigma$ on $M$.
While Kolmogorov complexity is unbounded, it is straight-forward to show that the thermodynamic complexity 
has a finite upper bound over the set of all $\sigma$~\cite{kolchinsky2020thermodynamic}. 


Intuitively, the calculation of \cref{eq:kolmogorov_correction} says that the minimal EF to
compute a string $\sigma$ is given by adding a ``correction'' term to the Kolmogorov
complexity, which consists of Chaitin's constant for the UTM, plus the 
log of the total prior probability
under the coin-flipping prior of all input strings that produce
$\sigma$. That correction arises from the fact that Kolmogorov
complexity is concerned with the smallest length input string,
out of those input strings which result in $\sigma$, whereas \cref{eq:kolmogorov_correction}
is concerned with the smallest amount of EF generated by running
the UTM, out of those input strings which result in $\sigma$.

The normalization constant $\Omega$ in \cref{eq:kolmogorov_correction}
is uncomputable, and the two functions in \cref{eq:kolmogorov_correction}
are nonrecursive~\cite{livi08}. So the sum of those three terms cannot be computed.
However, that sum is only a lower bound on EF in any case, given by assuming
zero residual EP. So if one can compute lower bounds on each of those
three terms for a given $\sigma$, then the sum of those three 
lower bounds provides us with a (computable) lower bound to the EF
needed by a system that implements the UTM $M$ to compute $\sigma$.


%

It is important to realize that the expression in \cref{eq:kolmogorov_correction}
reflects nonzero mismatch cost. In fact, such mismatch cost is 
unavoidable, in the sense that we have a fixed physical system implementing the UTM, and are varying
the distribution over its input strings (looking for the delta function distribution
that results in minimal EF). Indeed, if we had zero mismatch cost, then $p_0$, the
actual distribution over inputs would have to equal the coin-flipping prior. 
This would mean that the distribution over output strings produced by
running the UTM would not be restricted to $\sigma$ --- in fact, that
distribution would have full support over
the space of output strings (since $U$ is a UTM, by hypothesis). 
Moreover, the EF for that $p_0$ is infinite, by Levin's coding theorem~\cite{kolchinsky2020thermodynamic}. 

%

\subsection{Recent investigation of entropy dynamics of a logically reversible TM}
\label{sec:strasberg}

%

The recent paper~\cite{strasberg2015thermodynamics}
contains an analysis of the thermodynamics of a CTMC that obeys LDB and implements an arbitrary, fully-specified
TM. The authors use a variant of the three-tape prefix TMs described above,
which instead involves an input tape, a history tape, a working tape,
and an output tape. They require that their TM's head cannot back up when
reading the input tape, just like the conventional three-tape,
prefix TMs~\cite{livi08}. However, in contrast to such TMs, they require that the
input string on the input tape of their TM be finite and clearly delimited, by blanks. 

Unusually, they also require that the TM be ``logically reversible''
even though it is implemented using a CTMC that obeys LDB, and so
backward versions of every allowed forward transition between IDs are allowed. (They relate
this to earlier work on ``Brownian computers'' in general, and DNA TMs
in particular~\cite{bennett1979dissipation,benn82}.) To do this they require that the only
paths through the space of IDs that are allowed by the CTMC are those that
obey the update rule of the TM. However, any given transition along such a path can go backward, to an earlier
ID, rather than forward, to the next ID. ``Logical reversibility'' for
them means that the update rule of the TM is logically reversible. This implies
that there is no intersection between any two paths; even though the system 
evolves stochastically forward and backward on a path, 
each allowed path is uniquely specified by its starting state.

Note though that this is not the meaning of ``logical reversibility'' conventionally
considered in the literature (see \cref{sec:bit_erasure}). In their model of a TM, any
of the states along a particular path (after the first state)
has {two} possible predecessor states. So
the system dynamics involves two-to-one maps (albeit stochastic ones).
This means that the system is no more or less logically reversible than
simple bit erasure is, and no more or less thermodynamically reversible
than simple bit erasure is. In addition, each iteration of the system
can result in non-zero Landauer cost, just like bit erasure does. 

%

They have two major conclusions. First, they establish that in the
infinite time limit, when the probability distribution over IDs reaches
a steady state, the Landauer cost per iteration can be made arbitrarily
small, by making the bias of the CTMC in favor of forward rather than
backward transitions small enough. However, they also show that in
the finite-iteration regime, where the TM has only been running for
a finite number of steps, the Landauer cost per iteration will be negative.
This reflects the fact that the distribution over IDs starts as a delta
function when the TM begins, but (due to the stochastic nature of the
CTMC) diffuses as the iteration number grows. (See the discussion
at the end of \cref{sec:accounting}.)

Importantly, the authors consider a scenario
where both input strings and output strings persist after completion of
a computation, for an infinite number of following computations. So
their analysis does not obey standard accounting. Concretely, their system
requires that the user has an infinite set of initialized bits to use to
store both the inputs and the outputs of every computation that the user
runs. As discussed in
\cref{sec:fredkin,sec:accounting_section}, 
such a set of bits is an information battery. If we wanted to, we could ust
use such a battery directly to drive the EF of a conventional, irreversible TM,
rather than use it to store copies of the inputs and outputs of all
runs of the TM in the past.


%

\subsection{Early investigations of the Landauer cost of logically reversible Turing machines}
\label{sec:bennett}

The earliest work on the
thermodynamics of Turing machines was by Bennett~\cite{benn73,benn82}. This work was done 
before modern nonequilibrium statistical physics, and so had to rely entirely
on the original bound developed by Landauer involving bit erasure, without exploiting
modern nonequilibrium statistical physics.
(See the discussion motivating the term ``Landauer cost'', just
afer \cref{ex:3a}.) Working under the supposition that the only way to reduce
thermodynamic cost was to reduce the total number of bit erasures, these early papers
concentrated on how to convert a logically irreversible TM
into an equivalent logically reversible one. After the initial
work showing how to perform such a conversion, the focus shifted to how to minimize
the resources needed to run that converted system, i.e., on
how to minimize the growth as that system progresses
in the size of its buffer ``history tape'', which it uses
to ensure the computation stays logically reversible~\cite{bennett1989time}.

However, the procedure for converting a logically irreversible TM into an equivalent
logically reversible TM is similar to the procedure for converting
a logically irreversible circuit into an equivalent logically
reversible circuit, as described in \cref{sec:fredkin}. This means that
there are caveats concerning the entropic costs of running a logically reversible
TM which is constructed to emulate a given irreversible TM that 
are similar to caveats concerning the entropic costs of running a logically reversible circuit
constructed to emulate a given logically irreversible circuit.

Specifically, recall that 
the answer-reinitialization of a logically reversible
circuit will incur at least as much Landauer cost as is incurred in 
the answer-reinitialization of the 
equivalent (logically irreversible) AO device. In fact, in one model of how to perform
answer-reinitialization, the {extra} Landauer costs of answer-reinitialization
of a logically reversible circuit
will be at least as large as the {entire} Landauer cost of running the
equivalent AO device. 

Similar issues hold for relating the entropic costs of given logically irreversible TMs and
equivalent logically reversible TMs. In particular, in the case of circuits, the
extra answer-reinitialization costs arise due to the need to reinitialize an extra set of output
bits that contain a copy of $x^{IN}$. Analogously, in the case of TMs,
the extra answer-reinitialization costs arise due to 
the need to reinitialize an extra set of bits on the output tape of the TM that
contain a (history which, combined with the computation output, can
be used to construct a) copy of the initial input string on the TM's input tape,
$x^{IN}$.


\subsection{Early investigations of the Landauer cost of logically \textit{ir}reversible 
Turing machines}
\label{sec:zurek}

I now consider one of the most sophisticated of the early papers 
on the thermodynamics of computation, which considered the thermodynamics
of \textit{ir}reversible Turing machines~\cite{zure89a}.
This paper focused specifically on the connection between the minimal ``thermodynamic
price'' it takes to run a given TM to compute a desired output string $\sigma$ starting
from a given input string $x^{IN}$ on the one hand, and on the other hand,
the conditional Kolmogorov complexity of the inputs to the TM,
conditioned on the desired output string.

The analysis in~\cite{zure89a} considers a scenario in which one first runs
a Bennett-style reversible TM, but when that TM finishes, 
``one insists on replacement of the input with the
output in the computer memory''~\cite{zure89a}. So implicitly at least,
there is appreciation for the problems with not reinitializing
that input which were discussed in \cref{sec:fredkin}.

However, because of when~\cite{zure89a} was written, it
could not exploit the modern, exact equalities for the entropic costs of arbitrary processes,
but had to rely entirely on indirect arguments concerning the ``thermodynamic cost'' of bit erasure
given by Landauer's original bound.
As a result, it is not clear precisely how to interpret that analysis in modern terms.
For example, the analysis in~\cite{zure89a} does not distinguish among EF, EP and Landauer cost
(concepts that had not been defined in their modern form when the paper
was written). Instead it informally refers to ``thermodynamic price''. 
Confusing the issue of how to interpret~\cite{zure89a} even more is that
the discussion in~\cite{zure89a} repeatedly confounds logical irreversibility and thermodynamic
irreversibility. (See the discussion in \cref{sec:bit_erasure}.)

At a high level, the idea in~\cite{zure89a} seems to be as follows. Suppose
you run a Bennett-style TM based on an irreversible TM $U$, ending with the desired output 
$\sigma$ and a (history which, combined with the computation output, can
be used to construct a) copy of the input, $s$.
As discussed in~\cref{sec:fredkin}, you now need to reinitialize that copy
of the input, i.e., erase it, to allow you to use your TM again. Assume
that the ``thermodynamic price'' of each bit in that copy of the input that you erase is
just the original Landauer bound, $k_B T \ln[2]$, where $T$ is the temperature
of the (single) bath. Accordingly, before erasing
the copy of the input, you (reversibly) encode it in the shortest string you can
with $\sigma$ as side-information. So long as that encoding is lossless, carrying it out has
zero entropic cost. At this point, we erase that encoded version of $x^{IN}$, $y$,
to complete a cycle. 

Since by construction $y$ is not longer than $x^{IN}$, 
the thermodynamic price of erasing $y$ is not higher --- and
potentially far smaller --- than would have been the thermodynamic price of
erasing $x^{IN}$. In more detail, suppose we explicitly specify 
our encoding / decoding algorithm that translates losslessly between $(s, \sigma)$
and $(y, \sigma)$. We do that by specifying a 
Turing machine $E$ --- possibly different from $U$ --- that decodes 
$y$ into $s$, given the side information $\sigma$. 
Then the minimal length of $y$ would be $K_E(s \mid \sigma)$, the conditional Kolmogorov complexity
of $s$ given $\sigma$, using Turing machine $E$~\cite{livi08}. 
So, the minimal thermodynamic price of erasing the extra copy of the input
is $k_B T \ln[2] K_E(s \mid \sigma)$.




To make this interpretation of the analysis in~\cite{zure89a} somewhat 
more precise, it helps to express the computation we're interested in somewhat
differently. Suppose we have a system with countable state space $X$,
and that the system starts in a particular \textbf{input} state $x^{IN} \in X$.
(Such an initial, input state is sometimes called a \textit{program} in~\cite{zure89a}.) 
We then repeatedly apply a deterministic ``update''' function $g : X \rightarrow X$ to the state of the system.
In general, $X$ may be finite or not, and
the map $g$ may be logically reversible or not.
%

We interpret that sequence of iterations of $g$ as the computation. Note that
this computation overwrites the input to produce the output. So the requirement of
standard accounting that the input be reinitialized by the end of
the computation does not apply. (See \cref{sec:overwrite_inputs}.)

We allow the total number of iterations of $g$ in a computation to be either a
pre-fixed, finite value, or instead to be determined dynamically, in a way that depends
on the input $x^{IN}$ (e.g., as in a finite automaton, or a TM). 
In the second of those two cases, it may be that the sequence of maps
never halts at all. In light of this, I will write $X^H \subseteq X$ to mean 
the set of initial states such that the resultant sequence of states
does indeed ultimately halt. (If the number of maps is a pre-fixed value,
then $X^H = X$.) From now on I specialize this notation to the case of a prefix-free UTM $U$, identify $X$ with
its set of IDs, and $g$ with its deterministic update function over that set.

Paralleling the terminology in~\cite{benn73}, in~\cite{zure89a} a
\textbf{history} is used to mean any sequence of irreversible changes to the ID of 
$U$ that arise during the iterations of $g$ on some
particular input $x \in X^H$. I write $[x, x']$ to mean an (arbitrary, fixed, invertible)
encoding of any such ``irreversible change'' from $x$ to $x' = g(x)$, with the requirement that we can recover the original 
ID $x$ from any such encoded change, $[x, x']$, together with the value $x'$. 
(Formally, there is a single-valued map from $(x', [x, x'])$ to $x$
for all pairs $x \in X, x' = g(x)$ such that the pre-image $g^{-1}(x')$ contains more
than one element.)  As an example, if both $x$ and $x'$ are
binary strings of the same length, $[x, x']$ could be a specification of the precise bits in which
they differ.  I also write $g^h(x)$ for the history of all irreversible changes $[x', g(x')]$ produced by iterating the function $g$ on an $x \in X^H$,
e.g., using some implicit invertible code for concatenating all those $[x', g(x')]$. 
Finally, I write $f(x)$ to mean the single-valued partial function taking any
$x \in X^H$ to the resultant state of $X$ that the system is in
when the sequence of update functions halts.

Furthermore, suppose we have a space $Y$ with each $y \in Y$ interpreted as an encoding of 
$x$, conditioned on the information $f(x)$. Formally, I require that there is a map
$F: Y \times f(X^H) \rightarrow X^H$ such that for all $x \in X^H$, 
there exists a $y \in Y$ such that $F(y, f(x)) = x$. $F$ is the \textbf{decoding map}.
As an example, if $y$ is just a copy of $x$, then we can take $F(y, f(x)) = y$. If instead $y$ is
the history produced by iterating $g$ to produce $f(x)$ from $x$, then $F$ is more complicated.
To comport with the high-level description of the reasoning in~\cite{zure89a} presented above, suppose as well that 
the decoding map $F(., .)$ is run by a TM $E$.

Given these definitions, define an associated set-valued \textbf{encoding map} to be a function $G : X^H \rightarrow 2^Y$
such that for all $x \in X^H$, and all $y \in G(x)$, $F(y, f(x)) = x$. 
(Note that in general, if we change $F(., .)$, we must also change $G(.)$.) So $G$ takes any
halting input and produces a set of $y$'s for all of which
the decoding condition that defines $F$ is fulfilled. 
I will also write $D: Y \rightarrow \R^+$ for some (arbitrary) {\textbf{size}} function. 

As an example of these definitions, in~\cite{benn73} the function $[x', g(x')]$ is a prefix-free code that produces a bit string
for all irreversible changes $x' \rightarrow g(x')$ such that $x' = g^n(x)$ for some counting number $n$ and
some $x \in X^H$. 
$g^h(x)$ is the finite bit string produced by concatenating the successive bit strings $[x', g(x')]$
encoding the irreversible changes that result from iterating $g$ on any $x \in x^H$. $Y$ is the set of all such
finite bit strings, $\{g^h(x) : x \in X^H\}$.
So we can take $D(y)$ to just be $\ell(y)$. 
$F$ (i.e., the TM $E$) is the function that takes the output generated by running $U$ on $x^{IN}$, $f(x^{IN})$, together with any
associated encoded version of the history of producing $f(x^{IN})$ from $x^{IN}$, $y$, and reconstructs $x^{IN}$. It does this
by decoding the encoded version of the history and then running $U$ ``in reverse'' starting from the output of the forward run, $f(x^{IN})$,
using the entries in the history
to resolve any ambiguities in this reverse-evolution as they are encountered.\footnote{Note that for all possible
inputs to this $F$, we are guaranteed that the TM $E$ implementing $F$ halts after a finite number of steps.
Roughly, that number of steps is proportional to the sum of the number of bit strings $[x', g(x')]$ in the history $y$
 (since $E$ must decode all those encoded ID changes) and the number of
steps that $U$ took to halt (since $F$ must reverse all of those steps).}

\cite{zure89a} generalizes this setup of~\cite{benn73} in two ways. First, it assumes that 
$E$ is a universal TM. Second, it expands $Y$ to be all finite bit strings $u$ such that
running the UTM $E$ on $(u, f(x))$ for any $x \in X^H$ first produces the history $g^h(x)$ in a finite number of steps,
and then runs $U$ ``in reverse'' to generate $x$, exactly as in the setup considered in~\cite{benn73}.
With this modification, there is some
large (actually, infinite) set of strings ${\cal{Y}}_{x^{IN}} \subset Y$ such that $E$
will map all pairs of an ID $f(x^{IN})$ and a  $y \in {\cal{Y}}_{x^{IN}}$ to the same history, and so to the same initial string $x^{IN}$ on the tape of $U$.
$G $ is simply the map that takes any initial $x^{IN}$ to such a subset of $Y$, i.e., to 
the set of all encoded versions of the history $g^h(x^{IN})$ that are decodable by $E$.


The analysis in~\cite{zure89a} assumes that rather than implement the UTM $U$ directly in a physical system,
one implements the map $\Gamma : x \in X^H \rightarrow \big([ \argmin_{y \in G(x)} D(y)], f(x)\big)$. So $\Gamma$
takes the initial state $x$ to a combination of the final state $f(x)$, together with the encoding of the history
generated by producing $f(x)$ from $x$ that has smallest size out of all such encodings of that history.
By definition of $G$, $\Gamma$ is logically reversible. Therefore both its Landauer cost and its mismatch cost are zero,
no matter what the initial distribution over $X^H$ is, and no matter what the
prior over $X^H$ is. (See \cref{sec:fredkin}.) Assuming that the residual EP
of the system implementing $\Gamma$ is also zero, this means that the EF to run $\Gamma$ is zero.
This establishes Lemma 1 in~\cite{zure89a}.

%

After having run $\Gamma$, in order to meet the requirements
of standard accounting we must reinitialize $y$. Recognizing this, the central
question posed in~\cite{zure89a} is:
\begin{quote}
\textit{What is the minimal thermodynamic price to reinitialize the history (i.e., erase $y$)? }
\end{quote}
For simplicity, to try to provide a rigorous interpretation of the informal answer in~\cite{zure89a}  to this question, 
consider the situation that~\cite{zure89a} focuses on,
where $Y$ is the infinite set of bit strings described above, and $D(y)$ is the length of the bit string $y$.
There is no mention of EP in~\cite{zure89a}, and the formula
for mismatch cost had not been derived when that paper was written --- the proper tools
for analyzing the thermodynamics of computation had not been developed when the paper was written.
That is why ``thermodynamic price'' is not properly defined, from a modern perspective.
Nonetheless, we can try to translate 
the answer to the question given in~\cite{zure89a} into modern stochastic thermodynamics. Doing so, it seems
that it makes sense 
to interpret ``thermodynamic price'' as meaning total EF --- including mismatch cost --- for the case where the
residual EP is zero. 

This interpretation is based on several assumptions:
\begin{itemize}
\item 
Note that there is implicitly a prior $q(y)$ in the system that reinitializes (i.e., erases) $y$. Suppose that 
$q(y)$ is uniform over all strings of length less than or equal to some $L$,
and that it is zero for all longer encoded strings.\footnote{If we
did not build such an upper limit on the length of the strings into
this prior, then it would be un-normalizable.}
To ensure that there is no possibility of infinite mismatch cost with
this prior, we have to also assume that the actual distribution over
inputs to the TM $U$ has its support restricted to a set $\hat{I}_L$ 
containing only those $x^{IN} \in X^H$ that result in $y$'s that are not longer
than $L$ bits. In turn, to ensure reversibility of the map $\Gamma$, we assume that
$\hat{I}_L$ does not contain more than $2^L$ elements.

\item Suppose you are given a desired output of the TM, $\sigma$. Suppose
that $I(\sigma) \cap
\hat{I}_L \ne \varnothing$ and that the TM starts its computation of $\sigma$
starting from some particular $x^{IN} \in I(\sigma) \cap \hat{I}_L$.\footnote{Recall the definition
of $I(\sigma)$ from the beginning of \cref{sec:TM_me_old}.}
\end{itemize}
Making these assumptions, and adopting standard accounting, we ask:

\begin{quote}
\textit{Given some $x^{IN}$ where $f(x^{IN}) = \sigma$, what is the minimum value, over all $L$ 
such that $x^{IN} \in I(\sigma) \cap \hat{I}_L$, of the total EF that would be generated
by first running $\Gamma$ on $x^{IN}$, followed by 
erasure of the resultant $y$?}
\end{quote}

\noindent (It seems that this minimal EF is what~\cite{zure89a} refers to as the 
``thermodynamic price'' for computing $\sigma$ from $x^{IN}$.)

Since we're taking residual EP to equal zero, and since the EF 
incurred in running $\Gamma$ is zero, the total EF is the
drop in the cross-entropy between the actual distribution $p(y)$ and $q(y)$
that occurs from the beginning to the end of the process of erasing the $L$ bits of $y$. 
Since $q(y)$ is uniform, the associated drop in cross entropy is just $L \ln[2]$. 
Moreover, by definition of $G$, the minimal $L$ is just $K_E(x^{IN} \mid \sigma)$.
So the answer to the question is $K_E(x^{IN} \mid \sigma) \ln[2]$. This is Theorem 1 in 
\cite{zure89a}. 

There are also subtle aspects of the analysis in~\cite{zure89a} worth bearing in mind.
First, as mentioned above, the TM defining the Kolmogorov complexity function
$K_E(.)$ that appears in the results in~\cite{zure89a} 
need not equal the Kolmogorov complexity of the TM $U$ whose cost is being analyzed; there are two TMs
involved, and \textit{a priori}, they need not have any relation. 
At best, one could say that the difference between the Kolmogorov complexities
of the two TMs is bounded by an arbitrarily large constant, 
which is independent of the output string being computed.
In other words, the analysis gives ``the minimal thermodynamic price''
it takes to run a given TM to compute a desired output string
up to an unspecified additive term, a term that is bounded by a
constant that can be made arbitrarily large.
%
To avoid this issue, from now on require that the two TMs,
 $E$ and $U$, are identical.

Second, one shortcoming of the calculation based on finding a minimal $L$ 
is that it implicitly requires that we can vary the prior over $Y$, in order to find
the one that is optimal for the UTM to compute $\sigma$. This makes little physical sense.
However, we can make a semi-formal argument motivating a similar calculation in which we 
take $L$ to infinity, and also optimize the EF over the
set of all $x^{IN} \in I(\sigma)$, rather than only consider the entropic
cost for some specific arbitrary $x^{IN} \in I(\sigma)$. 

To present this semi-formal argument, first recall that we require that $E$ halts for all pairs $(\sigma, y) : y \in Y$,
and that $E$ is a prefix TM. So by taking $\sigma$ to be the empty string we see that $Y$ must be a prefix-free set.\footnote{Note that 
by the discussion above of the setup in~\cite{benn73}, in which each element of $Y$ is
a prefix-free encoding of a history, we are
guaranteed that
$Y$ contains an encoding of every history $g^h(x)$ that
can occur.}
Also assume that the prior of the history-erasing
process happens to be the coin-flipping prior restricted to $Y$, i.e., for any two bit strings
$y, y'$ , $q(y) / q(y') = 2^{\ell(y')-\ell(y)}$.\footnote{Note that we are not requiring the user
of the system to construct the system explicitly to have that prior --- in the real world, there
is almost never any consideration for what the prior of the dynamics of a physical system is when that
system is constructed. Here we're merely assuming
that the process just so happens to have the coin-flipping prior.} Write the normalization constant
of that prior as $Z$. In addition,
assume that the history-erasing process only erase the $\ell(y)$ bits that
comprise $y$.\footnote{Note that this implicitly relies on the fact that $y$ is an element in a prefix-free code, so
that the history-erasing process can know when to stop erasing.} 

Next, assume that the history-erasing process is run by a TM, and to ensure that this TM eventually halts, with $y$ set back to be all zeroes,
take the limit of running it for an infinite number of iterations (as in \cref{sec:TM_me_old}). Now \textit{any} distribution over those $y \in Y$ that can
be produced by $U$ is turned into
a delta function (about the sequence of all zeroes) by the history-erasing TM. In particular, this is true for whatever
the actual distribution is (determined implicitly by a distribution over $X^H$), and for the prior distribution.
So the ending cross-entropy after the history-erasing TM has halted between the actual distribution over $Y$ and the prior 
distribution goes to $1 \ln[1] = 0$.  Accordingly, the drop in cross-entropy between those two distributions during
the history-erasing process is just the initial cross-entropy between them just before the history-erasing process starts.

Next, suppose we have a single allowed initial state of $U$, $x^{IN} \in X^H$,
and define $y' \in \mathcal{Y}_{x^{IN}}$ to be the history produced by running $U$ on $x^{IN}$.
In this case  the actual distribution over $Y$ just before the history-erasing TM starts
is a delta function centered on $y$. So the value of the `` initial cross-entropy'' between the actual
distribution over $Y$ and the prior distribution over $Y$ just before the history-erasing process starts 
is $\ell(y') + \log[Z]$. As discussed above, the minimal value of $\ell(y')$ is
$K(x^{IN} \mid \sigma)$. Plugging this in and then minimizing over $x^{IN}$, we see
that the minimal EF is
\eq{
\min_{x^{IN} \in I(\sigma)} \bigg(K(x^{IN} \mid \sigma)\bigg) + \log[Z]
\label{eq:zurek_optimized}
}

To understand this result intuitively, note that $K(x^{IN} \mid \sigma)$ can be interpreted 
as the amount of information that is contained in $x^{IN}$ but not in $\sigma$.
Accordingly, the minimal EF given in \cref{eq:zurek_optimized} is just the least
possible amount of information that would be lost by the TM $U$ transforming one of the elements of $I(\sigma)$
into $\sigma$ (up to the additive constant of $\log[Z]$).

There are several points about this semi-formal argument that are worth emphasizing. First,
the ``thermodynamic price'' in \cref{eq:zurek_optimized} includes the dissipated work of the history-erasing
process due to mismatch cost. That term is the difference between the actual work expended
in the history-erasing process on the one hand,
and the minimal possible work which would need to be expended if the actual distribution over $Y$ and the prior over $Y$ were identical. 
Moreover, that ``actual distribution over $Y$'' is a delta function, reflecting the fact that this entire analysis assumes that the TM
$U$ and subsequent history-erasing process are run \textit{only} with the specified input $x^{IN}$. In
essence, the entire analysis is for a bespoke physical system which is only ever used
to implement $U$ on one specific input.

One might argue that it makes more sense to investigate the EF of the history-erasing process
that arises when the input distribution has nonzero support
over multiple inputs --- perhaps in fact for all $x \in X^H$ --- not just over one specific $x^{IN}$. Due to the nonlinearity of Shannon entropy, the
EF for that case would not just be the average over the EFs in~\cref{eq:zurek_optimized}. In the best possible
case, the actual distribution $p(x^{IN})$ over inputs would result
in any actual distribution over histories $y$ equals the prior over histories. In this case
there would be zero mismatch cost in the history-erasure process, its theoretical
minimum, and therefore there would be zero EP in the overall process. 

However, in this EP-minimizing scenario, the total EF would actually be lower-bounded
by the EF that would have been generated  if we had just run the irreversible UTM $U$ directly, without 
storing histories which must then be erased.\footnote{To see this, first note that
since EP is zero, the entire EF would be given by the generalized Landauer bound
of the process of erasing the history. That bound just equals the entropy of the distribution over histories, since
it ends with the history tape completely erased, no matter what the initial distribution over histories. Moreover,
since the running of $\Gamma$ is logically irreversible, the entropy of the joint distribution over the output
of the computation and of the history, i.e., over pairs $(f(x^{IN}), y)$, equals the entropy of the actual initial distribution over $x^{IN}$. 
Finally, note that the mutual information between the distribution over values $f(x^{IN})$ and associated histories $y$ is non-negative.
Combining, we conclude that the EF expended by the entire history-erasing process is lower-bounded by the difference in the
entropy over $x^{IN}$ and the entropy over $f(x^{IN})$.} In this sense, the entire complicated process
of storing and then erasing a history is superfluous, with no thermodynamic benefit at all.  

A second point worth emphasizing is that
%
the system considered in \cref{sec:TM_me_old}
is just (an implementation of) the single TM $U$. In contrast, assuming we take $U = E$, the system considered here first
runs a function $\Gamma$ which is \textit{constructed from} $U$, and then runs a history-erasing process.
Moreover, in the system considered in \cref{sec:TM_me_old}, the coin-flipping prior
arises as a distribution over $X^{IN}$, whereas in the system considered here,
the prior over $X^{IN}$ is arbitrary; instead, in the system considered here the coin-flipping prior arises 
as a distribution over $Y$ just before we start the history-erasing process. 

The price paid for changing the system so that we can move the coin-flipping prior from $X^{IN}$ to $Y$ is 
that that new system needs to be able to calculate the non-recursive function sending $x$ to the
shortest string $y$ such that $F_E(y, f(x)) = x$. In other words, the new system
has to have super-Turing capabilities, unlike
the original system, which consisted only of the TM $U$.

It is also worth emphasizing that the thermodynamic price of computing $\sigma$
from some specific $x^{IN}$ (as in~\cite{zure89a}) is not the same as the
minimal EF needed to run {the TM $U$} over {all} states $x^{IN}$ that result in the TM computing $\sigma$.
So it is not the quantity that is analyzed in \cref{sec:TM_me_old}. 
Instead, the analysis in~\cite{zure89a} fixes
$x^{IN}$ ahead of time, to an arbitrary element of $I(\sigma)$, and considers what
the minimal EF would be in running the UTM $U$ on $x^{IN}$, if we could implement
that UTM by implementing a system that runs the map $\Gamma$, followed by erasing $y$. 

As a final technical point, note that 
just like other results in Turing machine theory, all the results recounted here
hold only up to additive $O(1)$ terms. For infinite spaces $X$ and $Y$, that is fine,
but for finite $X$ and $Y$, such additive constants can swamp the other terms in these
results.\footnote{Zurek was well aware of this
problematic aspect of analysis summarized in this subsection, saying that the bound of Theorem 1
in \cite{zure89a} ``cannot \{actually
be met\} by recursive computation \{i.e., by running a TM guaranteed to halt\}'', that
at best it is met ``perhaps by sheer luck''.} 
So these results have little to say about computation with finite $X$. To give
a simple example, all of the results recounted here apply to straight-line circuits. Yet 
as recounted in \cref{sec:circuit_entropy_dynamics}, the formulas for the entropic costs of circuits
do not involve Kolmogorov complexity (in contrast to the results presented in this section).

As a closing, historical comment, an explicit goal of much of the early work on the thermodynamics
of TMs was to rederive statistical physics {with little if any use of 
probabilities}~\cite{livi08,caves1993information,caves1990entropy,zure89b,zupa94}.
Ultimately, perhaps the most fruitful way to consider the analysis in~\cite{zure89a}
is not to translate it into modern nonequilibrium statistical physics, so deeply grounded in
the use of probability distributions, but rather to view it as part 
of this research program which strived to expunge the use of probabilities
from statistical physics.
%

%
%

\section{The mimimal hidden computation occuring within a visible computation}
\label{sec:hidden}

Recall the construction summarized in \cref{sec:overwrite_inputs},
of an AO device that can implement any conditional 
distribution $\map$ over a set of ``visible'' states $X$ in a thermodynamically reversible manner --
even if the output distribution under $\map$ depends on the value of the input.
This construction works by expanding the original state space $X$, of size $|X|$,
into a state space of size $|X| \times |X'|$, and
defining a dynamics over $X \times X'$ that implements $\map$
over the subspace $\{(x, 0) : x \in X\}$. 

In that construction, $X'$ is the same
size as $X$, so the joint space $X \times X'$ has $|X| (|X| - 1)$ more ``hidden'' states
than the original space of ``visible'' states, $X$. That can be a \textit{huge} number
of extra states. For example, in a digital computer with a 64-bit address space,
$|X| = 2^{64}$ --- and so
the number of extra states needed to implement a nontrivial
map over $X$ using that construction is $\sim 2^{128}$.

This raises the question of what the minimal number of extra,
hidden states must be, in order to implement
a given distribution $\map$ in a thermodynamically reversible way. Some recent results
concerning this and related issues are summarized in this section.

First,  one doesn't need to restrict attention to
CTMCs that operate over Cartesian product spaces $X \times X'$.
In full generality, we are interested in CTMCs that operate over
a space of the form $X \cup Z$, for some appropriate $Z$. The construction
summarized in \cref{sec:overwrite_inputs} is just a special
case, since we can rewrite a space $X \times X'$
as $X \cup Z$ if we identify $Z$ as $X \times X' \setminus \{(x, 0) : x \in X\}$.

Bounds on the minimal value of $|Z|$ needed for a CTMC over $X \cup Z$
to both implement a given conditional distribution $\map$ over $\{(x, 0) : x \in X\}$
and to be thermodynamically reversible are derived in~\cite{owen_number_2017},
for the case of finite $X$. These bounds apply to any
$\map$, even ``maximally noisy'' ones where for every $x', x$, $\map(x \mid x')
\ne 0$.
That paper also contains some results for deterministic $\map$,
for the special case of a countably infinite $X$. 

However, in computer engineering we are typically interested in 
conditional distributions $\map$ that implement a deterministic, single-valued function
over a finite space $X$. 
There are several special properties of such scenarios.
First, perhaps surprisingly, it turns out that any nontrivial deterministic map
$x \rightarrow f(x)$ cannot be implemented by \textit{any} CTMC operating over
only the visible states $X$, \textit{even approximately}~\cite{wolpert_minimal_2017}. 
This is true independent of concerns about
thermodynamic reversibility or the like; it simply arises due 
to the mathematics of CTMCs~\cite{jia2016solution,lencastre2016empirical}. 
As a striking example, a simple bit flip over a space $X = \{0, 1\}$
cannot be implemented by any CTMC over $X$, even approximately,
no matter how much dissipation accompanies
the CTMC, or how the rate matrix varies in time, or how long we run the CTMC.

This means that in order to implement any nontrivial deterministic map,
thermodynamically reversibly or not, one \textit{must} use hidden states. So for
example, under the approximation that some real-world digital device implements a deterministic
Boolean map over a set of visible bits, and that the dynamics of the device
can be modeled using
stochastic thermodynamics, we know that that the CMTC going into
the stochastic thermodynamic analysis must use hidden states.

Perhaps even more surprisingly, it turns out that
any CTMC that implements a deterministic map
can be decomposed into a sequence of more than one ``hidden'' timesteps.
These timesteps are demarcated from 
one another by changes in what transitions are 
allowed under the rate matrix~\cite{wolpert_minimal_2017}.
In general, for a given set of visible states $X$, 
hidden states $Z$, and deterministic map $f : X \rightarrow X$, the minimal number of hidden
timesteps (for any CTMC over $X \cup Z$ to implement $f$ over $X$) is greater than $1$.
So any real-world digital device that implements some non-trivial Boolean
operation over its state space in each of its iterations must have a set
of multiple hidden timesteps that occur within each of those iterations
(assuming the device can be modeled as evolving under a CTMC
that implements a deterministic function).

Often there is a real-world cost for each additional hidden state, and also
a cost for each additional hidden
timestep. (For example, in systems that evolve while connected to a heat bath
and obeying LDB for some Hamiltonian,  at the end of a timestep either an infinite
energy barrier between elements of $Z$ is raised, or
an infinite energy barrier is lowered.) 
So there is a ``space/time'' tradeoff between the costs associated with the
number of hidden states used by any CTMC that implements a given $f(x)$
and the costs associated with the
number of hidden timesteps used by the CTMC to implement $f(x)$.

This tradeoff involving hidden states and hidden timesteps occurs within
any digital device, and in a certain sense
``lies underneath'' the more conventional space / time tradeoffs 
studied in computer science theory.
Indeed, consider the special case that there are no constraints on the CTMC
operating over $X \cup Z$, e.g., due to restrictions on the Hamiltonian
driving that CTMC. In this case, that CTMC
operates as an AO device over $X \cup Z$. For this special case,
the precise form of the hidden space/time tradeoff can be given in closed form~\cite{wolpert_minimal_2017}. As an example, it turns out that (if there
are no constraints on the CTMC) a bit flip
can be implemented using only one hidden state, but only if one uses three hidden timesteps.

As an example of this involving a natural rather than artificial system, suppose that we observe that a protein inside a cell goes from configuration $A$ to configuration $B$ with probability $1$, e.g., over several minutes. Suppose that if instead the protein was initially in configuration $B$, then in that same process it would evolve to configuration $A$ with probability $1$. So the protein configuration undergoes a ``bit flip''. Then we know that there must be at least one other configuration, different from both $A$ and $B$, that the protein adopts with nonzero probability at intermediate times. We also know that the whole process can be decomposed into at least three successive timesteps (assuming that the underlying process can be modeled as a CTMC).

\section{Future Directions}
\label{sec:future_work}

There has been a resurgence of interest recently in the entropic
costs of computation, extending far beyond the work in stochastic
thermodynamics summarized above. This work has taken place
in fields ranging from chemical reaction
networks~\cite{soloveichik2008computation,chen2014deterministic,murphy2018synthesizing,qian2011scaling,thachuk2012space} to cellular biology~\cite{benenson2012biomolecular,prohaska2010innovation} to neurobiology~\cite{laughlin2001energy,balasubramanian2001metabolically}, in addition to 
computer science~\cite{demaine2016energy}. There is
now a wiki serving as a community resource for workers 
from all these different disciplines
with a common interest in the entropic costs of 
computation (\texttt{http://centre.santafe.edu/thermocomp}).\footnote{Researchers 
are \textit{highly} encouraged to visit this site, not just
to find useful information, but also to improve
the site, e.g., by putting in announcements, adding references, adding
researcher web page information, etc.} In addition, there is a book coming
out in early 2019, that summarizes some of the insights of researchers
from all those fields~\cite{wolpert_book_2018}.

Almost all of the work to date on the entropic costs of computation
is concerned with \textit{expected} entropic costs,
averaging over all trajectories the microstates of a system might follow.
However, a very fruitful body of research in stochastic thermodynamics considers
the full distribution of the entropic costs of individual trajectories~\cite{seifert2012stochastic,van2015ensemble}.
The associated results allow
us to consider questions like what the probability is in a given
process of EP having a particular
value, e.g., as formalized in the famous ``fluctuation theorems''~\cite{crooks1999entropy,crooks1998nonequilibrium,jarzynski_equalities_2011}.
Some more recent research along these lines has resulted in ``uncertainty relations''~\cite{gingrich2017fundamental,gingrich2016dissipation,garrahan2017simple,pietzonka2017finite}, which relate how precise the dynamics of a system (e.g., of a computational
machine) can be on the one hand, to the EP that dynamics generates on the other hand. Other
recent research has resulted in classical ``speed limits'' to go 
along with the quantum ones~\cite{deffner2017quantum}. As an example, one such speed limit 
lower-bounds how confident
we can be that a desired state transition has occurred in a given amount of time,
as a function of EP~\cite{DONTUSE_shiraishi2018speed}. (See also~\cite{okuyama2018quantum}.)

Clearly these recent results are related to issues that play a fundamental role in
the entropic costs of computers, and in particular to the tradeoffs 
between
those costs and how confident we can be that a given physical system implements 
a desired computation exactly, how fast it can do so, etc. 
All of these issues remain to be explored. 

Restricted attention just to the issues touched on in this paper,
there are many avenues of future research that bear investigating.
%
One such set of issues concerns the question of how the entropic costs of probabilistic
Turing machines~\cite{arora2009computational} are related to those
of deterministic Turing machines (the kind considered in \cref{sec:TMs_us}). 
Similarly, the analysis in
\cref{sec:TMs_us} considered TMs that operate over the space of IDs of
the TM. So in each iteration, they act as an AO device. A natural line of
research is to explore how the entropic costs change if we instead 
model the dynamics of the TM in each iteration as a solitary process
over a finite subset of the variables in the TM, as given by the conventional definition of 
TMs in terms of heads that can only access one position on the tape at a time.
(See \cref{sec:TMs_def}.)

There are also many issues to investigate that are more closely tied
to the kinds of problems conventionally considered in computer science
theory. To give a simple example, for deterministic finite automata,
we could investigate the following issues, which are closely 
related to topics introduced (and solved)
in introductory courses in computer science theory:

\begin{enumerate}
\item Given a language $L$, does the (unique) minimal deterministic
FA that accepts $L$ also
result in the smallest total Landauer cost (conditioned on having the 
distribution over inputs produce some string in $L$) of any 
deterministic FA that accepts $L$?

\item Is the deterministic FA with minimal total Landauer cost (conditioned on having the 
distribution over inputs produce some string in $L$) unique (up to relabeling of states)?

\item What is the largest possible total Landauer cost of any deterministic FA that accepts $L$
(conditioned on having the distribution over inputs produce some string in $L$)?

\item Suppose we are given two deterministic FA, $M_1$ and $M_2$ and a shared distribution over input strings, 
where some of the strings accepted by $M_1$ are also accepted by $M_2$. What is 
the probability of a string that is accepted by both where the Landauer cost for
$M_1$ exceeds that for $M_2$? What is the probability of such a string for which the
two deterministic FAs have identical Landauer cost?

\item Suppose we have a deterministic FA that can get stuck in infinite loops, and / or accept
arbitrarily long strings. Then we can ask many ``entropy rate'' variants of these questions.
For example, does the Landauer cost per iteration of a given 
deterministoic FA approach an asymptotic value?
How many such asymptotic values could the same FA approach (depending on the
precise infinite input string)?

\item We can also consider variants of all these issues where 
consider mismatch costs instead of (or in 
addition to) Landauer costs.

\end{enumerate}

A core concern of computer science theory is how the tradeoffs among the
amounts of various resources needed to perform a computation scale with
the size of the computation. For example, a common question is
how the memory requirements of a computation trade off with the number
of iterations the computation requires, and how both of those scale
with the size of the computation. In fact, many textbooks are devoted to tradeoff
issues, instantiated over different kinds of computational machine. 

To date, none
of this analysis has included the \textit{thermodynamic} resources
need to perform the computation, and how they scale with the size of the
computation. In essence, every chapter in those textbooks can be revisited
using the tools of stochastic thermodynamics,
to see how they are extended when one includes these other kinds of resources.
%
%
%

$ $

\noindent \textbf{ACKNOWLEDGEMENTS:}
I would like to thank Nihat Ay and especially
Artemy Kolchinsky for many helpful discussions. This work was
supported by the Santa Fe Institute,   
Grant No. CHE-1648973 from the U.S. National Science Foundation and Grant No.
FQXi-RFP-1622 from the FQXi foundation. 
The opinions expressed in this paper are those of the author and do not necessarily 
reflect the view of National Science Foundation.

\newpage

\appendix

\section{Properties of multi-divergence}
\label{sec:multidivappendix}

Note that in \erf{eq:15a} I adopt the convention that we
subtract the sum involving marginal probabilities. In contrast, 
in conventional multi-information I subtract the term involving the joint probability.

To understand why I adopt this convention, 
consider the case where $X$ has two components, labelled $a$ and $b$,
and use the chain rule for KL divergence to write
\eq{
\IIDDf{\p}{\r} = \DDf{\p^{a \mid b}}{\r^{a \mid b}} -  \DDf{\p^a}{\r^a} 
\label{eq:kl_mut_alt}
}
with obvious notation,
where $\p$ and $\r$ are both distributions over the joint space $(X^a, X^b)$.

$\DDf{\p^{a}}{\r^{a}}$ quantifies how easy it is to tell that a sample $x_a$ came
from $\p(x_a)$ rather than $\r(x_a)$. Similarly, $\DDf{\p^{a \mid b}}{\r^{a \mid b}}$
tells us how easy it is, on average (according to $\p(x_b)$), 
to tell that a sample $x_a$ came from $\p(x_a \mid x_b)$ rather than $\r(x_a \mid x_b)$.

So \erf{eq:kl_mut_alt} says that the multi-divergence between $\p$ and $\r$ is the gain in
how easy it is for us to tell that a sample $x_a$ came from $\p$ rather than $\r$ if 
instead of only knowing the value $x_a$ I also know the value $x_b$. In other words, it quantifies how
much information $X^b$ provides about $X^a$ (on average), concerning the task of guessing whether
a given sample $x_a$ was formed by sampling $\p$ or $\r$.
(Note that since multi-divergence is symmetric among the components of $x$, $\IIDDf{\p}{\r}$ also quantifies how
much information $X^a$ provides about $X^b$.) However, this interpretation
requires our convention, of subtracting the sum of marginal divergences from the
joint divergence rather than the other way around.

In light of these properties of multi-divergence for variables with two components,
we can interpret the multi-divergence between two probability distributions 
both defined over a space $X$ that has an
arbitrary number of components as quantifying how much knowing one component
of $X$ tells us about the other. This is similar to what multi-information measures. 
Indeed, as mentioned in the text,
$\IIDDf{\p}{\r}$ reduces to conventional multi-information $\II(\p)$ in
the special case that $\r$ is a product distribution.

So under our convention, multi-divergence quantifies us how much knowing one component
of $X$ tells us about the other --- somewhat analogous to what multi-information measures. 
Indeed, $\IIDDf{\p}{\r}$ reduces to conventional multi-information $\II(\p)$ in
the special case that the reference distribution $\r$ is a product distribution. 

In addition to motivating our convention, this also means that the multi-divergence cannot be negative 
if $\r$ is a product distribution. 
More generally, however, if $\r$ is not a product distribution but the components of $X$ 
are correlated under $\r$
 in a manner ``opposite'' to how they are correlated
under $\p$, then $\IIDDf{\p}{\r}$ can be negative. In such a case, being given a value $x_b$ in
addition to $x_a$ makes it \emph{harder} to tell whether $x_a$ was formed by sampling from $\p$ or from $\r$. 

To illustrate this, consider a very simple scenario where
$X = \B^2$, and choose 
\eqn{
\p(x_b) &= \delta(x_b, 0) \\
\p(x_a \mid x_b = 0) &= 1/2 \qquad  \forall x_a \\
\r(x_b) &= \delta(x_b, 1) \\
\r(x_a \mid x_b = 0) &= 1/2 \qquad  \forall x_a \\ 
\r(x_a \mid x_b = 1) &= \delta(x_a, 0)
}
Plugging into \erf{eq:15} gives
\eqn{
\IIDDf{\p}{\r} &= - \sum_{x_b, x_a} \p(x_a \mid x_b) \p(x_b) \ln \bigg[ \frac{\r(x_a \mid x_b)}{\p(x_a \mid x_b)}\frac{\p(x_a)}{R(x_b)}  \bigg]   \\
&= -(1/2) \sum_{x_a} \ln \bigg[ \frac{\r(x_a \mid x_b = 0)}{(1/2)}\frac{(1/2)}{\r(x_b)}  \bigg]   \\
   &= -(1/2) \sum_{x_a} \ln \bigg[ \frac{(1/2)}{(1/2)}\frac{(1/2)}{\delta(x_a ,0)}  \bigg]  \\
   &= -\infty
}
So on average, if you are told a value of $x_b$ that unbeknownst to you came from $\p$, in addition to being told
a value of $x_a$ that unbeknownst to you came from $\p$, then you are less able to tell that that $x_a$ value
came from $\p$ rather than $\r$.

This phenomenon is loosely similar to what's sometimes known as Simpson's paradox. 
This can be seen by considering the instance of  that ``paradox'' 
where I have a distribution $\p(z, x_b, x_a)$ over three binary variables,
and simultaneously 
\eqn{
\p(z = 1 \mid x_b, x_a = 1) > \p(z = 1 \mid x_b, x_a = 0)
}
for any value of $x_b$, yet 
\eqn{
\p(z = 1 \mid x_a = 1) < \p(z = 1 \mid x_a = 0)
}
For such distributions,  if we are told the value of $x_b$ in addition to the value of $x_a$,
we conclude that $z$ if more likely to equal $1$ when $x_a = 1$ than
when $x_a = 0$. This is true
no matter what I am told the value of $x_b$ is. Yet I come to the opposite conclusion 
if we are only told the value of $x_a$, and are not told the value of $x_b$ (see \cite{pear00}).

\section{Proof of \cref{prop:subsystem_proc_props}}
\label{sec:subsystem_proc_proof}
%

\begin{proof}
The form of the rate matrix of a subsystem process
means that it is impossible to have a state transition
in which both $x_A$ and $x_B$ change simultaneously. Accordingly, we can write
$\bm{x} = (\bm{x}_A, \bm{x}_B)$, where $\bm{x}_A = 
(N^A, \vec{x}^A, \vec{\tau}^A, \vec{\nu}^A)$, and similarly for $\bm{x}_B$.
(Note that as shorthand, we do not
explicitly indicate in this decomposition that the value of $x_B$ doesn't
change when $x_A$ does and vice-versa.) 
In addition, for pedagogical clarity, in this appendix I express
rate matrices in the form $W_t^\nu(x' \rightarrow x)$ rather than $W_{x;x'}^\nu(t)$.
I modify the notation for survival probabilities similarly.

If subsystem $A$ undergoes a state transition 
$x^{A'} \rightarrow x^{A}$ at time $t$ while
system $B$ stays constant, the associated value of the rate matrix is $W_t(x^{A'}
 \rightarrow x^{A})$,
and similarly if it is subsystem $B$ that undergoes a transition. In addition,
for any three times $\tau < \tau' < \tau''$,
the survival probability of a subsystem not changing
state between $\tau$ and $\tau''$ equals the product of the
survival probability for going from $\tau$ to $\tau'$ and the survival probability for going from $\tau'$ to $\tau''$. These facts allow us to expand \cref{eq:trajweight}  
to evaluate the probability of a given trajectory $\bm{x}$ as
\begin{widetext}
\begin{align}
p(\bm{x}_A, \bm{x}_B \vert x^A_0, x^B_0) &=   \left(\prod_{i=1}^{N^A} S_{\tau^A_{i-1}}^{\tau^A_i}(x^A_{i-1}) W_{\tau^A_i}^{\nu^A_i}(x^A_{i-1}\!\shortrightarrow \! x^A_i) \right) \nonumber \\
& \qquad\qquad\qquad\qquad \times S_{\tau^A_{N^A}}^1(x^A_{N^A}) 
   \left(\prod_{j=1}^{N^B} S_{\tau^B_{j-1}}^{\tau^B_j}(x^B_{j-1}) W_{\tau^B_j}^{\nu^B_j}(x^B_{j-1}\!\shortrightarrow \! x^B_j) \right) 
             S_{\tau^B_{N^B}}^1(x^B_{N^B}) 
    \nonumber \\
& := p(\bm{x}^A \vert x^A_0) p(\bm{x}^B\vert x^B_0)
\label{eq:ctmc-rederiv}
\end{align}
\end{widetext}

If we marginalize both sides of \cref{eq:ctmc-rederiv} 
over all components of $\bm{x}_A$ except the starting and
ending values of $x^A$, do the same for $x_B$, and then translate to the notation of
\cref{prop:subsystem_proc_props}, we derive
\begin{align}
\map(a_1, b_1 \vert a_0, b_0) &= \map^A(a_1 \vert a_0)\map^B(b_1 \vert  b_0)
\end{align}
This formally establishes the intuitively
obvious fact, that the CTMC obeys \cref{prop:subsystem_proc_props}(\ref{enu:fracsystem1}).

Next, using \cref{eq:ctmc-ef}, we write the EF for the CTMC as
\begin{widetext}
\begin{align}
\WWW_\PP(p) &= \int p(x^A_0, x^B_0) p(\bm{x}^A \vert x^A_0) p(\bm{x}^B\vert x^B_0) \nonumber \\
& \times \bigg[
\sum_{i=1}^{N^A}  W_{\tau^A_i}^{\nu^A_i}(x^A_{i-1} \!\shortrightarrow x^A_{i})
\ln \frac{W_{\tau^A_i}^{\nu^A_i}(x^A_{i-1} \!\shortrightarrow x^A_{i})}
    {W_{\tau^A_i}^{\nu^A_i}(x^A_{i} \! \shortrightarrow x^A_{i-1})}
\;+\; \sum_{i=1}^{N^B}  W_{\tau^B_i}^{\nu^B_i}(x^B_{i-1} \!\shortrightarrow x^B_{i})
\ln \frac{W_{\tau^B_i}^{\nu^B_i}(x^B_{i-1} \!\shortrightarrow x^B_{i})}
    {W_{\tau^B_i}^{\nu^B_i}(x^B_{i} \! \shortrightarrow x^B_{i-1})}
\bigg]
 \; D\bm{x}^A D\bm{x}^B 		 \nonumber \\
&= \int p(x^A_0) p(\bm{x}^A \vert x^A_0) \sum_{i=1}^{N^A}  W_{\tau^A_i}^{\nu^A_i}(x^A_{i-1} \!\shortrightarrow x^A_{i})
\ln \frac{W_{\tau^A_i}^{\nu^A_i}(x^A_{i-1} \!\shortrightarrow x^A_{i})}
    {W_{\tau^A_i}^{\nu^A_i}(x^A_{i} \! \shortrightarrow x^A_{i-1})}  D\bm{x}^A	 \nonumber \\
&\qquad\qquad\qquad\qquad + \; \int p(x^B_0) p(\bm{x}^B \vert x^B_0) \sum_{i=1}^{N^B}  W_{\tau^B_i}^{\nu^B_i}(x^B_{i-1} \!\shortrightarrow x^B_{i})
\ln \frac{W_{\tau^B_i}^{\nu^B_i}(x^B_{i-1} \!\shortrightarrow x^B_{i})}
    {W_{\tau^B_i}^{\nu^B_i}(x^B_{i} \! \shortrightarrow x^B_{i-1})}  D\bm{x}^B	\nonumber \\
& := \WWW_A(p^A_0) + \WWW_B(p^B_0)
\label{eq:second_condition_holds}
\end{align}
\end{widetext}
This establishes the first part of \cref{prop:subsystem_proc_props}(\ref{enu:fracsystem2}).
Finally, note that $\WWW_{A}(p_A)$ is the EF that would be incurred 
for a CTMC over state space $X_A$ 
with rate matrices $W_{t}^{\nu}(a' \shortrightarrow a)$,
(i.e., if subsystem $B$ did not exist at all).<   
So
by the Second Law of Thermodynamics~\cite{seifert2012stochastic}, $\WWW_{A}(p_A) \ge 
S(p^A_0) - S(p_B^0)$, and similarly for $\WWW_{B}(p^B)$.
%
\end{proof}

\bibliographystyle{aipnum4-1}

\bibliography{../../../../BIB/refs.main.1.BIB.DIR}



\newcommand{\arXiv}[2]{\href{http://arxiv.org/abs/#1}{arXiv:#1 #2}}

\end{document}